%% file: paper_journal_v2_arxiv.tex
\documentclass[journal,11pt,onecolumn,draftclsnofoot]{IEEEtran}

\usepackage{cite}
\usepackage{graphicx,color,epsfig,rotating}
\usepackage{amsfonts,amsmath,amssymb,bbm}
\usepackage{setspace}
\usepackage{algorithm}
\usepackage[algo2e]{algorithm2e} 
\usepackage[compatibility=false]{caption}
\usepackage{mdwtab}
\usepackage{placeins}
\usepackage{psfrag}
\usepackage[utf8]{inputenc}
\usepackage{multirow}
\usepackage{stfloats}
\usepackage{tabularx} 
\usepackage{booktabs} 
\usepackage{url}
\usepackage{bm}
\usepackage{pstricks}
\usepackage{xspace}
\usepackage{geometry}
\input{macros}

\newcommand{\bl}[1]{{\color{blue} #1}}

\newcommand{\User}[1]{User~$#1$}

\newcommand{\Relay}[1]{Relay~$#1$}
\let \lef \left 
\let \rig \right

\SetKwInput{KwData}{Input}
\SetKwInput{KwResult}{Output}
\allowdisplaybreaks 
\graphicspath{{./images/}}

\begin{document}

\title{Fundamental Limits of Hierarchical Secure Aggregation with Cyclic User Association}

\input{author_journal.tex}

\maketitle

\begin{abstract} 
Secure aggregation is motivated by federated learning (FL) where a cloud server aims to compute an {aggregated} model (i.e., weights of deep neural networks) of the locally-trained models of numerous clients {through an iterative communication process}, while adhering to  data security requirements. 
Hierarchical secure aggregation (HSA)
extends this concept to a three-layer hierarchical network, where clustered users communicate with the server through an intermediate layer of relays. In HSA, beyond conventional server security, relay security is also enforced to ensure that the relays remain oblivious to the users' inputs (an abstraction of the local models in FL).
{Existing studies on HSA that jointly consider communication and secret key generation efficiency typically assume that each user is associated with only one relay,
limiting opportunities  for coding across inter-cluster users to achieve efficient communication and key generation.} In this paper, we consider HSA with a cyclic  association pattern where each user is  connected to $B$  consecutive relays in a  wrap-around manner.  
We propose an efficient aggregation scheme which includes a message design for the inputs inspired by gradient coding-a well-known technique for efficient communication in distributed computing-along with a highly non-trivial security key design.
We also derive novel converse bounds on the minimum achievable communication and key rates using information-theoretic arguments. Specifically, we characterize the optimal rate region for $B\le K-1$ ($K$ denotes the total number), as a result of the matching converse and achievable schemes, which shows that to securely compute one symbol of the desired input sum, each user needs to send at least $1$ symbol to the associated relays, each relay needs to send at least $1/B$ symbols to the server, each user needs to hold at least $1/B$  secret key symbols, and all users need to collectively hold at least $\max\{1, K/B-1\}$ independent key symbols. 
Our results reveal a fundamental inverse trade-off between 
the association number  $B$ and the communication and key rates. When $B=K$, we present a scheme that achieves the optimal communication and source key rates while attaining a near-optimal individual key rate.
\end{abstract}


\section{Introduction}
\label{sec: intro}
\subsection{Secure \Agg in Federated Learning}
Federated learning (FL) is a popular machine learning (ML) framework which utilizes datasets distributed across many users to train a centralized ML model such as deep neural networks~\cite{mcmahan2017communication,konecny2016federated,kairouz2021advances,rieke2020future}.
It finds broad applications such as improving query suggestion accuracy in Google virtual keyboard~\cite{yang2018applied}, on-device speech processing for Amazon Alexa \cite{chen2022self}, and collaborative medical research \cite{pfitzner2021federated}. Because the local datasets reside on each user's own device and are not shared with the parameter server or other users, FL possesses enhanced security features.  Taking the pioneering {FedAvg} algorithm~\cite{mcmahan2017communication} as an example, the  training process of FL alternates between a local training phase and a global \agg phase. In the local training phase, each user, independently, performs  multiple  stochastic gradient descent (SGD)-based optimization steps  on its own dataset. In the global \agg phase, the users upload their locally-trained ML models to a parameter/aggregation server which then  computes an averaged model and sends it back to the users. A new round of local training  will be performed by the users using the averaged model as an initialization point. The local training and the aggregation phases alternate until the global loss function converges.

Although the users' datasets are not directly  shared, FL still shows vulnerability to several types of security attacks~\cite{bouacida2021vulnerabilities, geiping2020inverting,mothukuri2021survey}. For example,  through the so-called model inversion attack~\cite{geiping2020inverting} in an image classification task, a considerable amount of information of the users' local datasets (\ie, images) can be inferred by the parameter server by solely observing the exchanged models. Therefore, additional security mechanisms  are  needed which has stimulated the study of secure \agg (SA) in FL~\cite{bonawitz2016practical, bonawitz2017practical, 9834981} where cryptographic techniques are used  to achieve (mostly  computational) security~\cite{bonawitz2016practical, bonawitz2017practical, 9834981,  wei2020federated,hu2020personalized,zhao2020local,andrew2021differentially,yemini2023robust,po2024differential,so2021turbo,kadhe2020fastsecagg, elkordy2022heterosag, liu2022efficient,jahani2023swiftagg+
}. 
In particular, Bonawitz \etal~\cite{bonawitz2017practical} proposed a secure \agg scheme which relies on pairwise random seed agreement between users to generate zero-sum keys to hide each \indiv user's local model from the server. 
Shamir's secret sharing~\cite{shamir1979share} is also used in~\cite{bonawitz2017practical} for security key recovery in cases of user dropouts. 
So \etal \cite{so2021turbo}  proposed an efficient secure \agg protocol which reduces the quadratic \comm overhead incurred by pairwise random seed  
agreement in~\cite{bonawitz2017practical}.
Differential privacy (DP)~\cite{wei2020federated,hu2020personalized,zhao2020local,andrew2021differentially,yemini2023robust} has also been investigated where small perturbation noises are added to protect the local models. A natural trade-off between model accuracy/convergence speed and protection level (\ie, the strength of noise relative to the local models) exists~\cite{wei2020federated}.
It should be noted that neither random seed agreement-based approaches nor DP-based approaches achieve perfect security due to Shannon's one-time pad encryption theorem~\cite{shannon1949communication} which states that, in order to fully protect a message with $L$ \indep symbols (from some finite field), the security key must be uniformly distributed and contains at least $L$ symbols. 

\subsection{\ITic Secure \Agg}
\Itic \secagg aims to achieve perfect security through the lens of zero mutual \info and has evolved into a standalone but theoretically interesting problem. \Ip, the locally trained models of the users are modeled as \iid random variables referred to as \emph{inputs} and denoted as $W_1,\cdots, W_K $ ($K$ being the total number of users). The \agg server wishes to compute the input sum $W_1+\cdots+W_K$ under the security constraint that it should not learn anything about the input set $W_1,\cdots , W_K$ beyond the input sum, which translates to the zero mutual information condition:
\be 
\label{eq:server security I=0, Intro}
I(\{X_k\}_{k=1}^K; \{W_k\}_{k=1}^K  |W_1+\cdots +W_K)=0
\ee
where $X_k$ denotes the message sent to the server by User $k$. Note that random seed-based security schemes, which rely on pseudorandomness, usually do not satisfy (\ref{eq:server security I=0, Intro}). \Itic \secagg usually focus on a \emph{two-fold} optimization of both the upload \comm rate $R_X \eqdef \sum_{k=1}^K{H(X_k)}/H(W_1)$ and the minimum security key sizes necessary to achieve input security. The \comm rate  has been widely adopted as  a performance metric by the FL community~\cite{bonawitz2016practical, bonawitz2017practical, 9834981,  wei2020federated,hu2020personalized,zhao2020local,andrew2021differentially,yemini2023robust,po2024differential,so2021turbo,kadhe2020fastsecagg, elkordy2022heterosag, liu2022efficient,jahani2023swiftagg+
} although the latter aspect of optimizing randomness consumption in key generation has been largely circumvented in these literature  due to the use of pseudorandomness or independently generated keys across the users. As already mentioned, pseudorandomness cannot achieve true security. The use of \indep keys eliminates the need for coordination among users in the key generation process, but unavoidably entails higher \comm overhead because the cancellation of the keys (and thus the recovery of the input sum) at the server becomes more challenging. For example, in the SwiftAgg+ algorithm of Jahani-Nezhad \etal~\cite{jahani2023swiftagg+}, inter-user \comm is needed on top of user-to-server \comm in order to fully cancel the $K$ \indep keys of the users. In contrast, a simple one-shot summation of the $X_k$'s  gets rid of the unknown keys at the server in~\cite{zhao2023secure} etc. due  to the use of \emph{correlated} security keys across the users. The use of correlated keys requires the aid of a trusted third-party entity which is responsible for generating and distributing the keys to \indiv users. Because the key \distn process also incurs extra \comm overhead, optimizing the key sizes is also an important consideration in \itic  secure \agg.

So far, \itic  secure \agg has been studied under a multitude of constraints such as user dropout and collusion resilience~\cite{9834981, so2022lightsecagg,jahani2022swiftagg,jahani2023swiftagg+}, groupwise keys~\cite{zhao2023secure,wan2024information,wan2024capacity}, user selection~\cite{zhao2022mds, zhao2023optimal},  weak security~\cite{li2023weakly,li2023weaklysecuresummationcolluding}, oblivious server~\cite{sun2023secure}, vector secure \agg~\cite{yuan2025vector}, and \hie secure \agg (HSA)~\cite{zhang2024optimal, 10806947,11154963, lu2024capacity}.
\Ip, Zhao \& Sun~\cite{9834981} presented an \itic formulation of  the secure \agg problem in a star-shaped network with the consideration of both user dropout and server collusion. To cope with potentially delayed \msg upload from the users, a two-round \comm protocol is used where in the first round masked inputs are uploaded and in the second round, additional masks are uploaded to cancel out the masks of the first round so that the input sum can be recovered. The optimal \comm rate region, consisting of all feasible rates over the two rounds, was characterized as $(R_1\ge 1,R_2\ge \frac{1}{U-T})$ where $U$ is the minimum number of surviving users and $T$ is the maximum number of colluding users.
Secure \agg schemes with improved key storage and \distn overhead were studied in~\cite{ so2022lightsecagg,so2021turbo}.
Groupwise keys were studied in~\cite{zhao2023secure,wan2024information,wan2024capacity} where each subset of $G$ users shares a common security key and the keys of \diff subsets are \indep. Groupwise keys enable the use of key agreement protocols for key generation and can eliminate the need for a trusted  third-party entity. Secure \agg with user selection was studied in~\cite{zhao2022mds, zhao2023optimal} where instead of computing the sum of inputs of all users, the server only needs to recover the sum of inputs of $U<K$ arbitrarily chosen users. For FL systems  with a very large number participating clients, user selection is necessary to reduce \agg latency and boost training speed~\cite{mcmahan2017communication}.
To capture systems with \het security \reqs, weak security was studied in~\cite{li2023weakly} where instead of protecting all $K$ inputs (see eq. (\ref{eq:server security I=0, Intro})),
only a selected subsets of inputs need to be protected against a selected subsets of colluding users. \Secagg with an oblivious server was studied in~\cite{sun2023secure}, where the server does not need to recover the input sum but instead acts merely as a \comm  facilitator, assisting each user in obtaining the input sum. In addition, vector secure \agg~\cite{yuan2025vector} extends to a more general scenario where the server aims to compute several \indep linear combinations of the inputs while being  prevented from learning another set of linear combinations.

\subsection{Secure \Agg in \Hie Networks}
To understand the fundamental impact of network topology on the \comm and key efficiency  of secure \agg, \hie \secagg (HSA)~\cite{zhang2024optimal} was  studied in a three-layer network, where clustered users connect to the \agg server through an  intermediate layer of relays. In this setup, two types of security constraints are enforced. First, \emph{\ssec} requires that the \agg server learns nothing about the input set $\{W_1,\cdots,W_K\}$ beyond  their summation. Second, \emph{\rsec} requires that each relay remains oblivious to the inputs, \ie, they should infer absolutely nothing about the input set. Given that the server and each relay can \resp collude with at most $T$ users, the optimal rate region, which consists of \comm rates over the two network layers and the secret key rates, has been characterized. A core insight from~\cite{zhang2024optimal} is that the mixing (\ie, summation) of the users' uploaded \msgs at the relays makes it harder  for the server to infer  the inputs which results in a smaller source key rate than in the original star network considered by Zhao \& Sun~\cite{9834981}.
HSA has also  been studied in a similar network architecture but with different \comm protocols and security models~\cite{egger2023private,11154963, lu2024capacity}. Specifically, these works rely on inter-relay \comm and focuses solely on \comm rate, which is \diff from~\cite{zhang2024optimal} which requires no \comm among relays and studies both \comm and key rates. 

One notable limitation of the network model considered in~\cite{zhang2024optimal} is that each user is associated with \emph{only one} relay, which may not be general in practice.
This simple single-association pattern also limits  the coding opportunities across relays, restricting the potential for more efficient \comm and secret key reuse among the outgoing links of each user. 
In this paper, we study HSA with a more general cyclic user-relay association pattern, as illustrated in Fig.~\ref{fig: system model}. The system consists of $K$ users and $K$ relays where each user is associated with $B$ consecutive relays in a cyclic wrap-around manner.\footnote{{Besides its theoretical interests in the context of \secagg, this network architecture is widely seen in many practically-motivated research literature such as the circular Wyner model for interference networks~\cite{7762234}, multi-access coded caching~\cite{10318146, peter2024multi}, and distributed computation\cite{namboodiri2025fundamentallimitsdistributedcomputing} etc.}} The \agg process consists of  two hops. In the first hop, each user sends a \msg to each of its associated relays. In  the second hop, upon receiving the \msgs from its associated users, each relay generates  and sends a \msg to the server. The server aims to recover the sum of inputs of all users subject to server and \rsec constraints whose definitions are consistent with that of~\cite{zhang2024optimal}.
We aim to find the optimal rate region $\Rc^*$, which is defined as the closure of all achievable tuples $(\rx,\ry, \rz,\rzsigma)$ where $\rx, \ry, \rz$ and $\rzsigma$ denotes \resp the per-user upload \comm rate, per-relay upload \comm rate, the \indiv key rate, and the source key rate (\ie, the total number of \indep key symbols  contained in all users' \indiv keys). 
Our major finding is that there exists a fundamental trade-off between the association number $B$ and the \comm and key rates: the larger $B$ is, the smaller the optimal \comm and key rates are. \Msp, when $B\le K-1$, we characterize the optimal rate region which  shows that \emph{to securely compute one symbol of the desired input sum, each user needs to send at least $\rx=1$ symbol  to the associated relays, each relay needs to send at $\ry=1/B$ symbols to the server, each user needs to hold at least $\rz=1/B$ independent key symbols, and all users need to collectively hold at least $\rzsigma=\max\{1,K/B-1\}$ key symbols.} This contains an achievable scheme and a novel converse proof that matches the scheme.
Compared with single-association HSA~\cite{zhang2024optimal} whose optimal rate region is
$ \Rc^*_{\rm single} =\{ \rx\ge 1, \ry\ge 1, \rz\ge 1, \rzsigma \ge K-1  \}$\footnote{This expression is derived by setting  the cluster size in~\cite{zhang2024optimal} as $V=1$ so that the number of users is equal to the number of relays, and the collusion level as $T=0$ since we do not consider user collusion in this paper.}, we observe the following: First, the result of~\cite{zhang2024optimal} can be recovered from this paper by setting $B=1$. Indeed,  the single-association HSA is partly a special case of the multiple-association HSA considered in this paper. Second, it can be seen that all $\ry$, $\rz$ and $\rzsigma$ are reduced by $B$ (or more than $B$) times. The reduction in $\ry$ is primarily due to the coding opportunities enabled by the \distn the symbols of the same input across multiple relays. Moreover, the reduction in $\rz$ is due to the \emph{reuse}  of the same \indiv key among the $B$ outgoing \comm links from each user. This causes no \info leakage because the relays do not talk to each other (by assumption) and thus key reuse will not expose the input symbols being protected.
When $B=K$, \ie, each user is associated with all relays, we present a scheme that achieves the optimal \comm and source key rates, as well as a near-optimal \indiv key rate. Moreover, we show that the \rts \comm rate cannot be reduced below $\ry\ge 1/(K-1)$ and, using the example of $B=K=2$, that the \indiv  key rate cannot be reduced below $\rz \ge 1/K$.

\subsection{Summary of Contributions}
\label{subsec: summary of contributions}
The major contributions of this paper are summarized \af:
\begin{enumerate}
    \item[\textbullet] We present an \itic formulation of HSA with cyclic wrap-around \assocn with the consideration of both server and relay security. We study both the \comm and key generation performance under a two-hop \comm protocol. 
    For an important regime $B\le K-1$, the optimal rate region is fully characterized and a fundamental trade-off between the \assocn number $B$ and the \comm and key rates is identified. 

    \item[\textbullet] We propose a secure \agg scheme which consists of a linear key generation scheme and a \comm scheme for input \agg. Given the source key, the \indiv keys of the users are generated as linear combinations of the  source key symbols.
    The proposed input \agg scheme efficiently exploits the inter-relay coding opportunities, \ie, coding across \diff inputs and the symbols of each input, to reduce the \rts \comm rate. The linear coefficients for the \indiv keys are then designed accordingly such that the server and relay security constraints can be satisfied.
    \item[\textbullet] We present a novel entropic converse proof and derive lower bounds on the achievable \comm and key rates for any $B\le K$. \Ip, we show that $\rx \ge 1, \ry\ge \max\{1/B, 1/(K-1)\}, \rz \ge 1/B$ and $\rzsigma \ge \max\{1, K/B-1\}$ for any $B\le K$.  
    When $B\le K-1$, the derived bounds match the achievable rates so that the optimal rate region can be established. When $B=K$, the scheme for $B=K-1$ can be applied by letting each user send an empty \msg over the last outgoing link (equivalent to disabling the link) so that each user effectively connects to $K-1$ relays. This results in the achievable rate  tuple of $(\rx=1, \ry=1/(K-1), \rz=1/(K-1), \rzsigma=1)$. Comparing with the converse bounds, we see that despite $\rz$ is within a factor of at most $1+1/(K-1)\le 2$ to its optimum, the achievable rates $\rx, \ry$ and $\rzsigma$ all reach  their individually minimum. As an intriguing discussion, we prove the impossibility of $(\rx=1,\ry=1,\rz=1/2, \rzsigma=1)$ using the example of $B=K=2$, which sheds light on understanding the optimal  solution for general $K$ where $B=K$. 
\end{enumerate}

\subsection{Related Work}
\label{subsec: related work}
Several works~\cite{zhang2024optimal, 11154963, lu2024capacity} are relevant to  this paper. The differences are clarified \af. As mentioned earlier, our previous work~\cite{zhang2024optimal} on HSA considered a \diff network architecture where each user is associated with only one relay, limiting the coding opportunities across relays as each input appears at only one relay.
{Egger \etal~\cite{11154963} studied the \comm efficiency of \secagg in two \hie  network settings. In the first setting, each user is connected to multiple base stations (BSs) which are then connected to the federator.  A sequential inter-BS \comm is necessary for input sum recovery due an extra secret key \agg phase following the initial masked input upload phase.
To get rid of the inter-BS \comm,  he second setting incorporates an additional layer of relays between the BSs and the federator.}
The work of Lu \etal~\cite{lu2024capacity} considered a network where each user is connected to all relays. For both~\cite{11154963} and~\cite{lu2024capacity}, inter-relay \comm is needed in order to cancel the \indeply generated keys at the users. The sequential key \agg among relays in ~\cite{11154963} incurs extra delay for input sum recovery, while the inter-relay \comm overhead is completely ignored in~\cite{lu2024capacity}.
In contrast, our work relies on  correlated key generation among users which enables one-shot input sum recovery and eliminates the need for a dedicated key \agg phase or inter-relay \comm.

\emph{Paper Organization}. The remainder of this paper is organized \af. The problem formulation is presented in Section~\ref{sec: problem description}. Section~\ref{sec:main result} gives the main result and its implications of this paper. In Section~\ref{sec: examples}, we provide an example to illustrate the basic design ideas. The general achievable schemes are presented in Section~\ref{sec: general scheme}. The converse proof is given in the following Section~\ref{sec: converse}. An insightful discussion regarding the optimal scheme at $B=K$ is provided in Section~\ref{sec:discussion}.
Finally, we conclude  the paper in Section~\ref{sec:conclusion} with possible future work.

\emph{Notation Convention}. Throughout the paper, the following notation is used: $[m:n] \eqdef  \{m,m+1,\cdots,n\}$ if $m\le n$ and $[m:n]=\emptyset$ if $m>n$. $[1:n]$ is written as $[n]$ for brevity. 
{$m \mid n$ means $m$ divides $n$.}
Calligraphic letters (\eg, $\Ac,\Bc$) denote sets. $\Ac \backslash \Bc \eqdef \{x\in \Ac: x\notin \Bc\}$.
Given a set of random variables $X_1,\cdots,X_m$, denote $ X_{\Sc}\eqdef \{X_i\}_{i\in \Sc}$ and $X_{1:m}\eqdef \{X_1,\cdots,X_m\}$.
 $\bm{0}_{m\times n}$ denotes the $m$-by-$n$ zero matrix and $\mbf{I}_m$ denotes the $m$-by-$m$ identity matrix.
For a matrix $\Am$, let $\Am_{i,:}$, $\Am_{:,j}$ and $\Am_{i,j}$ denote the \ith row, \jth column and the $(i,j)^{\rm th}$ entry of $\Am$ \resp. Moreover, $\Am_{i:j,:}$  denotes the submatrix consisting of the $i^{\rm th}$ to the \jth 
rows of $\Am$, and  $\Am_{:,i:j}$ denotes the submatrix consisting of the \ith to the \jth columns of $\Am$. $\spann(\Am)$ and $\nulll(\Am)$ represent the column span and null space of $\Am$ \resp. 
Denote  $[\av_1; \cdots; \av_m]\eqdef
[\av_1^T, \cdots,\av_n^T]^T \in \mbb{R}^{m\times n }  $  as the row stack of the row vectors $\av_1, \cdots, \av_m  \in \mbb{R}^{1\times n} $. $\modular(a,b)$ denotes the remainder of $a$ divided by $b$.

\section{Problem Description}
\label{sec: problem description}
\begin{figure}[t]
    \centering    \includegraphics[width=.5\textwidth]{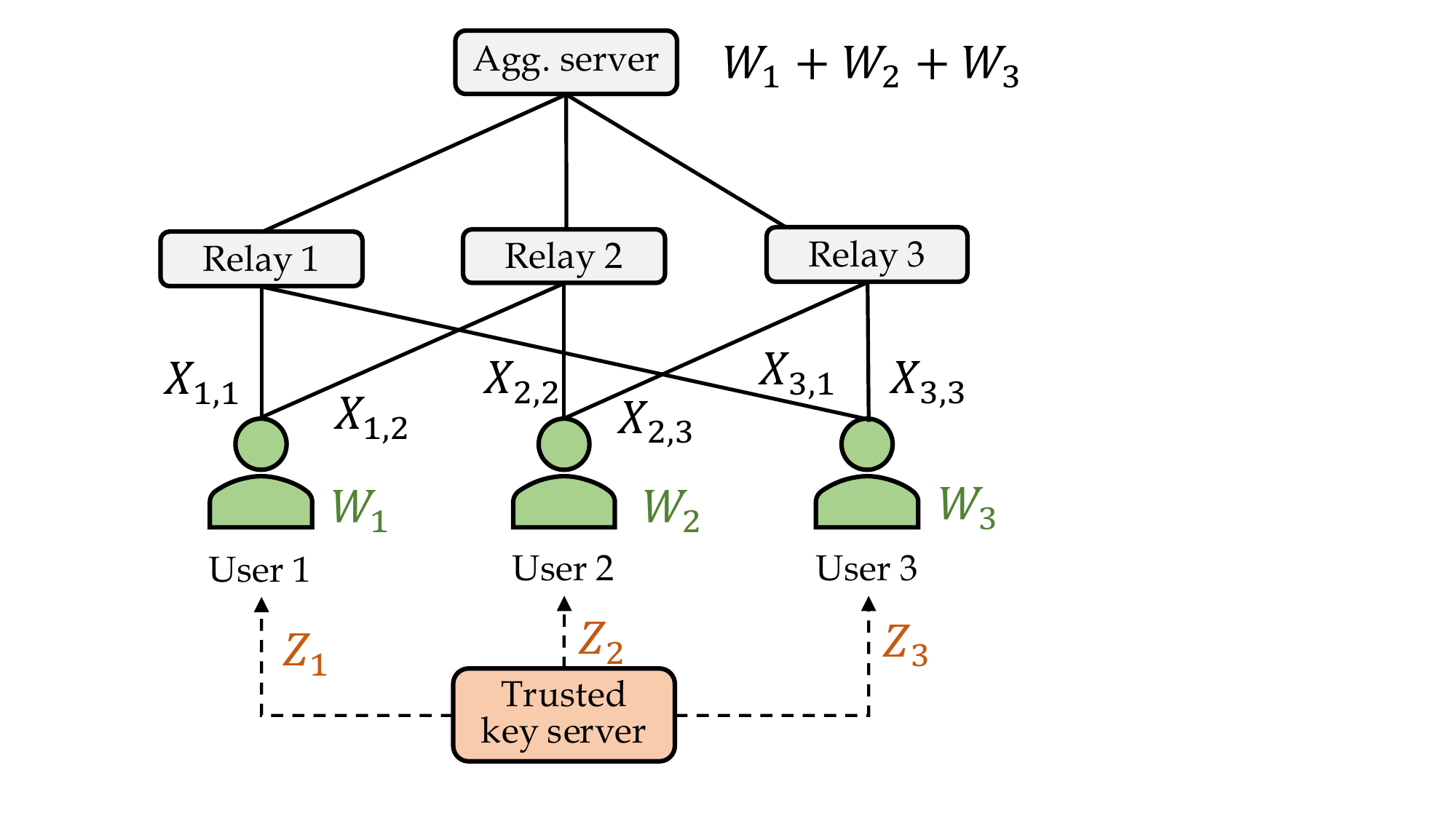}
    \caption{\small HSA with cyclic wrap-around user-relay association {with $3$ users and $3$ relays}.}
    \label{fig: system model}
\end{figure}

We study the \secagg problem in a three-layer \hie network which consists of an \agg server, $K$ relays and $K$ users as shown in Fig. \ref{fig: system model}. Each user is associated with $B$ relays in a cyclic wrap-around manner as shown in Fig.~\ref{fig: system model}. 
This network architecture is widely seen in the  \info theory literature such as the circular Wyner model for interference networks~\cite{7762234} where each receiver
is only  interfered by the neighboring transmitters, and multiple-access coded caching~\cite{10318146, peter2024multi,serbetci2019multi} where
each user has access to a number of consecutive cache memories.
Let $\Bc_i$ and $\Uc_i$ denote \resp the set of relays that the \ith user connects to and the set of users  associated with the \ith relay.
Under the cyclic association, we have 
\be 
\label{eq: def Bc_i, problem formulation}
\Bc_i  =\left[i:\min\{K,i+B-1  \}\right]\cup
[1 : B-K+i-1]
\ee
and 
\be 
\label{eq: def Uc_i, problem formulation}
\Uc_i  = [\max \{1, i-B+1\}:i] \cup [K-B+i+1:K],
\ee 
where $|\Bc_i|=|\Uc_i|=B, \forall i\in [K]$. Each User $k$ possesses an {input} $W_k$ (an abstraction of local gradients/models in FL) consisting of $L$ symbols from some finite field $\mbb{F}_q$. The inputs of  the users are assumed to be \iid uniform\footnote{{The
\iid uniform assumption on the inputs  is only
necessary for establishing the optimality of the proposed scheme. In fact,  the  proposed scheme works  for \tit{\arbily} distributed and correlated inputs among users.}}, \ie, 
\begin{align}
\label{eq: H(Wk)=L, indep of inputs}
 & H(W_k)  =L \; (\trm{in $q$-ary units}),\\
 & H(W_1,\cdots,W_K)  =\sum_{k=1}^KH(W_k).
\end{align}
To achieve security, each user  is also equipped with a secret key $Z_k$ (referred to as an \emph{\indiv key}), not known to the relays and the server, which consists of  $L_Z$ symbols from $\mbb{F}_q$ in order to protect the input. The \indiv keys $Z_1, \cdots,Z_K$ can be arbitrarily correlated and are generated from a \emph{source key $\zsigma$} containing $\lzsigma$ symbols, \ie, 
\be
\label{eq: indiv key gen based on source key}
H(Z_1,\cdots,Z_K|\zsigma)=0.
\ee 
The source key represents the number of \indep symbols held collectively by all the users. 
The \indiv keys are generated and distributed by a trusted third-party entity which does not leak information to the server and relays.
Moreover, the keys $Z_1,\cdots,Z_K$ are \indep of the inputs, \ie,
\be 
\label{eq: key input independence}
H\left(W_{1:K}, Z_{1:K}\right) =\sum_{k=1}^KH(W_k)+ H\left(Z_{1:K}\right).
\ee
The server wishes to compute the sum of all users' inputs $\sum_{k=1}^KW_k$ and should be prevented from learning anything about the input set $W_{1:K}$ beyond their summation. Relays should remain oblivious, that is, each relay should not infer any information about $W_{1:K}$ despite it has access to some messages uploaded by the associated users.

A two-hop \comm protocol is used. In the first hop,  each User $k\in[K]$ sends a  \msg $X_{k,i}$ to every associated relay $i\in  \Bc_k$. Let $L_{X,k,i}$ denote the number of symbols contained in $X_{k,i}$ and let
$L_X \eqdef \sum_{i\in \Bc_k} L_{X,k,i} $
denote the total number of symbols sent by each User $k\in [K]$\footnote{Due to symmetry, we assume that the total number of symbols sent by each user is identical across all users.}. In the second hop, each relay $i$ generates a \msg $Y_i$ which contains $L_{Y,i}$ symbols based on the received \msgs  $\{X_{k,i}\}_{k\in \Uc_i}$ from the associated users and sends $Y_i$ to the server.
The  above \msg generation process can be expressed as
\begin{align}
& H\left(\{X_{k,i}\}_{i\in \Bc_k}|W_k,Z_k  \right)=0,\; \forall k\in[K] \label{eq: H((X_ki)_i|W_k,Z_k)=0}\\
& H\left(Y_i| \{X_{k,i}\}_{k\in \Uc_i} \right)=0,\; \forall i\in[K]\label{eq: H(Yi|(X_{ki})_k)=0}
\end{align}
The server should be able to recover the desired input sum from the relay's \msgs, \ie, 
\be 
\label{eq: recovery constraint}
H\bigg(\sum_{k=1}^K W_k\Big|Y_{1:K}\bigg)=0.
\ee

\emph{Security model.}
Security refers to two constraints: \emph{1) server security:} the server should not infer anything about the input set $W_{1:K}$ beyond the knowledge of the desired input sum  $\sum_{k=1}^KW_k$, and \emph{2) relay security:} each relay should not infer anything about $W_{1:K}$, \ie, the relays should remain oblivious to the inputs. More specifically, server security translates to the statistical  independence between the \msgs  $Y_{1:K}$ and the inputs $W_{1:K}$, \ie, 
\begin{align}
\label{eq: server security constraint}
{\mrm{[Server\;\, \secty]}} \quad 
I\left(Y_{1:K};W_{1:K}\Big| \sum_{k=1}^KW_k  \right)=0,
\end{align}
where the conditioning on the input sum $\sum_{k=1}^K W_k$ is due to (\ref{eq: recovery constraint}).
Relay security translates  to the independence between the received \msgs by each relay and the inputs, \ie, 
\begin{align}
\label{eq: relay security constraint}
{\mrm{[Relay\;\, \secty]}} \quad 
I\left( \{X_{k,i} \}_{k\in \Uc_i};W_{1:K}\right)=0,\; \forall i\in [K]
\end{align}
where recall that  $\Uc_i \subseteq [K]  $ denotes the set of users associated with Relay $i$.

\emph{Performance metrics.} We study both the \comm and key generation aspects of the HSA problem. \Ip, the \utr \comm rate $\rx$ and \rts \comm rate $\ry$ \resp represent how many symbols that the \msgs $\{X_{k,i}\}_{i\in\Bc_k}$ sent by each User $k$ contains, and how many symbols on average that the \msg $Y_i$ sent by each Relay $i$ contains (normalized by the input size), \ie, 
\begin{align}
\label{eq: def comm rates}
\rx \eqdef \frac{L_X}{L}=
\frac{\sum_{i\in\Bc_k} L_{X,k,i}}{L},\;
\ry \eqdef \frac{1}{KL}{\sum_{k=1}^K L_{Y,k} }.
\end{align}
$\rx$ and $\ry$ characterize the \comm efficiency of  the secure \agg scheme and smaller rates imply more efficient  \comm.
The \indiv key rate $\rz$ and source key rate $\rzsigma$ represent  {the number of symbols that} each \indiv key $Z_k$ and the source key $\zsigma$ contain (normalized by the input size), \ie,
\begin{align}
\label{eq: def key rates}
\rz \eqdef \frac{L_Z}{L},\; \rzsigma \eqdef \frac{\lzsigma}{L}.
\end{align}
$\rz$ and $\rzsigma$ characterize the key generation efficiency of the \secagg scheme. Smaller key rates imply smaller \comm overhead incurred by the key \distn process (see Remark~\ref{remark:key distn overhead} below). 
A rate tuple $(\rx, \ry,\rz,\rzsigma)$ is said to be \emph{achievable} if there exists a \secagg scheme---\ie, a specific construction of the \msgs  $\{X_{k,i}\}_{k\in [K], i\in \Bc_k }$, $\{Y_k\}_{k\in  [K]}$, and the keys $\zsigma,\{Z_k\}_{k\in  [K]}$---that satisfies the constraints in (\ref{eq: indiv key gen based on source key}), (\ref{eq: H((X_ki)_i|W_k,Z_k)=0}), and (\ref{eq: H(Yi|(X_{ki})_k)=0}), and that \emph{simultaneously} achieves the \comm rates $\rx, \ry$ and key rates $\rz,\rzsigma$. Meanwhile, the scheme must ensure that the recovery constraint (\ref{eq: recovery constraint}) and the security constraints (\ref{eq: server security constraint}) and (\ref{eq: relay security constraint}) are all satisfied. 
The optimal rate region $\Rc^*$ is defined as the closure of the set of all achievable rate tuples. In addition, given $B$, we also denote $\rxstar,\rystar,\rzstar$ and $\rzsigmastar$ as the minimum individually  achievable rates in $\Rc^*$ \resp.

\begin{remark}[Key Generation and \Distn]
\label{remark:key distn overhead}
\emph{It is assumed that the secret keys are generated and distributed by a trusted third-party entity. Therefore, the key \distn process also incurs extra \comm overhead. \Ip, if the trusted entity distributes the keys through \indiv links to each user, the (normalized) \comm overhead is equal to $K\rz$. If the keys are distributed through a broadcast channel, the \comm overhead is then equal to  $\rzsigma$. Therefore, smaller key rates directly translate  to smaller overhead in the key \distn process.{
Note that the purpose of the above overhead analysis is to practically motivate the \indiv and source key rates, not meant to indicate only one of them needs to be studied at a time.
}
}
\end{remark}

{
For ease of reference, the problem-specific notation is summarized in Table \ref{tab:notations}.
\begin{table}[ht]
  \centering
  \caption{Table of notation.}
  \label{tab:notations}
  \begin{tabular}{c | c}
    \hline
    $K$ & Number of users/relays \\
    \hline
    $B$ &  Number of relays each user associated with   \\
    \hline
    $W_k$ &  Input of \User{k}   \\
    \hline
    $Z_k$ &  Secret key of \User{k}\\ 
    \hline
    $\zsigma$ & Source key \\
    \hline
    $X_{k,i}$ & Message of \User{k} to \Relay{i}    \\
    \hline
    $Y_i$ & Message of  \Relay{i}\\
    \hline
    $\Bc_k$ & Set of relays associated with \User{k}   \\
    \hline
    $\Uc_i$ & Set of users associated with \Relay{i}   \\
    \hline
  \end{tabular}
\end{table}

}

\section{Main Result}
\label{sec:main result}
\begin{theorem}
\label{thm: main result}
For \hie \secagg with cyclic wrap-around \assocn, when  $B\le K-1$,  the optimal rate region is given by
\be
\label{eq: thm 1}
\Rc^* =  \left\{
\left(\rx, \ry, \rz, \rzsigma\right)\left|
\begin{array}{c} 
\rx  \ge 1 ,\\
 \ry \ge 1/B ,\\
\rz\ge 1/B ,   \\
\rzsigma \ge \max\left\{1, {K}/{B}  -1\right\}
\end{array}\right.\right\}.
\ee
 When $B=K$, the rate  tuple
 \be 
 \label{eq: thm 1, B=K ach rate tuple}
 \left( \rx=1, \ry= \frac{1}{K-1}, \rz= \frac{1}{K-1}, \rzsigma=1   \right)
 \ee 
 is achievable. In  {(\ref{eq: thm 1, B=K ach rate tuple})}, the rates $\rx, \ry$ and $\rzsigma$ all reach their individually minimal values except that $\rz=1/(K-1)$ is within a multiplicative factor of $2$ to its minimum. 
\end{theorem}

Theorem \ref{thm: main result} has characterized the optimal rate region  when $B\le K-1$. This consists of a novel key generation scheme and an input \agg scheme presented in Section~\ref{sec: general scheme}, and a matching converse proof in Section \ref{sec: converse}.
When $B=K$, \ie, each user is associated with all relays, the optimal scheme for the case of $B=K-1$ can be applied by sending an empty \msg to the last relay associated with each user. This yields an achievable rate tuple $(1, 1/(K-1), 1/(K-1),1)$ given in Theorem \ref{thm: main result}, where $\rx,\ry$ and $\rzsigma$  reach their minimal values while $\rz=1/(K-1)$ is within a multiplicative factor of $K/(K-1)\le 2$ to its minimum value\footnote{It will shown in Section~\ref{sec: converse} that $ \rz \ge 1/B,\forall B\in [K]$ so that the achievable \indiv key rate $1/(K-1)$ is at most twice of its minimum given that $K\ge 2$.}. In Section~\ref{sec:discussion}, we provide an insightful discussion on the infeasibility of $(1,1/(K-1),1/K,1)$ when $B=K$ using the minimal example of $B=K=2$. For general cases of $B=K$, the optimal rate region remains  unsolved.
The achievable rate region is illustrated in Fig.~\ref{fig: rates, main result}.
\begin{figure}[ht]
    \centering
    \includegraphics[width=0.5\textwidth]{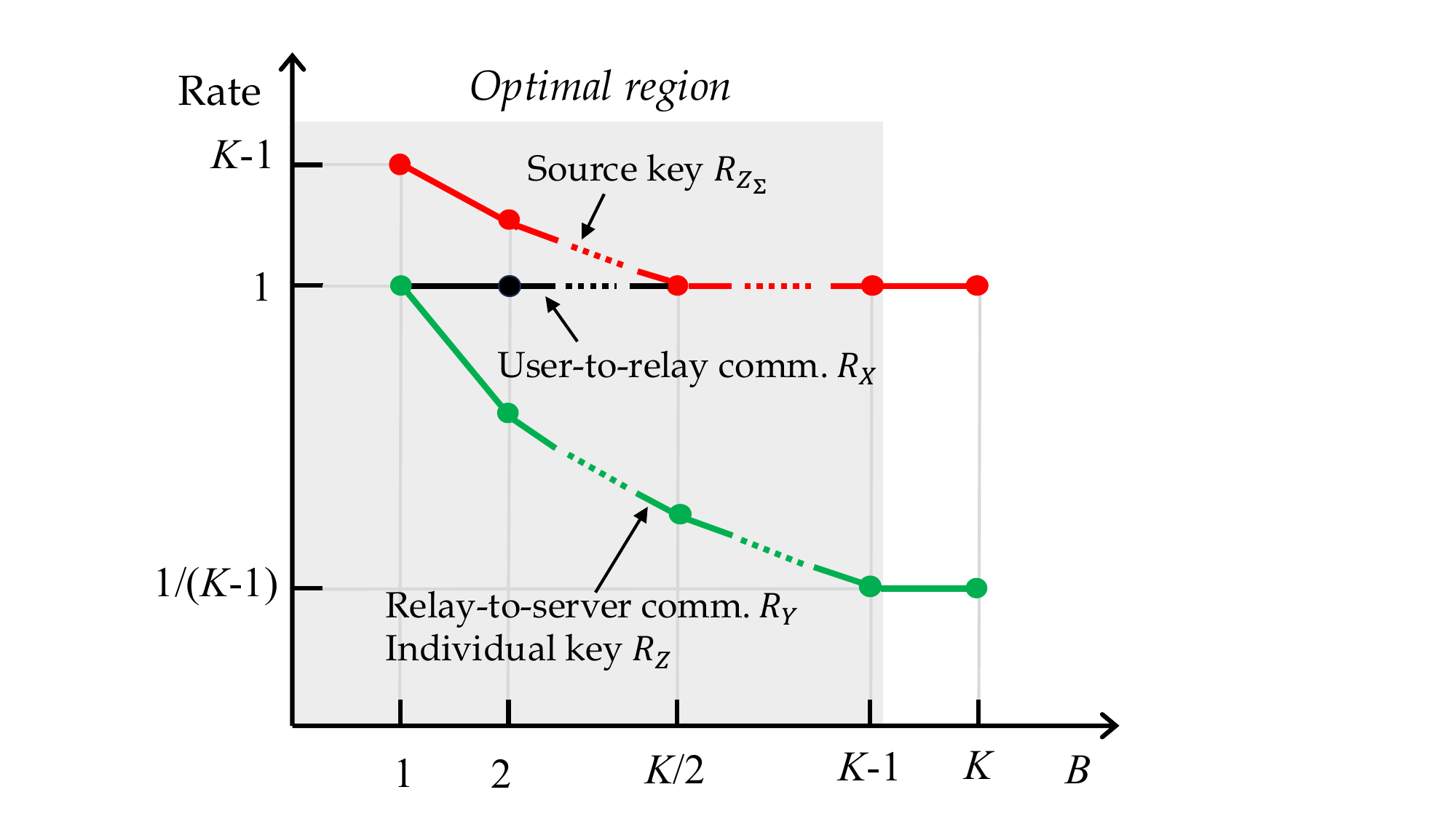}
    \caption{\small Achievable region for HSA with cyclic association. The optimal rate region (the shaded part) has been characterized when $B\le K-1$. Note that in the proposed scheme, $\ry $ and $\rz$ have identical values so they are both marked by the green curve.  }
    \label{fig: rates, main result}
\end{figure}

We highlight the implications of Theorem \ref{thm: main result} as follows:
\begin{enumerate}
\item \emph{\Comm Efficiency}: When $B\le K-1$,  
the optimal \comm rates are given by $\rystar=1/B$ and $\rxstar=1$ which is equivalent to an average upload rate of $1/B$ on each \utr link.  The per-link rates  on both hops are inversely proportional to $B$, revealing a fundamental trade-off between \comm efficiency and the association number $B$. Therefore, \Thm~\ref{thm: main result} suggests that for federated learning systems with limited link capacity, the overall \agg delay can be reduced by establishing parallel connection to a larger number of relays.

\item \emph{Key Efficiency}: Theorem \ref{thm: main result} also suggests that when $B\le K-1$, the optimal \indiv key rate $ \rzstar=1/B$ is inversely proportional to $B$. The optimal source key rate  $\rzsigmastar=\max\{1,K/B-1\}$ scales inversely with $B$ and is lower bounded by 1. It can be seen that the optimal key rates decrease as $B$ increases. This is because distributing  \comm over multiple relays creates an opportunity for key reuse across the $B$ outgoing \msgs sent by each user.
Because the relays do not  talk to each other, key reuse does not leak  \info about the input symbols contained in these \msgs.
It is then only necessary to ensure that the \diff input symbols sent by \diff users to the same relay are each protected by an \indep key.
Therefore, a larger $B$ not only reduces \comm rates, but also reduces the key rates, making the key \distn process more efficient.

\item \emph{Impact of Multiple \Assocn:}
HSA has been previously studied in the single-association setting~\cite{zhang2024optimal} where each user is associated with one relay. Fixing the total number of users to $K$ (\ie, each cluster contains only one user), the optimal rates were given by $\widetilde{R}_X^*=\widetilde{R}_Y^*= \widetilde{R}_Z^*=1$ and $\widetilde{R}_{Z_{\Sigma}}^*=K-1$, which can be recovered by choosing $B=1$ in eq. (\ref{eq: thm 1}). Therefore, our work can be viewed as an extension of~\cite{zhang2024optimal}. Theorem~\ref{thm: main result} reveals a  twofold trade-off between \comm, key efficiency, and the association number $B$. \Ip, suppose each user  can maintain association with an $\alpha$ portion of the relays (\ie, $B= \alpha K$), then the  \comm overhead incurred by the key \distn process is at most $\max\{
K\rzstar, \rzsigmastar
\}=1/\alpha$ which does not scale with $K$. In contrast, the scheme of \cite{zhang2024optimal} has an linear overhead of   $\max \{\widetilde{R}_{Z_{\Sigma}}^*,K\widetilde{R}_Z^*\}  =K$ which becomes impractical for systems with a large number of users.

\item \emph{Connection to Gradient Coding~\cite{ye2018communication}}:
A similar trade-off to $R_Y^*=1/B$ was previously identified in gradient coding (GC)~\cite{tandon2017gradient} by Ye and Abbe~\cite{ye2018communication}, albeit \emph{without} the security \reqs. The connection of \Thm~\ref{thm: main result} to~\cite{ye2018communication} is illustrated  \af. In GC, a user wishes to retrieve the sum of the gradients on $N$ datasets which are stored distributedly across $N$ computing nodes. Each node can store up to $d\le N$ datasets and thus can compute $d$ distinct gradients. GC aims to design the dataset placement and the retrieval scheme so that the download \comm rate $R_{\rm GC}$ can be minimized. Under the cyclic data placement where each computing node stores $d$ consecutive datasets in a wrap-around manner, Ye and Abbe identified the optimal rate as $R_{\rm GC}^*=1/d,\forall d\in[N]$. Setting $N=K,d=B$ and replacing
the datasets in GC with the inputs in HSA,
the user in GC with the server in HSA, the computing nodes in GC with the relays in HSA, and the dataset placement in GC with user uploads to the relays in HSA, an equivalence between GC and HSA in terms of optimizing $\ry$ can be established. Our contribution lies in showing that with the incorporation of security, this optimal inverse trade-off $1/B$ can be preserved except that when $B=K$, $\ry$ cannot be reduced below $1/(K-1)$ due to security.
\end{enumerate}

\section{Examples}
\label{sec: examples}
We first provide an example to highlight the ideas behind the proposed design. The general scheme is presented in Section \ref{sec: general scheme}.


\begin{example}
\label{example 1}
Consider $K=3$ users and relays where each user is connected to $B=2$ relays as shown in Fig. \ref{fig: system model}. Suppose each input consists of $L=2$ symbols in $\mathbb{F}_3$, \ie, $W_k=(W_k^{(1)},W_k^{(2)}),k\in[3]$. Let the source key be  $ \zsigma=(N_1,N_2)$ where
$N_1,N_2\in \mathbb{F}_3$ are two \iid uniform random variables. 
The \indiv keys of the users are chosen as
\be
\label{eq: indiv keys, example 1}
 Z_1= N_1,\;Z_2=N_2,\;Z_3=N_1+N_2.
 \ee 
 As a result,  the key rates are $\rzsigma=1,\rz =1/2$. 
The user-to-relay \msgs are given by
\begin{align}
\label{eq: X msgs, example 1}
& \trm{User 1:}\; \left\{
\begin{array}{l}
  X_{1,1} = -2W_1^{(1)}- Z_1,  \\
   X_{1,2}   =-\big(W_1^{(1)} + W_1^{(2)}\big) + Z_1.  
\end{array}\right.\\
&
\trm{User 2:}\; \left\{
\begin{array}{l}
X_{2,2}  =W_2^{(1)} - W_2^{(2)} + 2Z_2,  \\
X_{2,3}   =2W_2^{(1)} + Z_2. 
\end{array}\right.\\
& \trm{User 3:}\; \left\{
\begin{array}{l}
X_{3,3}  =W_3^{(1)} + W_3^{(2)} + Z_3,  \\
X_{3,1}   =W_3^{(2)} - W_3^{(1)} + 2Z_3. 
\end{array}\right.
\end{align}
It can be seen that each  \indiv key $Z_k$ is reused by the two \msgs $\{X_{k,i}\}_{i\in \Bc_k}$ sent by each User $k\in [3]$. 
Each relay then sends a \msg to the server which is a sum of the received \msgs from the users, \ie,
$Y_1 = X_{1,1} + X_{3,1},Y_2 = X_{1,2} + X_{2,2}$  and $Y_3 = X_{2,3} + X_{3,3}$. \Ip,
\begin{align}
\label{eq: Y msgs, example 1}
Y_1 & = -2W_1^{(1)}+ W_3^{(2)} - W_3^{(1)} + N_1+ 2N_2, \notag\\
Y_2 & = -\big(W_1^{(1)} + W_1^{(2)}\big) + W_2^{(1)} - W_2^{(2)} + N_1+ 2N_2, \notag\\
Y_3 & = 2W_2^{(1)}+ W_3^{(1)} + W_3^{(2)} + N_1+ 2N_2.
\end{align}
Since each of the above \msgs contains one symbol, the achieved rates are $\rx=1,\ry=1/2$.

\tbf{Input sum recovery.} To recover the input sum, the server computes $(Y_3-Y_1)/2$ to obtain $W_1^{(1)} + W_2^{(1)} +W_3^{(1)}$, \ie, sum of the first symbol of all inputs, and 
$(Y_1-2Y_2 + Y_3)/2$ to obtain $ W_1^{(2)} + W_2^{(2)} +W_3^{(2)}$. Then  $W_1+W_2+W_3$ can be recovered by concatenation.

\tbf{Proof of relay security.} From (\ref{eq: X msgs, example 1}), we note  that  each of the two \msgs received by any relay  is  protected by an \indep key variable. For example, for Relay 1, $X_{1,1}$ is protected by $-N_1$ and $X_{3,1}$ is protected by $2(N_1+N_2)$. Because $-N_1$ is \indep of $2(N_1+N_2)$, none of them can be eliminated and therefore Relay 1 infers nothing about the input symbols $W_1^{(1)}$ and $W_3^{(2)} - W_3^{(1)} $  contained in $X_{1,1}$ and $X_{3,1}$ \resp.
More formally, \rsec can be proved \af:
\begin{subequations}
\label{eq: proof of rs for relay 1, example 1}
\begin{align}
 & I(X_{1,1},X_{3,1}; W_1,W_2,W_3) \notag \\
 & \;= H(X_{1,1},X_{3,1}) - H(X_{1,1},X_{3,1}|W_1,W_2,W_3)\\
& \;=H\big(-2W_1^{(1)}- N_1, W_3^{(2)} - W_3^{(1)} + 2(N_1+N_2)\big )\notag \\
& \quad \;\; - H(-N_1, 2(N_1+N_2)|W_1,W_2,W_3   )\\
& \; \le 2 - H(-N_1, 2(N_1+N_2))\label{eq: step 0, proof of rs for relay 1, example}\\
& \; = 2  - H(N_1,N_2)\\
& =0, \label{eq: step 1, proof of rs for relay 1, example}
\end{align}
\end{subequations}
where (\ref{eq: step 0, proof of rs for relay 1, example}) is due to  the fact that the uniform distribution maximizes the joint entropy and the independence of the inputs from the keys. (\ref{eq: step 1, proof of rs for relay 1, example}) is due to the uniformity of the key variables.
Since mutual \info is non-negative, we have $I(X_{1,1},X_{3,1}; W_1,W_2,W_3)=0$, proving relay security (\ref{eq: relay security constraint}). The proof is similar for other relays.

\tbf{Proof of server security.} Intuitively, \ssec is guaranteed because each \rts \msg is protected by a  key variable and \emph{any key-canceling linear combination\footnote{A key-canceling linear combination $(\ell_1, \ell_2, \ell_3)$ is one for which the linear expression $\ell_1 Y_1 + \ell_2 Y_2 + \ell_3 Y_3$ results in complete elimination of the key variables, leaving only the underlying message content.} of $Y_1,Y_2$ and $Y_3$ reveals nothing beyond the desired input sum}. \Msp, by (\ref{eq: Y msgs, example 1}), any key-canceling vector must be in the form of  $\bm{\ell}=(\ell_1,\ell_2,-(\ell_1+\ell_2))$ where $\ell_1,\ell_2\in \mathbb{F}_3$. We have  $\sum_{k=1}^3\ell_iY_i  = -(2\ell_1+\ell_2)\sum_{k}W_k^{(1)} - \ell_2 \sum_{k}W_k^{(2)}$ which is a function of  $W_1+W_2+W_3$.
Therefore, the server learns nothing beyond what  it is supposed to know and \ssec is achieved.
Rigorously, we have 
\begin{subequations}
\label{eq: proof of ss, example}
\begin{align}
& I(Y_1,Y_2,Y_3; W_1,W_2,W_3| W_1+W_2+W_3) \notag\\
& \; = H(Y_1,Y_2,Y_3| W_1+W_2+W_3) \notag\\
& \quad\;\; -H(Y_1,Y_2,Y_3| W_1,W_2,W_3)\\
&\; = H(Y_1,Y_2,Y_3| W_1+W_2+W_3) \notag\\
& \quad\;\; -H(N_1 +2N_2| W_1,W_2,W_3)\\
& \;= H(Y_1,Y_2,Y_3| W_1+W_2+W_3) -H(N_1 +2N_2)\label{eq: step 0, proof of ss, example 1}\\
&\; = H(Y_1,Y_2,Y_3| W_1+W_2+W_3) -1\\
&\; = H(Y_1| W_1+W_2+W_3)  \notag \\
& \quad \;\;+ \underbrace{H(Y_2,Y_3| W_1+W_2+W_3,Y_1)}_{=0}  -1\label{eq: step 1, proof of ss, example 1}\\
& \; \le H(Y_1)-1\le0, \label{eq: step 2, proof of ss, example 1}
\end{align}
\end{subequations}
where (\ref{eq: step 0, proof of ss, example 1}) is due to the independence of the inputs and the key variables; (\ref{eq: step 1, proof of ss, example 1}) is because $Y_2,Y_3$ can be recovered from $W_1+W_2+W_3$ and $Y_1$ given the \msg design in (\ref{eq: Y msgs, example 1}). In particular, 
$Y_2 =Y_1+ \sum_kW_k^{(2)} - \sum_kW_k^{(1)}$, 
$Y_3 =Y_1+2\sum_kW_k^{(1)}$; (\ref{eq: step 2, proof of ss, example 1}) is because $Y_1$ contains one symbol and uniform distribution maximizes the entropy. As a result, \rsec is proved.

{
\begin{remark}[High-level Design Idea of Example \ref{example 1}]
\label{remark:high-level idea of example 1}
We briefly explain the general ideas behind the design of Example \ref{example 1}. We first determine the optimal \skr  which determines the minimum number  of required \iid key symbols ($N_1,N_2$ in this case). The  \indiv keys are then generated as linear combinations of the source keys symbols with correlation among them. We then choose the \coeffts for  both  the input and key components in the  \msgs \suth under a linear transform at the server, the desired input sum can be recovered while unintended \info about the inputs remains concealed.
\end{remark}
}

\tbf{Converse.} A  set of lower bounds $\rx\ge1, \ry \ge  1/2,  \rz\ge  1/2 $, and $\rzsigma \ge1$ can  also be derived (see Example~\ref{example 2} in Section~\ref{subsec: indiv key rate}). Because these  lower bounds coincide with the achievable rates, the optimality of the above example is  thereby established. The derivation of the converse bounds, which involves several technical lemmas and intermediate steps, is deferred to the converse section.
\hfill $\lozenge$
\end{example}

\section{General Scheme}
\label{sec: general scheme}
The general scheme consists two parts: a \msg \genn scheme and a key \genn scheme,  both of which are linear. 
We first focus on the case  of $B\le K-1$ to illustrate the design process. Then we look at the special case of $B=K$ where each user is connected to all relays. This ubiquitous connection makes it more challenging to achieve security and requires higher \comm and \indiv key rates than $1/B$ as for $B\le K-1$.

\subsection{Scheme for $B\le K-1$}
\label{subsec: scheme for B<=K-1}
Suppose each input contains $L=B$ symbols from some finite field $\mbb{F}_q$,\footnote{{The field size $q$ is assumed to be sufficiently large in order to ensure the existence of relevant matrices (\ie, $\Hm$ and $\bmLambda$ in (\ref{eq: indiv ley gen, general linear transform}) and (\ref{eq: def of Lambda})) over $\Fmbb_q$ with  certain rank constraints in Section \ref{subsec: scheme for B<=K-1}.}}
\ie, $W_k=\big(W_k^{(1)}, \cdots,W_k^{(B)} \big )$. Let the source key be $\zsigma =(N_1,\cdots, N_{\lzsigmastar })$ where $N_1,
\cdots, N_{\lzsigmastar}$ are $\lzsigmastar\eqdef \max\{B, K-B\}$ \iid uniform random variables from $ \mathbb{F}_q$. The \indiv keys are generated through a linear map
\be
\label{eq: indiv ley gen, general linear transform}
 {
\mrm{[\Indiv\;\,key\;\,\genn]} \quad 
}
\lef[Z_1,\cdots, Z_K\rig]^T = \Hm Z_{\Sigma}^T,
\ee 
where $\Hm  \in \mbb{F}_q^{K\times \lzsigmastar}$ is the encoding matrix  referred to as \emph{\indiv key generation matrix}. 
Denoting $\hv_k$ as    the \kth row of $\Hm$, we have $Z_k=\hv_kZ_{\Sigma}^T\in \mbb{F}_q$. It can be seen that $\rz=1/B,\rzsigma=\max\{1, K/B-1\}$.
The \utr \msgs are chosen as  
\be 
\label{eq: user-to-relay msg, general linear form}
 {\mrm{[User\text{-}to\text{-}relay\;\, \msg] }  }\quad
X_{k,i} =\sum_{j=1}^B\alpha_{k,i}^{(j)}  W_k^{(j)} + \lambda_{k,i}Z_k, \; \forall k\in[K],i\in\Bc_k 
\ee 
where $ \alpha_{k,i}^{(j)} $ and $\lambda_{k,i}$ represent the linear \coeffts for the  input and key symbols \resp. The \rts \msg are chosen as 
\be 
\label{eq: relay-to-server msg, general form}
 {\mrm{[Relay\text{-}to\text{-}server\;\, \msg] }  }\quad
Y_i = \sum_{k \in  \Uc_i}X_{k,i},\; \forall i\in [K]
\ee 
Hence, $\rx=1, \ry=1/B$.

Key \genn refers to the design of the encoding matrix $\Hm$ in (\ref{eq: indiv ley gen, general linear transform}).
Message \genn refers to the design of the linear \coeffts $\{\alpha_{k,i}^{(j)}\}_{k\in[K], i\in \Bc_k,j\in[B] }$ and $\{\lambda_{k,i}\}_{ k\in[K], i\in \Bc_k  }$ for the input and
\indiv key symbols in (\ref{eq: user-to-relay msg, general linear form}). We aim to \emph{jointly} design the keys and the \msgs so that the input sum  $\sum_{k=1}^KW_k$ can be correctly recovered by the server, subject to the security constraints (\ref{eq: server security constraint}) and (\ref{eq: relay security constraint}). Specifically, suppose the server applies a linear transform 
\be 
\label{eq: def R, general scheme}
\Rm \eqdef [\rv_1,\cdots,\rv_B] \in  \mbb{F}_q^{K\times B}
\ee
to the \rts \msgs to recover the input sum, \ie, $\sum_{k=1}^KW_k =  [Y_1,\cdots,Y_K]\Rm$ where $ \rv_i\in \mbb{F}_q^{K}$ recovers the sum of the \ith symbol of the inputs, \ie,
$
\sum_{k=1}^KW_k^{(i)} = [Y_1,\cdots,Y_K ]\rv_i, i\in[K]
$. The following two tasks need to be fulfilled:

\tbf{Task 1: Input sum recovery.} The \coeffts $\{\alpha_{k,i}^{(j)}\}_{k\in[K], i\in \Bc_k,j\in[B] }$ should be chosen such that the input sum 
$ \sum_{k=1}^K W_k =   \big(\sum_{k=1}^KW_k^{(i)}\big)_{i=1}^B  $ can be decoded under $\Rm$, \ie,
\be
\label{eq: input sum recovery under R}
 {\mrm{[Input\;\,sum\;\,recovery]}}
\quad 
\sum_{k=1}^KW_k =  \lef[Y_1,\cdots,Y_K\rig]\Rm.
\ee

\tbf{Task 2: Key generation.} 
The \indiv key \genn matrix $\Hm$ in (\ref{eq: indiv ley gen, general linear transform}) and the \coeffts
$\{\lambda_{k,i}\}_{ k\in[K], i\in \Bc_k  }$ should be chosen such that the keys in (\ref{eq: user-to-relay msg, general linear form}) cancel out  under $\Rm$ during \agg, which is a necessary condition for correct recovery of the input sum. \Ip, plugging (\ref{eq: user-to-relay msg, general linear form}) into (\ref{eq: relay-to-server msg, general form}), we have
\be
\label{eq: Y_i = W^Y_i + Z^Y_i, gen scheme}
Y_i =  \underbrace{\sum_{k\in \Uc_i} 
\sum_{j=1}^B{\alpha_{k,i}^{(j)}}  W_k^{(j)} }_{{\eqdef W_i^Y}} +  \underbrace{\sum_{k\in \Uc_i}{\lambda_{k,i}\hv_k}  Z_{\Sigma}^T}_{{\eqdef Z^Y_i}},\; i\in [K]
\ee
where $W_i^Y$ and $Z_i^Y$ denote \resp the input  and key component in $Y_i$.

For Task 1, we design the \coeffts drawing inspiration from the \comm-efficient gradient coding (GC) scheme by Ye \& Abbe \cite{ye2018communication}. Without the security constraints, we observe that there exists an 
equivalence in terms of input sum recovery between the no-straggler GC with storage 
capacity $B$  and the considered HSA problem with \assocn number $B$. More specifically, the \ith relay can be viewed as a computing node/worker in GC which stores $B$ datasets $\{W_k\}_{k\in \Uc_i}$. Therefore, minimizing the \rts \comm rate $\ry$ is the same as minimizing the \comm rate in GC.

\bl{The} major contribution of this work lies in the design of $\Hm$ and the \coeffts $\{\lambda_{k,i}\}_{ k\in[K], i\in \Bc_k  }$ in Task 2, which ensures both cancellation of the keys during  \agg for input sum recovery, and the server and relay security. Specifically, given the optimal source key $\zsigma$ which consists of $\lzsigmastar=\max \{B,K-B\}$ \iid symbols, we choose  $\Hm$ as an $(K,\max \{B,K-B\})$-Maximal Distance Separable (MDS) matrix so that any subset of $\lzsigmastar$ \indiv keys are mutually \indep. $\lzsigmastar \ge  B$ ensures that each of the $B$ \utr \msgs received by any relay is protected by an \indep \indiv key, so that the relay infers nothing about the input symbols encoded in these \msgs (\rsec). When $B \le K/2$, besides \rsec, $\lzsigmastar=K-B\ge B$ \indep source key symbols are needed to ensure \ssec. That is, to recover the $L=B$ symbols of the input sum while not leaking anything else to the server, at least $\sum_{k=1}^KH(Y_k)-  H\big(\sum_{k=1}^KW_k\big)=K-B $ \indep key symbols are needed. We propose two specific constructions of the keys and \coeffts  depending on whether $B\ge K/2$.

A detailed description of the \msg and key designs is presented  in the following two subsections.

\subsubsection{{Input Coefficient Design for Task 1}} 
\label{subsubsec: design for task 1} We follow the design of Ye \& Abbe~\cite{ye2018communication}. Security is achieved through the proposed security key design in the following section. \Ip, let $ \theta_1,\cdots, \theta_K$ be $K$ distinct numbers from $\mbb{F}_q$. 
We define  a \poly $p_k(x)$ of degree $K-B$---as an indicator of the user-relay \assocn---\af:
\be 
\label{eq: def p_k(x)}
p_k(x) \eqdef \prod_{i\in [K]\backslash \Bc_k    }(x- \theta_i), 
\ee
where each $\theta_i$ is associated with Relay $i$, and each $p_k$ is associated with input $W_k$ (or User $k$). Note that $p_k(\theta_j)\ne 0$ if $j\in \Bc_k$, meaning Relay $j$ has access to input $W_k$ (\ie, User $k$ is associated with Relay $j$);  $p_k(\theta_j)=0$ if $j\in [K]\backslash \Bc_k$, meaning Relay $j$ does not have access to $W_k$ (\ie, User $k$ is not associated with Relay $j$). By (\ref{eq: def p_k(x)}), we can see that each Relay $i$ has access to the inputs $\{W_k\}_{k\in  \Uc_i}$, or equivalently, each input $W_k$ is accessible by relays in $\Bc_k$.

Based  on $p_k(x)$, we further generate $B$ \polys $p_k^{(1)}(x), \cdots,p_k^{(B)}(x) $ recursively:
\begin{align}
\label{eq: def, recursive polys}
p_k^{(1)}(x) & = p_k(x), \notag\\
p_k^{(b)}(x) &  = xp_k^{(b-1)}(x) - p_{k,K-B-1}^{(b-1)}p_k^{(1)}(x),\; b\in [2:B]
\end{align}
where $\{ p^{(b)}_{k,j} \}_{j=0}^{K-1}$ denote the \coeffts  of $ p^{(b)}_k(x)$ 
if we write $p^{(b)}_k(x) =\sum_{j=0}^{K-1} p_{k,j }^{(b)} x^j   $.  It can be seen that  $ p_k(x)$  divides every $ p_k^{(b)}(x),b\in [B]$ so that the \assocn indicator property of $ p_k(x)$ is still preserved in the generated \polys, \ie,
$ 
p_k^{(b)}( \theta_j) \ne 0
$ 
if $j\in \Bc_k$  and $ 
p_k^{(b)}( \theta_j)= 0
$
if 
$j\in [K]\backslash \Bc_k$ for all $b\in[B]$.
$p_k^{(b)}(x)$ has a degree of $K-B+b-1$ and its leading \coefft is 1, \ie, 
\begin{align}
\label{eq: p_k^{(b)}(x) leading coefficient}
 &  p_{k, K-B+b-1}^{(b)}  =1, \;\forall  b\in [B] \notag\\
 & p_{k, K-B+b-1 + l}^{(b)}  =0 ,\; \forall l  \in  [B-b]
\end{align}
It can also be easily verified that $\forall b\in [B]$,
\begin{align}
p^{(b)}_{k,K-B+b-1-l} =0, \; \forall l\in [1:b-1]
\end{align}

We then define a matrix $\Bm \in  \mbb{F}_q^{BK \times K} $
whose entries are the \coeffts of the \polys $\{p_k^{(b)}(x) \}_{k\in [K], b\in[B]}$, \ie,
\begin{align}
\label{eq: B matrix def}
\Bm_{(i-1)B+b,j}  = p_{i, j-1}^{(b)}, \;
\forall i\in[K],b\in[B],j\in[K]
\end{align}
It can be verified that the last $B$ columns of $\Bm$ is equal to
\be
\label{eq: last B columns of B matrix}
\Bm_{:,(K-B+1): K} = \left [\mbf{I}_{B}; \mbf{I}_{B}; \cdots ;\mbf{I}_{B} \right ].
\ee 
Denote $ 
\wv \eqdef \big[ W_1^{(1)}, \cdots,  W_1^{(B)}, \cdots, W_K^{(1)}, \cdots,  W_K^{(B)}\big] \in \mbb{F}_q^{1\times BK}
$ 
as the concatenation of all the input symbols. 
The input component of $Y_i$ (see (\ref{eq: Y_i = W^Y_i + Z^Y_i, gen scheme})) is then given by
\be
\label{eq: W_i^Y in Y_i}
W_i^Y =  \wv\Bm \big[ 1, \theta_i,\cdots,\theta_i^{K-1}  \big]^T, \; i\in [K]
\ee

Note that $W_i^Y$ depends only on the inputs  of Relay $i$'s associated users. To see this,  we can write 
$$\Bm \big[  1,x, \cdots, x^{K-1} \big]^T =\left[p_1^{(1)}(x),\cdots, p_1^{(B)}(x), \cdots, p_K^{(1)}(x),\cdots, p_K^{(B)}(x)  \right     ]^T $$ 
so that $\Bm \bm{\theta}_i=\big[p_1^{(1)}(\theta_i),\cdots, p_1^{(B)}(\theta_i), \cdots, p_K^{(1)}(\theta_i),\cdots, p_K^{(B)}(\theta_i)   \big]^T   $. Due to the specific construction of the recursive \polys (see (\ref{eq: def p_k(x)}) and (\ref{eq: def, recursive polys})), only the entries  $\{p_k^{(b)}(\theta_i)\}_{k\in \Uc_i, b\in[B]  } $ in $\Bm \bm{\theta}_i$  are nonzero while all others are zero. This means $W_i^Y = \wv \Bm \bm{\theta}_i$ depends only on $\{W_k\}_{k\in \Uc_i}$.
As a result,  the input symbol \coeffts $\{\alpha_{k,i}^{(j)}\}_{k\in[K], i\in \Bc_k,j\in[B] }$ in (\ref{eq: Y_i = W^Y_i + Z^Y_i, gen scheme}) can be correspondingly determined. 
Denote $
\bm{\theta}_k \eqdef  [1, \theta_k,\cdots, \theta_k^{K-1}]^T
,k\in [K]$ and 
\be 
\label{eq: def Theta matrix}
\mbf{\Theta} \eqdef [\bm{\theta}_1, \cdots, \bm{\theta}_K],
\ee
then the input components of the \rts  \msgs can be written as
\be
\label{eq: all W_i^Y from Theta}
\big [W_1^Y, \cdots,W_K^Y\big] = \wv \Bm \mbf{\Theta}.
\ee

According to~\cite{ye2018communication}, the decoding process of the input sum is \af.
Let the column vectors $\ev_i, \cdots, \ev_K$ be the standard basis of $\mbb{F}_q^{K}$, \ie, all the coordinates of $\ev_i$ are 0 except the \ith coordinate which is 1. By (\ref{eq: last B columns of B matrix}), we have
\be
\label{eq: wBe_{K-B+i}}
\wv \Bm \ev_{K-B + i} = \sum_{k=1}^K W_k^{(i)}, \; i\in [B]
\ee
Because $\mbf{\Theta}$ is  invertible (since it is a Vandermonde matrix with distinct elements), the server can calculate every
$\wv \Bm \ev_{K-B + i}$ by applying a  linear transform
\be
\label{eq: def ri=Theta*e(K-B+i)}  
\rv_i =   \mbf{\Theta}^{-1} \ev_{K-B+i}  \in \mbb{F}_q^K
\ee
to the \msgs $Y_1, \cdots, Y_K$. \Ip,  
\begin{subequations}
\label{eq: input sum decoding from rv_i's}
\begin{align}
 \left ([Y_1, \cdots, Y_K] \rv_i       \right)_{i=1 }^B  & =
  \left ([Y_1, \cdots, Y_K] \mbf{\Theta}^{-1} \ev_{K-B+i}      \right)_{i=1 }^B \\
 &   \overset{(\ref{eq: Y_i = W^Y_i + Z^Y_i, gen scheme})}{=}  \left (\big [W_1^Y, \cdots,W_K^Y\big] \mbf{\Theta}^{-1} \ev_{K-B+i}      \right)_{i=1 }^B  +  \left( \big[Z_1^Y, \cdots,Z_K^Y\big ] \mbf{\Theta}^{-1} \ev_{K-B+i}        \right)_{i=1 }^B\label{eq: step 0, server input sum decoding final} \\
 &   \overset{(\ref{eq: wBe_{K-B+i}}), (\ref{eq: def ri=Theta*e(K-B+i)})}{=}   \left( \wv \Bm \ev_{K-B + i}      \right)_{i=1 }^B\\
&   = \left( \sum_{k=1}^K W_k^{(i)}     \right)_{i=1 }^B = \sum_{k=1}^K W_k.
\end{align}
\end{subequations}
Note that the above decoding process requires the \emph{prefect cancellation of the key components $\{Z^Y_i\}_{i\in [K]}$  in (\ref{eq: step 0, server input sum decoding final}) under the linear transforms $\rv_1, \cdots, \rv_B  $}, \ie, 
\be
\label{eq: key cancellation under input decoding linear transforms, each ri}
\left[Z_1^Y, \cdots ,Z_K^Y \right] \rv_i = 0, \; \forall  i \in [B]
\ee
which can be written more compactly as
\be 
\label{eq: key cancellation under input decoding linear transforms}
\left[Z_1^Y, \cdots ,Z_K^Y\right] \mbf{R} = \bm{0}_{1\times B},
\ee 
where $\mbf{R} \eqdef
[\rv_1,
\cdots, \rv_B] \in \mbb{F}_q^{K \times B}$ denotes the \emph{recovery matrix} for the input sum.  


\subsubsection{{Individual Key and Coefficient Design for Task 2}} 
\label{subsubsec: design for task 2}
From the analysis in the previous section, to correctly recover the input sum, the \indiv keys $Z_1,\cdots, Z_K$ and the \coeffts in $\{\lambda_{k,i} \}_{k\in [K],i\in \Bc_k}$ in (\ref{eq: user-to-relay msg, general linear form}) must be  carefully  chosen such that the following two conditions are satisfied:
\begin{enumerate}
    \item[] \tbf{\tit{Condition 1}}: The key symbols must cancel out during \agg as implied by (\ref{eq: key cancellation under input decoding linear transforms}). 
    
    \item[] \tbf{\tit{Condition 2}}: The security constraints (\ref{eq: server security constraint}) and (\ref{eq: relay security constraint}) must be satisfied.
\end{enumerate}

Regarding Condition 1, by (\ref{eq: Y_i = W^Y_i + Z^Y_i, gen scheme}), we have $Z_i^Y= \sum_{k\in \Uc_i}\lambda_{k,i}\hv_k  Z_{\Sigma}^T$ (recall that $\lambda_{k,i}$ denotes the \coefft for $Z_k$ in the \msg that User $k$ sends to Relay $i$) and
\be 
\label{eq: (Z_1^Y,...,Z_K^Y), general scheme}
\left[Z_1^Y, \cdots, Z_K^Y \right]^T= \mbf{\Lambda}^T \Hm \zsigma^T, 
\ee 
where 
\be 
\label{eq: def of Lambda}
\setlength{\arraycolsep}{3pt}
\renewcommand{\arraystretch}{0.8}
 {\mrm{[Key\;\, \coefft\;\, matrix]}}
\quad 
\mbf{\Lambda} \eqdef 
\begin{bmatrix}
 \lambda_{1,1} & \lambda_{1,2} & \cdots  &  \lambda_{1,B} & 0 & 0 &\cdots & 0 & 0 \\
0 &  \lambda_{2,2} &  \lambda_{2,3} & \cdots &   \lambda_{2,B+1}  & 0  & \cdots & 0 & 0\\
 \vdots & \vdots & \vdots & \ddots & \vdots & \vdots & &  \vdots & \vdots\\
  \lambda_{K,1} & \lambda_{K,2} & \cdots & \lambda_{K,B-1} & 0  & 0 & \cdots & 0 & \lambda_{K,K}
\end{bmatrix} \in \mbb{F}_q^{K\times  K}
\ee
denotes the  \emph{key \coefft matrix}, {the \kth row of which represents the key \coeffts used by \User{k} in the uploaded \msgs to the \assoced relays. 
Note that $\lambda_{k,i}\ne 0, \forall k\in[K], \forall i  \in \Bc_i$. Otherwise, the \msg \compos in $X_{k,i}$ (see (\ref{eq: user-to-relay msg, general linear form})) will be exposed to the relays, violating \rsec.

}
Therefore, 
(\ref{eq: key cancellation under input decoding linear transforms}) can be written as
\be
\label{eq: Zsigma*HT*Lambda*R=0}
 \zsigma \Hm^T \mbf{\Lambda }\Rm = \bm{0}_{1\times B}.
\ee 
Since the source key $\zsigma=(N_1,\cdots, N_{\lzsigmastar})$ consists of  \iid  uniform random variables from $\mbb{F}_q$, and (\ref{eq: Zsigma*HT*Lambda*R=0}) holds true for every possible realization of $ \zsigma$, it must follow that 
\be
\label{eq: HT*Lambda*R=0}
 \Hm^T \mbf{\Lambda }\Rm = \bm{0}_{ \lzsigmastar \times B}.
\ee
Our goal is to design the \indiv key generation matrix $\mbf{H}$ and the key \coefft matrix $\mbf{\Lambda}$ such that the columns of $\mbf{R}$ span the \emph{entire} null space of $ \mbf{H}^T\mbf{\Lambda}$. This \req  ensures that the keys can be canceled \emph{if and only if} the decoding vector lies in $\spann(\Rm)$. Consequently, 
no vector $\rv \notin \spann(\mbf{R})$ can satisfy  $\mbf{H}^T\mbf{\Lambda}\rv= \bm{0}_{\lzsigmastar  \times 1   }$, implying that any unintended linear combination (beyond the desired input sum) of the input symbols  will necessarily be entangled with the keys, thereby remaining protected from the server. 
Equivalently, (\ref{eq: HT*Lambda*R=0}) can be written  as
\be
\label{eq: RT*LambdaT*H=0}
\Rm^T \mbf{\Lambda }^T\Hm = \bm{0}_{  B\times  \lzsigmastar}.
\ee
Since $\mbf{\Lambda }^T\Hm  $ has $ \lzsigmastar=\max\{B, K-B\}$ columns---which is at least as large as $K-B$, the dimension of the null space of $\Rm^T$---it is possible to select  $\mbf{\Lambda}$ and $\Hm$ such that
the columns of $\mbf{\Lambda }^T\Hm$ expand the entire null space of $\Rm^T$.

Regarding Condition 2,  an intuitive proof of  security  {based on linear decoding} is provided \af. A rigorous proof is presented at the end of this section. 
\begin{enumerate}
    \item[i)] \tit{\Rsec}:  \Rsec  is achieved by ensuring the key symbols $\{Z_k\}_{k\in \Uc_i}$ embedded in the \msgs received by each Relay $i$ are mutually \indep. This guarantees that each  input symbol in  the \msgs $\{X_{k,i}\}_{k\in \Uc_i}$  is protected by an \indep key symbol, thereby preventing the relay from inferring any information about the original inputs $W_{1:K}$.
    Note that each $ \lambda_{k,i}$ in $\mbf{\Lambda}$ must be nonzero; otherwise, the input symbols contained in $X_{k,i}$ would be exposed to Relay $i$, violating the \rsec constraint. Choosing $\Hm$ as a $(K,\lzsigmastar)$-MDS generator matrix ensures any subset of $\lzsigmastar\ge B$ \indiv keys are mutually \indep and is thus sufficient for \rsec. 
    
    \item[ii)] \tit{\Ssec}: \Ssec is guaranteed by the fact that the columns of $\Rm$ (i.e., $\rv_1, \cdots, \rv_B$) span the entire null space of $ \Hm^T\mbf{\Lambda}$. \Aar, any nonzero key-canceling vector  $\dv \in \mbb{F}_q^K\backslash \{\bm{0}_{K\times 1}\}$ must be expressible as a  linear combination of $\rv_1,\cdots, \rv_B$. 
    \Ip, suppose $\dv=\sum_{i=1}^Bd_i \rv_i$ for some $d_1,\cdots,d_B \in \mbb{F}_q$, then
    \begin{subequations}
    \begin{align}
    \left[Y_1,\cdots,Y_K \right] \dv  & =
      \left[W^Y_1,\cdots,W^Y_K\right] \dv +
     \underbrace{\left[ Z^Y_1,\cdots,Z^Y_K\right] \dv}_{ =0  } \\
    & =\sum_{i=1}^Bd_i\left[W^Y_1,\cdots,W^Y_K\right]\rv_i\\
     & \overset{(\ref{eq: input sum decoding from rv_i's})}{=}  \sum_{i=1}^Bd_i \sum_{k=1}^K W_k^{(i)}, 
    \end{align}
    \end{subequations}
    which is a function of the input sum $\sum_{k=1}^K W_k$. This implies that any linear transformation of $Y_1,\cdots,Y_K$ reveals no \info about the inputs beyond their sum. Therefore, server security is guaranteed.  
\end{enumerate}

In the following, we propose two schemes for designing  $\Hm$ and $\mbf{\Lambda}$, \crspdg  to the cases $ 2\le B\le K/2 $ ($\lzsigmastar=K-B$) and $K/2<B\le K-1$ ($\lzsigmastar=B$), \resp.
Note that when $B=1$, the  considered problem degenerates to a special case of the single-\assocn HSA problem studied in~\cite{zhang2024optimal}, where each user is connected to only  one relay and the optimal rates are given by  
$\rxstar=\rystar=\rzstar=1$ and $\rzsigmastar=K-1$. In this case, $\Hm$ is chosen as an $K \times (K-1)$ extended Vandermonde matrix~\cite{zhang2024optimal} and $\mbf{\Lambda}$ is simply the $K\times K$ identity matrix. 
{\Bef proceeding, 
we first introduce several necessary notations. Let $\bm{\theta}^{(i)} \eqdef [\theta_1^{i-1},\theta_2^{i-1}, \cdots, \theta_K^{i-1}], i\in [K]$. Then (\ref{eq: def Theta matrix}) can also be written as 
\be
\label{eq: def Theta matrix, theta^i}
\mbf{\Theta} = \big[ \bm{\theta}^{(1)};\bm{\theta}^{(2)}; \cdots; \bm{\theta}^{(K)}\big]. 
\ee 
Denoting  $ \mbf{\Theta}^{-1} \eqdef \big [ \widehat{\bm{\theta}}_1, \cdots, \widehat{\bm{\theta}}_K \big ]$, then  $\mbf{\Theta}\mbf{\Theta}^{-1} =\mbf{I}_K $ implies
\be
\label{eq: theta^i*theta_j=1,0}
\bm{\theta}^{(i)} \widehat{\bm{\theta}}_j =
\left \{
\begin{array}{ll}
 1 ,   & \trm{if } i=j  \\
 0 ,   & \trm{if } i\ne j 
\end{array}
\right. 
\ee 
Recall that $ \Rm =[\rv_1,\cdots,\rv_B]$ where $ \rv_i = \widehat{\bm{\theta}}_{K-B+i},i\in [B] $, \ie, $\Rm$ consists of the last $B$ columns of $\mbf{\Theta}^{-1}$.

} 

\tbf{\emph{\underline{Scheme 1}: Key generation for $ 2 \le B\le K/2$}}.
\label{subsubsec: key gen for B<=K/2}
In this case,  $ \rzsigmastar=K/B-1$ and  $\lzsigmastar= K-B$.
We choose $\mbf{\Lambda} $ and $ \Hm$ such that
\be
\label{eq: Lambda^T * H choice}
\mbf{\Lambda}^T\Hm = \Qm \eqdef
\left[\big(\bm{\theta}^{(1)}\big)^T, \big(\bm{\theta}^{(2)}\big)^T, \cdots, \big(\bm{\theta}^{(K-B)}\big)^T  \right]   \in  \mbb{F}_q^{K\times  (K-B)},
\ee 
where $\Qm$ is a Vandermonde matrix with distinct \eval points $\theta_1,\cdots, \theta_K$  \soth every $(K-B)\times (K-B)$ submatirx of $\Qm$ has full rank. This means $\rank(\mbf{\Lambda}^T\Hm )=K-B$ so that the columns of $\mbf{\Lambda}^T\Hm $ span the entire null space of $\Rm^T$.
{Given $\Qm$, we aim to find a proper $\bmLambda$ \suth 1) $\bmLambda^T$ is invertible \soth $ \Hm$ can be computed as $\Hm= (\bmLambda^T)^{-1}\Qm$, and 2) the MDS property of $\Qm$ carries over  to $\Hm$, which is \nece to ensure \rsec.
To this end, we  first  parameterize the circulant matrix $\bmLambda$ (\ref{eq: def of Lambda}) through an element  $g\in \Fmbbq \bkslsh \{0\}$ not lying in the subgroup of $K$-th roots of unity, 
\ie, $g^K \ne 1$, proving the existence of proper $g$ and $\Fmbbq$ such that $\bmLambda$ is invertible, and every $(K-B)\times (K-B)$ submatrix of $\Hm$ has full rank.

\tit{-- Choice of $\bmLambda$:}
Assume $K \mid (q-1)$ so that there exists a primitive $K$-th root $\omega$ of unity in $\Fmbbq$.
Fix $g \in \Fmbbq\bkslsh \{0\}$. 
Define $a_j \eqdef g^{j-1}\ne 0, \forall j\in [B]$  and parameterize $\bmLambda$ by $g$ as
\be 
\label{eq:bmLambda=circ(q1,...,qB), scheme 1}
\bmLambda_g \eqdef \mrm{circ}(a_1,  \cdots,a_B, 0, \cdots, 0   )\in \Fmbb_q^{K\times K},
\ee 
where $\mrm{circ}(\av)$ denotes the circulant matrix generated by the vector $\av$. Lemma  \ref{lemma:existence of bmLambda and H,scheme 1} shows the existence of  a proper $g$ and field size $q$ so that $\bmLambda_g $ is invertible and $\Hm$ has t the required MDS property. 

\begin{lemma}
\label{lemma:existence of bmLambda and H,scheme 1}
\tit{
Assume $K  \mid  (q-1)$ and choose distinct  $\theta_1,\cdots, \theta_K  \in  \Fmbbqnozero $ to form the \Vander matrix $\Qm$ in (\ref{eq: Lambda^T * H choice}). Then there exists $g\in \Fmbbqnozero, g^K\ne  1 $ \suth  $\bmLambda_g$ defined in (\ref{eq:bmLambda=circ(q1,...,qB), scheme 1}) is invertible and  $\Hm_g \eqdef (\bmLambda_g^T)^{-1} \Qm $ satisfies the MDS condition that every $(K-B)\times(K-B)$ submatrix of it has full rank over $\Fmbbq$, provided that $q$ is sufficiently large. 
Let $ \varphi(K,B) \eqdef \binom{K}{B}(K-B)(K-1)(B-1)+BK+2$.
Then $q\ge \varphi(K,B)$ guarantees the existence of at least one  valid $g$.
If $g$ is drawn uniformly at random from   $\Fmbbq$, the probability that  $g$ is valid is at least
\be
\label{eq:g valid prob lower bound, lemma, scheme 1}
\mrm{Pr}\{\trm{\emph{$g$ is valid}}\}
\ge 1- \frac{\varphi(K,B)}{q},
\ee 
which converges to $1$ as $q\to \infty$.
}
\end{lemma}
\begin{IEEEproof}
See Appendix \ref{sec:proof of lemma:existence of bmLambda and H,scheme 1,appendix}. The general idea of the proof is that, by parameterizing $\bmLambda$ with $g$, we can show that there is a limited set of `bad' values of $g$  under which $\bmLambda_g$  is not invertible. \Thf, avoiding such values would ensure an invertible $\bmLambda_g$. Moreover, given that   $\bmLambda_g$ is invertible, we can compute $\Hm_g = (\bmLambda_g^T)^{-1}\Qm$, where $\Hm_g$ is a function of $g$. 
To ensure the MDS property on $\Hm_g$, we can define a product polynomial $P(g)$ that  contains the determinant of every $(K-B)\times (K-B)$  submatrix of $\Hm_g$. 
Since $P(g)$ has at most $\deg(P(g))$ roots, there is a limited number of values of $g$ that render $P(g)=0$, implying that at least one submatrix does not have full rank. \Thf, if we avoid such values,  $P(g)\ne 0$ can be guaranteed, ensuring the MDS property on $\Hm_g$.  Overall, by the Schwartz-Zippel lemma~\cite{Schwartz, Zippel}, with a  sufficiently large field size $q$, the existence of valid  choices of $g$ is guaranteed. \end{IEEEproof}

\tit{-- Choice of $\Hm$:} Given  any valid $g$ and $\bmLambda_g$ found in (\ref{eq:bmLambda=circ(q1,...,qB), scheme 1}), $\Hm$ is chosen as
\be
\label{eq: H, case B<=K/2}
\Hm = \big(\bmLambda_g^T\big)^{-1}\Qm.
\ee 

}   

Next, we show that the above design satisfies the key cancellation condition
(\ref{eq: RT*LambdaT*H=0}), \ie, $\Rm^T\mbf{\Lambda}^T \Hm=\bm{0}_{B\times (K-B)}$.
Recall that the recovery matrix $ \Rm$ consists of the last $B$ columns of $\mbf{\Theta}^{-1}$, \ie, $\Rm =[ \widehat{\bm{\theta}}_{K-B+1}, \cdots, \widehat{\bm{\theta}}_{K} ]$. Plugging in (\ref{eq: Lambda^T * H choice}) and due to
(\ref{eq: theta^i*theta_j=1,0}), we have
\begin{align}
\Rm^T\mbf{\Lambda}^T \Hm & 
=\left [ \widehat{\bm{\theta}}_{K-B+1}, \cdots, \widehat{\bm{\theta}}_{K}\right]^T
\left[  \big(\bm{\theta}^{(1)}\big)^T ,\big(\bm{\theta}^{(2)}\big)^T,\cdots,  \big(\bm{\theta}^{(K-B)} \big)^T  \right]\notag\\
&=
\renewcommand{\arraystretch}{0.8}
\begin{bmatrix}
 \bm{\theta}^{(1)}\widehat{\bm{\theta}}_{K-B+1} &  \bm{\theta}^{(2)}\widehat{\bm{\theta}}_{K-B+1} & \cdots & \bm{\theta}^{(K-B)}\widehat{\bm{\theta}}_{K-B+1}\\
 \bm{\theta}^{(1)}\widehat{\bm{\theta}}_{K-B+2} &  \bm{\theta}^{(2)}\widehat{\bm{\theta}}_{K-B+2} & \cdots & \bm{\theta}^{(K-B)}\widehat{\bm{\theta}}_{K-B+2} \\
\vdots & \vdots & \ddots & \vdots \\
  \bm{\theta}^{(1)}\widehat{\bm{\theta}}_{K} &  \bm{\theta}^{(2)}\widehat{\bm{\theta}}_{K} & \cdots & \bm{\theta}^{(K-B)}\widehat{\bm{\theta}}_{K} 
\end{bmatrix} = \bm{0}_{B\times (K-B)}.
\end{align}

\tbf{\emph{\underline{Scheme 2}: Key generation for $ K/2<B \le K-1$}}.
\label{subsubsec: key gen for B>K/2}
In this case, we have  $\rzsigmastar= 1,\lzsigmastar=B$.
{We first choose $\Hm$ as a \Vander matrix with the proper MDS property, and then design $\bm{\Lambda}$ correspondingly \suth the  key \coeffts $\lambda_{i,j}$'s are nonzero.

\tit{-- Choice of $\Hm$}:
$\Hm$ is chosen as
\be 
\label{eq: H, case B>=K/2}
\Hm =\left[ \big(\bm{\theta}^{(1)}\big)^T, \big(\bm{\theta}^{(2)}\big)^T, \cdots, \big(\bm{\theta}^{(B)}\big)^T     \right]\in \mbb{F}_q^{K\times B},
\ee 
where recall that  $\bm{\theta}^{(i)}\eqdef \big[\theta_1^{i-1},\theta_2^{i-1}, \cdots, \theta_K^{i-1}\big]  \in \Fmbb_q^{1\times K}  , i\in [K]$. It can be seen that every $B\times B$ submatrix of $\Hm$ has full rank because $ \Hm$ is a Vandermonde matrix with distinct elements $\theta_1, \cdots, \theta_{B}$.

\tit{-- Choice of $\bmLambda$}:
Consider the key \coefft $\{\lambda_{i,k}\}_{i\in  \Uc_k}$ design  for Relay $k$. Suppose $\Uc_k=\{i_1,\cdots, i_B\}$ where $i_1<i_2<\cdots <i_B$ (refer to (\ref{eq: def Uc_i, problem formulation})). Denote
\be
\label{eq: ThetaUk, scheme 2}
\renewcommand{\arraystretch}{0.7}
\mbf{\Theta}_{\Uc_k} \eqdef 
\begin{bmatrix}
1  & \theta_{i_1} & \cdots &  \theta_{i_1}^{B-1}\\
1  & \theta_{i_2} & \cdots &  \theta_{i_2}^{B-1}\\
\vdots  & \vdots & \ddots &  \vdots\\
1  & \theta_{i_B} & \cdots &  \theta_{i_B}^{B-1} 
\end{bmatrix}\in  \mbb{F}_q^{B\times B}, 
\ee
which is a $B\times B$ submatrix corresponding  to the $i_1^{\rm th}, \cdots, i_B^{\rm th}$ rows 
of $\Hm$. 
Note that $\rank( \mbf{\Theta}_{\Uc_k})=B$. 
Let
$\bm{\lambda}_{\Uc_k} \eqdef [\lambda_{i_1,k}, \cdots,\lambda_{i_B,k}] {\in  \Fmbb_q^{1\times B}} $
denote the key  \coefft vector for Relay $k\in[K]$ (\ie, the \nz elements in the \kth row of $\bmLambda$). Then $\bmLambda$ is completely determined by $\bmlambda_{\Uc_1}, \cdots, \bmlambda_{\Uc_K}$.

To determine $\bmlambda_{\Uc_k}$, we let
\be
\bm{\lambda}_{\Uc_k}\mbf{\Theta}_{\Uc_k}= 
\left[\beta, \theta_k, \theta_k^2,\cdots, \theta_k^{K-B-1}, \bm{0}_{1\times (2B-K)}  \right], 
\ee
where $\beta \in \Fmbbq \bkslsh \{0\}  $ is  a parameter to be designed. Since $\mbf{\Theta}_{\Uc_k}$ has full rank,  $\bmlambda_{\Uc_k}$ can be solved as
\be
\label{eq: lamba_Uc_k, key gen, B>=K/2}
\bm{\lambda}_{\Uc_k} = 
\lef[\beta, \theta_k, \theta_k^2,\cdots, \theta_k^{K-B-1}, \bm{0}_{1\times (2B-K)}  \rig]\mbf{\Theta}_{\Uc_k}^{-1},  \;  k \in [K]
\ee 
{Suppose $ \mbf{\Theta}_{\Uc_k}^{-1}\eqdef \big[ \widehat{\bm{\theta}}^{(k)}_1; \cdots; \widehat{\bm{\theta}}^{(k)}_B \big ]     $.}
We have
\be
\bmlambda_{\Uc_k}= \beta \widehat{\bm{\theta}}^{(k)}_1 + 
\underbrace{\sum_{i=1}^{K-B-1} \theta_k^i   \widehat{\bm{\theta}}^{(k)}_{i+1} }_{\eqdef \bm{\gamma}^{(k)}  },
\ee 
which can be written entrywise as
\be 
\bmlambda_{\Uc_k}(j) =
\beta \widehat{\bm{\theta}}^{(k)}_1(j) + \bmgamma^{(k)}(j), \; j \in [B]
\ee 
Now, if $ \widehat{\bm{\theta}}^{(k)}_1(j)\ne 0$,
$
\beta= -\frac{\bmgamma^{(k)}(j)}{\widehat{\bm{\theta}}^{(k)}_1(j)}
$ 
will render $ \bmlambda_{\Uc_k}(j)=0$, which needs to be avoided. It can be shown that choosing $ \theta_1, \cdots, \theta_K$ as nonzero and distinct \eles from $\Fmbbq $ ensures $ \widehat{\bm{\theta}}^{(k)}_1(j)\ne 0,\forall j\in[B], k\in [K]$\footnote{See Lemma \ref{lemma:proof of lambdaUkj not zero, appendix} in Appendix \ref{sec:proof of lambdaUkj not zero, appendix} for proof.}.
Define the `bad' set for \Relay{k} as
\be 
\label{eq:bad set Mk, scheme 2}
\Mc_k \eqdef 
\lef\{ 
-\frac{\bmgamma^{(k)}(j)}{\widehat{\bm{\theta}}^{(k)}_1(j)},   j\in [B]
\rig\}, \; \forall k \in [K]
\ee
which suggests at most $|\Mc_k|\le B$ forbidden values for \Relay{k} that may cause at least one entry of $\bmlambda_{\Uc_k}$ to be zero.
\Thf, 
if we choose $\beta \in \Fmbbq \bkslsh  \Mc_k$, then automatically $ \bmlambda_{\Uc_k}(j)\ne 0$ for every \coord $j\in [B]$.

Since the same $\beta$ is used  for all relays, define the global bad set as
\be 
\label{eq:global bad set M, scheme 2}
\Mc \eqdef \bigcup_{k=1}^K \Mc_k,
\ee 
satisfying $|\Mc|\le \sum_{k=1}^K|\Mc_k|\le KB $ by the union bound. So if the field size $q>KB$, then $ \Fmbbq \bkslsh \Mc \ne \emptyset$, and we can always pick a single $\beta \in \Fmbbq \bkslsh (\Mc \cup \{0\})  $ that guarantees all
$\bmlambda_{\Uc_k},k\in[K]$ have no zero entries \simuly. 

}  

With the above design of $\bmLambda$, we  have
\begin{align}
\label{eq: choice of Lambda^T*Hm, B>=K/2}
\setlength{\arraycolsep}{3pt}
\renewcommand{\arraystretch}{0.6}
\mbf{\Lambda}^T \Hm   = 
\begin{bmatrix}
\bm{\lambda}_{\Uc_1} \mbf{\Theta}_{\Uc_1}\\
\bm{\lambda}_{\Uc_2} \mbf{\Theta}_{\Uc_2}\\
\vdots\\
\bm{\lambda}_{\Uc_K} \mbf{\Theta}_{\Uc_K}\\
\end{bmatrix}
& =
\begin{bmatrix}
\beta &  \theta_1 & ... & \theta_1^{K-B-1}  & \bm{0}_{1\times (2B-K)} \\
\beta &  \theta_2 & ... & \theta_2^{K-B-1} & \bm{0}_{1\times (2B-K)} \\
\vdots & \vdots&  \ddots & \vdots & \vdots \\ 
\beta &  \theta_K & ... & \theta_K^{K-B-1} & \bm{0}_{1\times (2B-K)} 
\end{bmatrix}\notag\\
& =
\left[ \beta \big(\bm{\theta}^{(1)}\big)^T ,\big(\bm{\theta}^{(2)}\big)^T,\cdots,  \big(\bm{\theta}^{(K-B)}\big)^T ,\bm{0}_{K\times (2B-K)}  \right],
\end{align}
where recall  that $ \bm{\theta}^{(i)} \eqdef \big[\theta_1^{i-1}, \cdots,\theta_K^{i-1}\big], i\in[K]   $.
Since the submatrix consisting of the first $ K-B$ columns of $\mbf{\Lambda}^T \Hm$ forms a Vandermonde matrix (up  to a nonzero scaling factor $\beta$ in the first column), it follows that $\rank(\mbf{\Lambda}^T \Hm)=K-B$, implying that the columns  of $\mbf{\Lambda}^T \Hm$ span the entire null space of $\Rm^T$.

Next, we prove that the above design of $\Hm$ and $\bmLambda$ (see (\ref{eq: lamba_Uc_k, key gen, B>=K/2}) and (\ref{eq: H, case B>=K/2}))
satisfies
the key cancellation condition
(\ref{eq: RT*LambdaT*H=0}). 
Recall that the recovery matrix $ \Rm$ consists of the last $B$ columns of $\mbf{\Theta}^{-1}$, \ie, $\Rm =\big[ \widehat{\bm{\theta}}_{K-B+1}, \cdots, \widehat{\bm{\theta}}_{K} \big]$. By (\ref{eq: theta^i*theta_j=1,0}), we have
\begin{align}
\setlength{\arraycolsep}{3pt}
\renewcommand{\arraystretch}{0.6}
\Rm^T\mbf{\Lambda}^T \Hm & =\left [ \widehat{\bm{\theta}}_{K-B+1}, \cdots, \widehat{\bm{\theta}}_{K}\right]^T
\left[ \beta   \big(\bm{\theta}^{(1)}\big)^T ,\big(\bm{\theta}^{(2)}\big)^T,\cdots,  \big(\bm{\theta}^{(K-B)}\big)^T ,\bm{0}_{K\times (2B-K)}  \right]\notag\\
&=
\begin{bmatrix}
\beta \bm{\theta}^{(1)}\widehat{\bm{\theta}}_{K-B+1} &  \bm{\theta}^{(2)}\widehat{\bm{\theta}}_{K-B+1} & \cdots & \bm{\theta}^{(K-B)}\widehat{\bm{\theta}}_{K-B+1} & \bm{0}_{1\times  (2B-K)}\\
\beta \bm{\theta}^{(1)}\widehat{\bm{\theta}}_{K-B+2} &  \bm{\theta}^{(2)}\widehat{\bm{\theta}}_{K-B+2} & \cdots & \bm{\theta}^{(K-B)}\widehat{\bm{\theta}}_{K-B+2} & \bm{0}_{1\times  (2B-K)}\\
\vdots & \vdots & \ddots & \vdots & \vdots\\
\beta \bm{\theta}^{(1)}\widehat{\bm{\theta}}_{K} &  \bm{\theta}^{(2)}\widehat{\bm{\theta}}_{K} & \cdots & \bm{\theta}^{(K-B)}\widehat{\bm{\theta}}_{K} & \bm{0}_{1\times (2B-K)}
\end{bmatrix} = \bm{0}_{B\times B}.
\end{align}

In the following, we formally prove that the two proposed key designs satisfy the relay and server security constraints (\ref{eq: relay security constraint}) and (\ref{eq: server security constraint}).

\emph{\tbf{Proof of Relay Security.}}
\Rsec is rather straightforward owning to the MDS property of $\Hm$. \Ip, by the proposed design (see (\ref{eq: H, case B<=K/2}), (\ref{eq: H, case B>=K/2})), any $\lzsigmastar= \max\{B,K-B\}\ge B$ rows of $\Hm$ are linearly \indep, 
which ensures that any collection of $B$ \indiv keys are mutually \indep. Consequently, for each Relay $i$, each of the $B$ received \msgs $\{X_{k,i}\}_{k\in \Uc_i}$  is masked by an \indep key symbol, thereby preventing Relay $i$ from obtaining any \info about the embedded input symbols. More formally, for any $i\in[K]$
\begin{subequations}
\label{eq: proof of relay security}
\begin{align}
I\left( \{X_{k,i}\}_{k\in  \Uc_i}; W_{1:K} \right) & = H\left( \{X_{k,i}\}_{k\in  \Uc_i}    \right) - H\left( \{X_{k,i}\}_{k\in  \Uc_i}|W_{1:K}    \right) \\
& \overset{(\ref{eq: user-to-relay msg, general linear form})}{=}   H \left( \{X_{k,i}\}_{k\in  \Uc_i}    \right)  -
H \left( \left\{ \lambda_{k,i}Z_k\right\}_{k\in \Uc_i}    \right)  
\\
& \overset{(\ref{eq: H, case B<=K/2}), (\ref{eq: H, case B>=K/2}) }{=}
H \left( \{X_{k,i}\}_{k\in  \Uc_i}    \right)  -B \\
& \le \sum_{k\in \Uc_i}  H \left( X_{k,i}  \right)  -B \le 0,
\end{align}
\end{subequations}
where the last step is because each $X_{k,i}$ contains  one symbol and uniform \distn maximizes the entropy. Since mutual \info is non-negative, we conclude $I \{X_{k,i}\}_{k\in  \Uc_i}; W_{1:K}  )=0 $, proving  \rsec.

\emph{\tbf{Proof of Server Security.}}
By (\ref{eq: (Z_1^Y,...,Z_K^Y), general scheme}), the key components contained in the \rts \msgs are 
\be 
\label{eq: (Z_1^Y,...,Z_K^Y), general scheme, repeated}
\lef[Z_1^Y, \cdots, Z_K^Y \rig ]   = \zsigma \Hm^T \mbf{\Lambda},
\ee 
where each  $Y_i=W_i^Y + Z_i^Y$ is masked by a key symbol
$Z_i^Y= \sum_{k\in \Uc_i} \lambda_{k,i}  Z_k = (\sum_{k\in \Uc_i}\lambda_{k,i} \hv_k )\zsigma^T $. Note that $Z_i^Y \ne 0$ because $  \sum_{k\in \Uc_i}\lambda_{k,i} \hv_k  \ne \bm{0}_{1\times \lzsigmastar}  (\lambda_{k,i}\ne 0, \forall k,i) $ due to the MDS property of $ \Hm$. 
By the  joint design of  $\mbf{\Lambda}$ and $\Hm$, we know that $\rank(\Hm^T\mbf{\Lambda})=K-B$ for any $B\in [2:K-1]$ (see (\ref{eq: Lambda^T * H choice}), (\ref{eq: choice of Lambda^T*Hm, B>=K/2})). 
Since the recovery matrix $\Rm=[\rv_1,\cdots, \rv_B]$ has rank $B$, the key cancellation condition $\Hm^T \mbf{\Lambda }\Rm = \bm{0}_{ \lzsigmastar \times B}$ (see (\ref{eq: HT*Lambda*R=0})) implies that the null space of $\Hm^T\mbf{\Lambda}$ coincides with the column span of $\Rm$, provided that 
$\rank(\Hm^T\mbf{\Lambda})=K-B$. This structure guarantees that any nonzero vector $ \rv \in  \mbb{F}_q^{K}$ capable of canceling the keys symbols in $Y_1, \cdots, Y_K$---\ie, satisfying $\Hm^T \mbf{\Lambda}\rv= \bm{0}_{\lzsigmastar \times 1}$---must lie in the column span of $\Rm$.
Since each column of $\Rm$ enables the recovery of a distinct  dimension of the input sum, \ie, 
$[Y_1,\cdots, Y_K]\rv_i= \sum_{k=1}^K W_k^{(i)},i\in[B] $ (see (\ref{eq: input sum decoding from rv_i's})), any  \lcb $[Y_1,\cdots, Y_K]\rv$ must be a function of $\big\{ \sum_{k=1}^K W_k^{(i)}     \big  \}_{i\in [B]}$. \Aar, the server learns nothing about $W_{1:K}$ beyond the intended input sum $\sum_{k=1}^K W_k$, thereby ensuring \ssec.
More formally,  
\begin{subequations}
\label{eq: proof of server security, general scheme}
\begin{align}
I \left(Y_{1:K}; W_{1:K} \Big| \sum_{k=1}^K W_k    \right) & = H\left(Y_{1:K} \Big| \sum_{k=1}^K W_k    \right)- H \left(Y_{1:K} | W_{1:K}   \right)\\
& \overset{(\ref{eq: Y_i = W^Y_i + Z^Y_i, gen scheme})}{=} H\left(Y_{1:K} \Big| \sum_{k=1}^K W_k    \right)- H \left( \{Z_k^Y\}_{k\in [K]} | W_{1:K}   \right)\\
& \overset{(\ref{eq: key input independence})}{=} H\left(Y_{1:K} \Big| \sum_{k=1}^K W_k    \right)- H \left( \{Z_k^Y\}_{k\in [K]} \right)\\
& \overset{(\ref{eq: (Z_1^Y,...,Z_K^Y), general scheme, repeated})}{=} H\left(Y_{1:K} \Big| \sum_{k=1}^K W_k    \right)- H \left( \zsigma \Hm^T \mbf{\Lambda} \right)\\
& \overset{(\ref{eq: Lambda^T * H choice}),(\ref{eq: choice of Lambda^T*Hm, B>=K/2})}{=} H\left(Y_{1:K} \Big| \sum_{k=1}^K W_k    \right)- (K-B)\\
& = H\left(Y_{1:K}, \sum_{k=1}^K W_k    \right)- H\left(\sum_{k=1}^K W_k    \right) -(K-B)\\
& \overset{(\ref{eq: recovery constraint})}{=} H\left(Y_{1:K}\right)- B -(K-B)\\
& \le \sum_{k=1}^K H\left(Y_k\right)- B -(K-B)\\
& \le K-B -(K-B)=0,
\end{align}
\end{subequations}
where the last step is because each $Y_k $ contains one symbol and uniform \distn maximizes the entropy. Since mutual \info  is non-negative, we conclude $I(Y_{1:K}; W_{1:K} | \sum_{k=1}^K W_k )=0$, proving \ssec.

\subsection{Scheme for $B=K$}
\label{subsec: scheme for B=K}
We notice that the scheme described in the previous section for $B \in [K-1]$ cannot be directly extended to the case of
$B=K$ to achieve $(\rx=1,\ry=\frac{1}{K-1}, \rz=\frac{1}{K} ,\rzsigma=1)$\footnote{Note that when $B=K$, $\max\{1, K/B-1\}=1$.}. Nevertheless, we can adopt the scheme for $B=K-1$ by disabling the last outgoing link from each user, which yields the achievable rate tuple $(1, \frac{1}{K-1}, \frac{1}{K-1}, 1)$. Comparing this with the converse bounds derived in Section~\ref{sec: converse}, we find that both the \comm rate and the source key rate are optimal, while the \indiv key rate incurs at most a constant-factor gap of $\frac{K}{K-1} \le 2$ from its lower bound. This indicates that applying the $B=K-1$ scheme directly to the case $B=K$ results in only a mild performance loss in the \indiv key rate.
In Section~\ref{sec:discussion}, we present a concrete proof of impossibility of $\rz=1/2$ using the example of $B=K=2$. It  is conjectured that  the rate $(\rx=1, \ry=\frac{1}{K-1}, \frac{1}{K},1)$ is \emph{not} \achvb  for any $K\ge 3$ when $B=K$. Formal proof of this conjecture, along with the characterization of the optimal rate tuple for the case $B = K$, is left for future work.

\section{Converse: Lower Bounds on \Comm and Key Rates}
\label{sec: converse}

In this section, we present the converse proof of Theorem \ref{thm: main result}, which establishes the optimality of the proposed scheme. To facilitate the proof,  we first define the corresponding entropic versions of the \comm and key rates as follows:
\be 
\label{eq:def entropic rates, converse}
\rxh \eqdef \frac{\sum_{i\in \Bc_k}H(X_{k,i})}{L}, \; 
\ryh \eqdef \frac{\sum_{k=1}^K H(Y_k)}{L},\;  \rzh \eqdef \frac{H(Z_k)}{L}, \; \rzsigmah \eqdef \frac{H(\zsigma)}{L}.
\ee  
These entropic rates serve as lower bounds for the rates defined in Section~\ref{sec: problem description} (see (\ref{eq: def comm rates}), (\ref{eq: def key rates})), \ie, 
$\rx  \ge \rxh, \ry\ge \ryh, \rz\ge \rzh $ and $\rzsigma \ge \rzsigmah$. We derive lower bounds on these  entropic rates, which automatically serve as lower bounds on the original \comm and key rates defined in (\ref{eq: def comm rates}) and (\ref{eq: def key rates}).\footnote{The rates defined in (\ref{eq: def comm rates}) and (\ref{eq: def key rates}) use the number of symbols contained in each key and \msg as the numerator. This quantity can be made arbitrarily large by padding zeros into the keys and messages. However, such zero padding does not affect the entropy of the keys and messages. \Aar,
the optimal rate region $\Rc^*$, defined using  (\ref{eq: def comm rates}) and (\ref{eq: def key rates}), can be conveniently expressed as a box region. This avoids complications arising from potential correlations among the entropic \comm and key rates defined in (\ref{eq:def entropic rates, converse}).}

We begin with a useful lemma that provides a lower bound on the joint entropy of the messages sent by each user, as well as the messages sent by any set of $B$ consecutive relays.

\begin{lemma}
\label{lemma: input cutset bound on X}
\emph{For every $k\in [K]$, it holds that
\begin{align}
& H\big(  \left\{ X_{k,i}   \right\}_{i\in \Bc_k} |\{W_{k'},Z_{k'}\}_{k'\in [K]\backslash\{k\}}   \big)
\ge  L,\label{eq: {X_ij}_j cutset bound, converse}\\
& H\big( \{Y_i\}_{i\in \Bc_k} |\{W_{k'},Z_{k'}\}_{k'\in [K]\backslash\{k\}} \big) \ge L.\label{eq: Y_i cutset bound, converse}
\end{align}
}
\end{lemma}
\begin{IEEEproof}
This lemma  is derived from a simple cut-set bound: for the server to recover $\sum_{k=1}^K W_k$, each $W_k$ must propagate through the corresponding \utr links $\{(k, i)\}_{i\in \Bc_k}$ and the \rts links $\{(i, \trm {server})_{i\in \Bc_k}\}$. Hence, a cut between User $k$ and its associated relays $\Bc_k$ yields (\ref{eq: {X_ij}_j cutset bound, converse}),  and the cut between the relays in $\Bc_k$ and the server yields (\ref{eq: Y_i cutset bound, converse}). \Msp,
\begin{subequations}
\label{eq: proof X cutset bound, converse}
\begin{align}
& H\left(  \left\{ X_{k,i}   \right\}_{i\in \Bc_k} |\{W_{k'},Z_{k'}\}_{k'\in [K]\backslash\{k\}}  \right) \notag\\
 &  \; \ge I\left( 
 \left\{ X_{k,i}   \right\}_{i\in \Bc_k}; \sum_{i=1}^K W_i \Big|\{W_{k'},Z_{k'}\}_{k'\in [K]\backslash\{k\}}
 \right)\\
 & \;= H\left( \sum_{i=1}^K W_i  \Big|\{W_{k'},Z_{k'}\}_{k'\in [K]\backslash\{k\}} \right) -
H\left( \sum_{i=1}^K W_i \Big |\{W_{k'},Z_{k'}\}_{k'\in [K]\backslash\{k\}},   \left\{ X_{k,i}   \right\}_{i\in \Bc_k}    \right)\\
 & \; \overset{(\ref{eq: def Bc_i, problem formulation} ), (\ref{eq: H((X_ki)_i|W_k,Z_k)=0})}{=} H\left( W_k \right) -
H\left( \sum_{i=1}^K W_i \Big |\{W_{k'},Z_{k'}\}_{k'\in [K]\backslash\{k\}}, \left\{ X_{k',i} \right\}_{ k'\in [K]\backslash \{k\}, i\in \Bc_{k'}  } ,  \left\{ X_{k,i} \right\}_{i\in \Bc_k}  \right)\label{eq: proof X cutset bound, converse, step 0}\\
 & \; \ge  H\left( W_k \right) -
H\left( \sum_{i=1}^K W_i  \Big | \left\{ X_{k',i} \right\}_{k'  \in [K],  i\in \Bc_k}  \right)\\
 & \;  \overset{(\ref{eq: H(Yi|(X_{ki})_k)=0})}{=} H\left( W_k \right) -
H\left( \sum_{i=1}^K W_i  \Big | \left\{ X_{k',i} \right\}_{k'  \in [K],  i\in \Bc_k}, \{Y_i\}_{i\in [K]} \right)\label{eq: proof X cutset bound, converse, step 1}\\
& \; \ge  H\left( W_k \right) -
\underbrace{H\left( \sum_{i=1}^K W_i  \Big | \{Y_i\}_{i\in [K]} \right)}_{ \overset{(\ref{eq: recovery constraint})}{=} 0     }\\
& \; = H(W_k)=L,
\end{align}
\end{subequations}
where (\ref{eq: proof X cutset bound, converse, step 0}) is due to  $\{W_{k'},Z_{k'}\}_{k'\in [K]\backslash\{k\}}\Rightarrow \left\{ X_{k',i} \right\}_{ k'\in [K]\backslash \{k\}, i\in \Bc_{k'}}$ and (\ref{eq: proof X cutset bound, converse, step 1})
is due to $ \left\{ X_{k',i} \right\}_{k'  \in [K], i\in \Bc_k}\Rightarrow \{Y_i\}_{i\in [K]}  $ by the \comm protocols (\ref{eq: H((X_ki)_i|W_k,Z_k)=0}), (\ref{eq: H(Yi|(X_{ki})_k)=0}).

Similarly, 
\begin{subequations}
\label{eq: proof Y cutset bound, converse}
\begin{align}
& H\left( \{Y_i\}_{i\in \Bc_k} |\{W_{k'},Z_{k'}\}_{k'\in [K]\backslash\{k\}} \right)  \notag\\
& \; \ge  I\left( \{Y_i\}_{i\in \Bc_k}; \sum_{i=1}^K W_i \Big|\{W_{k'},Z_{k'}\}_{k'\in [K]\backslash\{k\}}   \right)\\
&  \;  =  H\left( \sum_{i=1}^K W_i \Big|\{W_{k'},Z_{k'}\}_{k'\in [K]\backslash\{k\}} \right) - H\left( \sum_{i=1}^K W_i  \Big|\{W_{k'},Z_{k'}\}_{k'\in [K]\backslash\{k\}} ,\{Y_i\}_{i\in \Bc_k}\right)\\
&  \; \overset{(\ref{eq: def Bc_i, problem formulation} ),(\ref{eq: H((X_ki)_i|W_k,Z_k)=0}), (\ref{eq: H(Yi|(X_{ki})_k)=0})}{=}  H\left( W_k\right) - H\left( \sum_{i=1}^K W_i \Big |\{W_{k'},Z_{k'}\}_{k'\in [K]\backslash\{k\}} ,
\{Y_i\}_{i\in [K] \backslash \Bc_k}, \{Y_i\}_{i\in \Bc_k}    \right)\label{eq: proof Y cutset bound, converse, step 0}\\
& \; \ge 
H\left( W_k\right) -  \underbrace{H\left( \sum_{i=1}^K W_i \Big | \{Y_i\}_{i\in [K] }    \right)}_{ \overset{(\ref{eq: recovery constraint})}{=}0    } \\
& \; = H(W_k)=L.
\end{align}
\end{subequations}
Note that Lemma~\ref{lemma: input cutset bound on X} follows  solely from the recovery constraint (\ref{eq: recovery constraint}) at the server; neither the \ssec constraint (\ref{eq: server security constraint})  nor   the \rsec constraint (\ref{eq: relay security constraint}) was invoked in the proof.
\end{IEEEproof}

\subsection{Lower Bounds on \Comm Rates}
\label{subsec: comm rates}

\subsubsection{Proof of $\rx \ge 1$}
We have
\begin{subequations}
\label{eq: proof of Rx>=1, converse}
\begin{align}
\rx  &   \ge     {\sum_{i\in\Bc_k} L_{X,k,i}}/{L}
\\
& \ge {\sum_{i\in \Bc_k}H(X_{k,i})   }/ {L}\\
& \; \ge {H\left(\{X_{k,i}\}_{i\in \Bc_k}\right)   } /{L}\\
& \; \ge  H\left(\{X_{k,i}\}_{i\in \Bc_k}| \{W_{k'},Z_{k'}  \}_{k'\in [K]\backslash  \{k\}  }   \right)/L\\
& \; \overset{(\ref{eq: {X_ij}_j cutset bound, converse})}{\ge}  L/L=1,
\end{align}
\end{subequations}
where Lemma \ref{lemma: input cutset bound on X} is applied in the last step. Note that the above proof follows a cut-set-like argument which does not rely on the security constraints.

\subsubsection{Proof of $\ry \ge \max\{ 1/B, 1/(K-1)\}  $}
We prove  $R_Y \ge  1/B$ and $R_Y \ge 1/(K-1), \forall B \in[K]$ \resp.

\tbf{\emph{Proof of $R_Y \ge 1/B, \forall B\in [K]$}}.
The proof of $R_Y \ge 1/B$ is rather straightforward once equipped with Lemma \ref{lemma: input cutset bound on X}:
\begin{subequations}
\label{eq: proof of Ry>=1/B, converse}
\begin{align}
R_Y \ge \ryh  & \eqdef  \frac{\sum_{k=1}^KH(Y_k) }{KL}
\\
&  = \frac{B\sum_{k=1}^K H(Y_k)  }{BKL}\\
&  = \frac{\sum_{k=1}^K\sum_{i\in \Bc_k}H(Y_i)   }{BKL}\\
&  \ge \frac{\sum_{k=1}^KH\left(\{Y_i\}_{i\in \Bc_k}\right)   }{BKL}\\
&   \ge \frac{\sum_{k=1}^KH\left(\{Y_i\}_{i\in \Bc_k}| \{W_{k'},Z_{k'}\}_{k'\in [K]\backslash  \{k\}  }    \right)   }{BKL}\\
&  \overset{(\ref{eq: Y_i cutset bound, converse}) }{\ge} {KL}/(BKL)= {1}/{B},\label{eq: step 0, proof of Ry>=1/B, converse}
\end{align}
\end{subequations}
where Lemma~\ref{lemma: input cutset bound on X} is used in the last step. Note that the security constraints are not used in the proof  of $\ry \ge 1/B$. 
A direct consequence is that the lower bound $\ry\ge 1/K$ is not tight when $B=K$ as shown in the following section.

\tbf{\emph{Proof of $R_Y \ge 1/(K-1)$}}.
We first present a useful lemma before proceeding to the proof of $R_Y \ge 1/(K-1)$.
\begin{lemma}
\label{lemma: H(Y1:K|W1+...+WK)>=(KRy-1)L}
\emph{
For every $B\in[K]$, it holds that
\be
\label{eq: lemma, H(Y1:K|W1+...+WK)>=(KRy-1)L}
H\left(Y_{1:K} \Big | \sum_{k=1}^KW_k\right) \le (K\ryh-1)L.
\ee 
}
\end{lemma}
\begin{IEEEproof}
It follows that
\begin{subequations}
\label{eq: proof of lemma H(Y1,...,YK|W1+...+WK)<=(KRy-1)L}
\begin{align}
K\ryh L & \overset{(\ref{eq:def entropic rates, converse})}{=} H(Y_1) + \cdots+H(Y_K)\\
& \; \ge H(Y_1,\cdots,Y_K)\\
& \; \overset{(\ref{eq: recovery constraint})}{=}H\left(Y_{1:K}, \sum_{k=1}^K W_k\right)\label{eq: step 0, proof of lemma H(Y1,...,YK|W1+...+WK)<=(KRy-1)L}\\
& \; =H\left(\sum_{k=1}^K W_k\right) +H\left(Y_{1:K} \Big| \sum_{k=1}^K W_k\right) \\
& \; =L +H\left(Y_{1:K} \Big| \sum_{k=1}^K W_k\right), 
\label{eq: step 1, proof of lemma H(Y1,...,YK|W1+...+WK)<=(KRy-1)L}\\
&\Rightarrow H\left(Y_{1:K} \Big| \sum_{k=1}^K W_k\right)   \le \lef(K\ryh-1\rig)L,
\end{align}
\end{subequations}
where (\ref{eq: step 0, proof of lemma H(Y1,...,YK|W1+...+WK)<=(KRy-1)L}) is due to the input sum recovery constraint (\ref{eq: recovery constraint}),   and (\ref{eq: step 1, proof of lemma H(Y1,...,YK|W1+...+WK)<=(KRy-1)L}) is because the inputs are \iid uniform. 
\end{IEEEproof}

Following Lemma \ref{lemma: H(Y1:K|W1+...+WK)>=(KRy-1)L}, we have
\begin{subequations}
\label{eq: proof of Ry>=1/(K-1), converse}
\begin{align}
K \ryh L &  \overset{(\ref{eq:def entropic rates, converse})}{=}H(Y_1) + \cdots + H(Y_K)\\
& \; = \sum_{k=1}^K H(Y_k|W_{1:K}) + I(Y_k;W_{1:K})\\
& \; \le  \sum_{k=1}^K H(Y_k|W_{1:K}) + I\left(Y_k, \{ X_{i,k}\}_{i\in \Uc_k}   ;W_{1:K}\right)   \\
&\; = \sum_{k=1}^K H(Y_k|W_{1:K}) + 
\underbrace{I\left(\{ X_{i,k}\}_{i\in \Uc_k}   ;W_{1:K}\right)}_{ \overset{(\ref{eq: relay security constraint})}{=}0   }  + \underbrace{ I\left(Y_k   ;W_{1:K} |\{ X_{i,k}\}_{i\in \Uc_k}\right)}_{ \overset{(\ref{eq: H(Yi|(X_{ki})_k)=0})}{=}0  }   
\label{eq: step 0, proof of Ry>=1/(K-1), converse}\\
&\; = \sum_{k=1}^K H(Y_k|W_{1:K},W_1+\cdots+W_K)\\
&\; \le \sum_{k=1}^K H(Y_k|W_1+\cdots+W_K) \\
&\; \le \sum_{k=1}^K H(Y_{1:K}|W_1+\cdots+W_K) \\
& \; \overset{(\ref{eq: lemma, H(Y1:K|W1+...+WK)>=(KRy-1)L})}{\le }
K\lef(K\ryh-1\rig)L\label{eq: step 1, proof of Ry>=1/(K-1), converse}\\
& \Rightarrow K\ryh L \le K\lef(K\ryh-1\rig)L \\
& \Rightarrow R_Y \ge \ryh \ge \frac{1}{K-1}.
\end{align}
\end{subequations}
In (\ref{eq: step 0, proof of Ry>=1/(K-1), converse}), $I\left(\{ X_{i,k}\}_{i\in \Uc_k}   ;W_{1:K}\right)=0$ is due to  the security of Relay $k$, and $I\left(Y_k   ;W_{1:K} |\{ X_{i,k}\}_{i\in \Uc_k}\right)=0$ is because $Y_k$ is generated from $\{X_{i,k}\}_{i\in \Uc_k} $ (see (\ref{eq: H(Yi|(X_{ki})_k)=0})).  In (\ref{eq: step 1, proof of Ry>=1/(K-1), converse}), Lemma \ref{lemma: H(Y1:K|W1+...+WK)>=(KRy-1)L} is applied. As a result, we have proved $R_Y \ge 1/(K-1)$. Note that only the \rsec constraint (\ref{eq: relay security constraint}) is invoked in the proof of $\ry\ge 1/(K-1)$.
Combining (\ref{eq: proof of Ry>=1/B, converse}) and (\ref{eq: proof of Ry>=1/(K-1), converse}), we conclude that $R_Y \ge \max \{ \frac{1}{B}, { \frac{1}{K-1}   }  \},\forall B\in [K]$.

\subsection{Lower Bound on Source Key Rate} 
\label{subsec: source key rate} 
Let $Z_{k,i}$ denote the subset of key symbols from $Z_k$ that appear in the \msg $X_{k,i}$, \ie,  $H(Z_{k,i})=H(X_{k,i}|W_k), k\in[K], i\in \Bc_k$.  We have
\be
\label{eq: H(Zki,all i|Zk)=0}
H(\{Z_{k,i}\}_{i\in \Bc_k}|Z_k)=0,\; \forall k\in [K]
\ee 
Equation~(\ref{eq: H((X_ki)_i|W_k,Z_k)=0}) is thus equivalent to  
\be 
\label{eq: H(Xki|Wk,Zki)=0}
H(X_{k,i}|W_k,Z_{k,i})=0,\;\forall k\in[K], \forall i\in \Bc_k.
\ee
We first provide a useful  lemma which states that the joint entropy of the key symbols contained in the \msgs received by any relay should  not be smaller than the sum of the conditional entropy of the \msgs received by that relay.
\begin{lemma}
\label{lemma: H(Zki,all k) lower bound}
\emph{
Suppose  $\Uc_i=\{k_{i,1},\cdots,k_{i,B}\}$ where $k_{i,1}<\cdots<k_{i,B}$. For any $B\in [K]$, it holds that 
\begin{align}
H\left( \{Z_{k,i}\}_{k\in \Uc_i}   \right)  & \ge
\sum_{j=1}^B H\left( X_{k_{i,j},i} \big | \left\{
W_{k_{i,j'}},  Z_{k_{i,j'}}   
\right\}_{j' \in [j-1]}    \right),\;  \forall i\in[K]
 \label{eq: lemma, H(Zki,all k) lower bound}
\end{align}
}
\end{lemma}
\begin{IEEEproof}
Recall that  $\Uc_i=\{k_{i,1},\cdots,k_{i,B}\}$ represents the the set of users associated with Relay $i\in[K]$ as given by (\ref{eq: def Uc_i, problem formulation}). For any $i\in[K]$, we have
\begin{subequations}
\label{eq: proof H(Zki,all k) lower bound, lemma, converse}
\begin{align}
H\left( \{Z_{k,i}\}_{k\in \Uc_i}   \right)  & \ge  
H\left( \{Z_{k,i}\}_{k\in \Uc_i} | \{W_k\}_{k\in \Uc_i}  \right) \\
 &   \ge I\left( \{Z_{k,i}\}_{k\in \Uc_i};\{X_{k,i}\}_{k\in \Uc_i} | \{W_k\}_{k\in \Uc_i}  \right) \\
  &  = H\left( \{X_{k,i}\}_{k\in \Uc_i}| \{W_k\}_{k\in \Uc_i}  \right)   -
 \underbrace{H\left( \{X_{k,i}\}_{k\in \Uc_i}| \{W_k\}_{k\in \Uc_i} ,\{Z_{k,i}\}_{k\in \Uc_i} \right)}_{\overset{(\ref{eq: H(Xki|Wk,Zki)=0})}{=}0   }\label{eq: step 0, proof H(Zki,all k) lower bound, lemma, converse}\\
&  = H\left( \{X_{k,i}\}_{k\in \Uc_i} \right) - 
\underbrace{I\left( \{X_{k,i}\}_{k\in \Uc_i};\{W_k\}_{k\in \Uc_i}  \right)}_{\overset{(\ref{eq: relay security constraint})}{=}0  }\label{eq: step 1, proof H(Zki,all k) lower bound, lemma, converse}\\
& = \sum_{j=1}^B H\left( X_{k_{i,j},i} \big | \left\{X_{k_{i,j'},i} \right\}_{j'\in [j-1]}  \right)\\
& \ge  \sum_{j=1}^B H\left( X_{k_{i,j},i}  \big| \left\{X_{k_{i,j'},i} \right\}_{j'\in [j-1]}, \left\{
W_{k_{i,j'}},  Z_{k_{i,j'}}   
\right\}_{j' \in [j-1]}   \right)\\
& \overset{(\ref{eq: H(Xki|Wk,Zki)=0}) }{=}  \sum_{j=1}^B H\left( X_{k_{i,j},i} \big|  \left\{
W_{k_{i,j'}},  Z_{k_{i,j'}}   
\right\}_{j' \in [j-1]}   \right),\label{eq: step 2, proof H(Zki,all k) lower bound, lemma, converse}
\end{align}
\end{subequations}
where (\ref{eq: step 0, proof H(Zki,all k) lower bound, lemma, converse}) and (\ref{eq: step 2, proof H(Zki,all k) lower bound, lemma, converse}) are due to the dependence of the \msg $X_{k,i}$ on the input $W_k$ and key $Z_{k,i}$ for every $i\in \Bc_k$ (see (\ref{eq: H(Xki|Wk,Zki)=0}); (\ref{eq: step 1, proof H(Zki,all k) lower bound, lemma, converse}) is due to the security constraint (\ref{eq: relay security constraint}) of Relay $i$. As a result, Lemma \ref{lemma: H(Zki,all k) lower bound} is proved.
\end{IEEEproof}

We derive a lower bound on the source key rate 
$\rzsigma \ge \max \{1,K/B-1\}$, which takes the maximum of two terms: the first term $\rzsigma\ge 1$ arises from \rsec, while  the second  term $\rzsigma\ge K/B-1$ follows from the \ssec \req. In the following, we prove these two lower bounds \resp.
Again, denote $\Uc_i=\{k_{i,1},\cdots,k_{i,B}\}$ where $k_{i,1}<\cdots<k_{i,B},\forall i\in[K]$. We have
\begin{subequations}
\label{eq: proof of source key rate, Rzsigma>=1}
\begin{align}
KH(\zsigma)  &  \overset{(\ref{eq: indiv key gen based on source key}), (\ref{eq: H(Zki,all i|Zk)=0})}{\ge}     \sum_{i=1}^K H\left( \{Z_{k,i}\}_{k\in \Uc_i}    \right)\label{eq: step 0, proof of source key rate}\\
& \overset{(\ref{eq: lemma, H(Zki,all k) lower bound})}{\ge} \sum_{i=1}^K\sum_{j=1}^B H\left( X_{k_{i,j},i} \big | \left\{
W_{k_{i,j'}},  Z_{k_{i,j'}}   
\right\}_{j' \in [j-1]}    \right)\label{eq: step 1, proof of source key rate}\\
& = \sum_{k=1}^K \sum_{i\in \Bc_k} H\left( X_{k,i}|\{W_j,Z_j \}_{j\in \Uc_i:j\le k-1} \right)\label{eq: step 10, proof of source key rate}\\
& \ge \sum_{k=1}^K  H\left( \{X_{k,i}\}_{i\in \Bc_k}|\{W_j,Z_j \}_{j\in \Uc_i:j\le k-1} \right)\\
& \ge \sum_{k=1}^K  H\left( \{X_{k,i}\}_{i\in \Bc_k}|\{W_j,Z_j \}_{j\in [K]\backslash \{k\}} \right)\label{eq: step 2, proof of source key rate}\\
& \overset{(\ref{eq: {X_ij}_j cutset bound, converse})}{\ge} KL, \label{eq: step 3, proof of source key rate}\\
& \Rightarrow \rzsigma \ge \rzsigmah \eqdef \frac{H(\zsigma)}{L} \ge 1,
\end{align}
\end{subequations}
where (\ref{eq: step 0, proof of source key rate}) is because all the \indiv keys are generated from the source key (see (\ref{eq: indiv key gen based on source key}), (\ref{eq: H(Zki,all i|Zk)=0})); In (\ref{eq: step 1, proof of source key rate}), Lemma \ref{lemma: H(Zki,all k) lower bound} is applied; In (\ref{eq: step 10, proof of source key rate}), the terms are rearranged in the summation. 
(\ref{eq: step 2, proof of source key rate}) is because $\{j\in \Uc_i:j\le k-1\} \subseteq [K]\backslash \{k\} $; In (\ref{eq: step 3, proof of source key rate}), Lemma \ref{lemma: input cutset bound on X} is applied. Therefore, we have proved $\rzsigma\ge 1$. 

{
\begin{example}[Converse for Example \ref{example 1}: $\rzsigma\ge 1$]
\label{example 2} 
To illustrate the above converse proof, let us revisit Example \ref{example 1} and prove $\rzsigma \ge 1$. 
Note that $ \Uc_1=\{1,3\},  \Uc_2=\{1,2\}$ and $\Uc_3 =\{2,3\}$. Lemma \ref{lemma: H(Zki,all k) lower bound} suggests
\begin{align}
H(Z_{1,1}, Z_{3,1}) & \ge H(X_{1,1}) + H(X_{3,1}|W_1,Z_1),\\
H(Z_{1,2}, Z_{2,2}) & \ge H(X_{1,2}) + H(X_{2,2}|W_1,Z_1),\\
H(Z_{2,3}, Z_{3,3}) & \ge H(X_{2,3}) + H(X_{3,3}|W_2,Z_2).
\end{align}
Therefore,
\begin{subequations}
\label{eq: example, proof of Rzsigma>=1, converse}
\begin{align}
3H(\zsigma) & \ge H(Z_{1,1}, Z_{3,1})+H(Z_{1,2}, Z_{2,2})+H(Z_{2,3}, Z_{3,3})\\
&  \ge H(X_{1,1}) + H(X_{1,2}) + H(X_{2,2}|W_1,Z_1)+H(X_{2,3})\notag  \\ 
&\quad  +H(X_{3,1}|W_1,Z_1)+H(X_{3,3}|W_2,Z_2)\\
&  \ge H(X_{1,1}) + H(X_{1,2}) + H(X_{2,2}|W_1,Z_1)+H(X_{2,3}|W_1,Z_1)\notag  \\ 
&\quad  +H(X_{3,1}|W_1,Z_1,W_2,Z_2)+H(X_{3,3}|W_1,Z_1,W_2,Z_2)\\
&  \ge H(X_{1,1},X_{1,2}) + H(X_{2,2},X_{2,3}|W_1,Z_1)  +H(X_{3,1},X_{3,3}|W_1,Z_1,W_2,Z_2)\\
&  \ge H(X_{1,1},X_{1,2}|W_2,Z_2,W_3,Z_3) + H(X_{2,2},X_{2,3}|W_1,Z_1,W_3,Z_3 )\notag \\
&  \quad +H(X_{3,1},X_{3,3}|W_1,Z_1,W_2,Z_2)\\
& \overset{(\ref{eq: {X_ij}_j cutset bound, converse})}{\ge } 3L,\\
&  \Rightarrow \rzsigma \ge  1,
\end{align}
\end{subequations}
which establishes the optimality  of Example \ref{example 1}.
\hfill $\lozenge$
\end{example}

}

Next, we prove $ \rzsigma\ge K/B-1$, which is due to \ssec. Let $Z^Y_i$ denote the set of key symbols in $Z_i$ that influence the \msg $Y_i,i\in [K]$, \ie, $H(Z^Y_i)= H(Y_i|\{W_k\}_{k\in  \Uc_i}  )$. We have
\begin{align}
H\left(Y_i|\{W_k\}_{k\in  \Uc_i}, Z^Y_i\right) &=0,\; \forall i\in[K]
\label{eq: H(Yi|Wk,k in Ui, Z^Y_i)=0}\\
H\left(Z^Y_i| \{Z_{k,i}\}_{k\in \Uc_i}   \right)&=0,\; \forall i\in[K]\label{eq: H(Z^Y_i|Zki,k in Ui)=0}
\end{align}
An immediate consequence with the definition of $\{Z^Y_i\}_{i\in[K]}$ is Lemma~\ref{lemma: H(Z^Y_1,...,Z^Y_K)>=H(Y1,...,YK)-L}, which states that the joint entropy of the key variables $\{Z^Y_i\}_{i\in[K]}$ must be at least the size of the joint entropy of the \msgs $\{Y_i\}_{i\in[K]}$ minus the input size $L$.

\begin{lemma}
\label{lemma: H(Z^Y_1,...,Z^Y_K)>=H(Y1,...,YK)-L}
\emph{
With $Z^Y_i$'s defined in (\ref{eq: H(Yi|Wk,k in Ui, Z^Y_i)=0}) and (\ref{eq: H(Z^Y_i|Zki,k in Ui)=0}), it holds that
\be
\label{eq: lemma, H(Z^Y_1,...,Z^Y_K)>=H(Y1,...,YK)-L}
H\left(Z^Y_1,\cdots, Z^Y_K\right) \ge 
H(Y_1,\cdots,Y_K)-L.
\ee 
}
\end{lemma}
\begin{IEEEproof}
The \ssec constraint (\ref{eq: server security constraint}) requires  that  the server must learn only the input sum $W_1+  \cdots + W_K$ from the \msgs $Y_1,\cdots,Y_K$, and nothing beyond that. Consequently, any information not contained in the input sum must be concealed by the key variables. This implies that the key size must be at least the entropy of the protected input symbols, \ie, $H(Y_{1:K})-H(\sum_{k=1}^KW_k)$.
More specifically,
\begin{subequations}
\label{eq: proof, lemma, H(Z^Y_1,...,Z^Y_K)>=H(Y1,...,YK)-L}
\begin{align}
H\left(Z^Y_1,\cdots, Z^Y_K\right) & 
\overset{(\ref{eq: key input independence})}{=}H\left(Z^Y_1,\cdots, Z^Y_K|W_{1:K}\right)\label{eq: step 0, proof, lemma, H(Z^Y_1,...,Z^Y_K)>=H(Y1,...,YK)-L}\\
&  \ge  I\left(Z^Y_1,\cdots, Z^Y_K; Y_{1:K}|W_{1:K}\right)\\
&  \ge  H\left( Y_{1:K}|W_{1:K}\right) -\underbrace{H\left( Y_{1:K}|W_{1:K}, Z^Y_1,\cdots, Z^Y_K\right)}_{\overset{(\ref{eq: H(Yi|Wk,k in Ui, Z^Y_i)=0})}{=}0}\label{eq: step 1, proof, lemma, H(Z^Y_1,...,Z^Y_K)>=H(Y1,...,YK)-L}  \\
& = H\left( Y_{1:K}\right) -
I\left( Y_{1:K}; W_{1:K}, \sum_{k=1}^KW_k \right)  \\
& = H\left( Y_{1:K}\right) -
I\left( Y_{1:K}; \sum_{k=1}^KW_k \right) - \underbrace{I\left( Y_{1:K}; W_{1:K}\Big| \sum_{k=1}^KW_k \right)}_{\overset{(\ref{eq: server security constraint})}{=}0  }\label{eq: step 2, proof, lemma, H(Z^Y_1,...,Z^Y_K)>=H(Y1,...,YK)-L}  \\
& \ge H\left( Y_{1:K}\right) -
H\left( \sum_{k=1}^KW_k \right)\\
& \ge H\left( Y_{1:K} \right)-L,
\end{align}
\end{subequations}
where (\ref{eq: step 0, proof, lemma, H(Z^Y_1,...,Z^Y_K)>=H(Y1,...,YK)-L})  is due to the independence between the inputs and the keys (see (\ref{eq: key input independence})); (\ref{eq: step 1, proof, lemma, H(Z^Y_1,...,Z^Y_K)>=H(Y1,...,YK)-L}) is due to the \msg generation protocol (\ref{eq: H(Yi|Wk,k in Ui, Z^Y_i)=0}); (\ref{eq: step 2, proof, lemma, H(Z^Y_1,...,Z^Y_K)>=H(Y1,...,YK)-L}) is due to the \ssec constraint (\ref{eq: server security constraint}). The last step is due to the uniformity and independence of the inputs (see (\ref{eq: key input independence})). 
\end{IEEEproof}

Equipped with Lemma \ref{lemma: H(Z^Y_1,...,Z^Y_K)>=H(Y1,...,YK)-L}, we prove $\rzsigma \ge  K/B-1$ \af. 
Denote
\be 
\delta \eqdef\left \lfloor \frac{K}{B} \right\rfloor \in  \{1, \cdots, K\}
\ee 
Also denote $Y_{\Sc}\eqdef\{Y_i\}_{i\in \Sc}$ and $ \Uc_{\Sc} \eqdef  \cup_{i\in \Sc} \Uc_i, \forall \Sc \subseteq [K]$. We define $B$ permutation vectors $\pv_1,\cdots, \pv_B$, each of length-$K$. The first vector is defined as $\pv_1\eqdef (1,2,\cdots,K)$, 
and for each $k\in [2:B]$, 
$\pv_i\eqdef(p_{k,1}, p_{k,2},\cdots,p_{k,K})$ is  obtained  by circularly shifting $\pv_1$    to the left by $k-1$ positions. For example, $\pv_2=(2,3,
\cdots,K,1)$, and so on.  
We have
\begin{subequations}
\label{eq: proof of Rzsigma>=K/B-1, converse}
\begin{align}
BH\left(\zsigma \right) & \ge
BH\left(Z^Y_1,\cdots, Z^Y_K\right) \\
& \overset{(\ref{eq: lemma, H(Z^Y_1,...,Z^Y_K)>=H(Y1,...,YK)-L})}{\ge} BH\left(Y_1,\cdots, Y_K\right) - BL\\
& = \sum_{i=1}^B H(Y_{p_{i,1}},Y_{p_{i,2}}, \cdots, Y_{p_{i,K}}  )-BL\\
& = H(Y_1,Y_2,\cdots,Y_K) +\sum_{k=2}^B H(Y_{p_{k,1}},Y_{p_{k,2}}, \cdots, Y_{p_{k,K}}  )-BL\\
& = \sum_{i=1}^{\delta  } H\left(Y_{(i-1)B+1:iB}|Y_{1:(i-1)B} \right) +  \sum_{i=1}^{K-\delta  B  }H\left( Y_{\delta   B+i }| Y_{ 1: \delta  B+i-1  }   \right) \notag\\
& \quad + \sum_{k=2}^B \left( \sum_{i=1}^{K- \delta B   }  H\left(Y_{p_{k,K-i+1 }}  | Y_{{\{p_{k,K-i+2 },\cdots, p_{k,K} \}  } } \right)\right.\notag \\
& \quad\quad \quad \quad \quad +  \left. \sum_{i=1}^{ \delta    } H\left( Y_{\left\{  
 p_{k, \left(\delta-i\right)B+1},\cdots , p_{k, \left(\delta+1-i\right)B}\right \} }  
 | Y_{\left\{  p_{k, \left(\delta+1-i\right)B+1},\cdots , p_{k,K}\right\}  }  \right)   \right)-BL \label{eq: step -2, proof of Rzsigma>=K/B-1, converse} \\
& \ge \sum_{i=1}^{\delta  } H\left(Y_{(i-1)B+1:iB}|Y_{1:(i-1)B},  \{W_{k'}, Z_{k'}\}_{k'\in[K] \backslash \{(i-1)B+1\}   } \right) \notag \\
 & \quad + \sum_{k=2}^B \sum_{i=1}^{ \delta    } H\left( Y_{\left\{ 
 p_{k, \left(\delta-i\right)B+1},\cdots , p_{k, \left(\delta+1-i\right)B}\right \}  }  
 | Y_{\left\{ p_{k, \left(\delta+1-i\right)B+1},\cdots ,p_{k,K}\right\}   }  
  , \{W_{k'}, Z_{k'}\}_{k'\in[K] \backslash \left \{ p_{k, \left(\delta-i\right)B+1    }  \right\}   }   \right) \notag\\
  &\quad  + \sum_{i=1}^{K-\delta B  }H\left( Y_{\delta  B+i }| Y_{ 1: \delta B+i-1    }   \right)  + \sum_{k=2}^B  \sum_{i=1}^{K- \delta B   }  H\left(Y_{p_{k,K-i+1 }}  | Y_{\left\{ p_{k,K-i+2 },\cdots , p_{k,K} \right\} } \right)-BL \label{eq: step -1, proof of Rzsigma>=K/B-1, converse}\\
  & \overset{(\ref{eq: H((X_ki)_i|W_k,Z_k)=0}), (\ref{eq: H(Yi|(X_{ki})_k)=0})}{=}    \sum_{i=1}^{\delta } H\left( 
\underbrace{
Y_{(i-1)B+1:iB}}
_{=Y_{\Bc_{(i-1)B+1}  }}
| \{W_{k'}, Z_{k'}\}_{k'\in[K] \backslash \{(i-1)B+1\}   } \right) \notag \\
 & \qquad + \sum_{k=2}^B \sum_{i=1}^{ \delta  } H\left( 
 \underbrace{
 Y_{\left\{  
 p_{k, \left(\delta-i\right)B+1},\cdots , p_{k, \left(\delta+1-i\right)B}\right \}  }
 }_{=Y_{\Bc_{p_{k, \left(\delta-i\right)B+1}   }}  }
 |\{W_{k'}, Z_{k'}\}_{k'\in[K] \backslash \left \{ p_{k, \left(\delta-i\right)B+1    }  \right\}   }   \right) -BL\notag\\
  &\qquad  + \sum_{i=1}^{K-\delta B  }H\left( Y_{\delta  B+i }| Y_{ 1: \delta B+i-1 }   \right)  + \sum_{k=2}^B  \sum_{i=1}^{K- \delta B   }  H\left(Y_{p_{k,K-i+1 }}  | Y_{\left\{  p_{k,K-i+2 },\cdots , p_{k,K} \right\}  } \right) \label{eq: step 0, proof of Rzsigma>=K/B-1, converse}\\
& \overset{(\ref{eq: Y_i cutset bound, converse})}{\ge} \sum_{i=1}^{K-\delta B  }H\left( Y_{\delta  B+i }| Y_{ 1: \delta B+i-1 }   \right)+ \sum_{k=2}^B  \sum_{i=1}^{K- \delta B   }  H\left(Y_{p_{k,K-i+1 }}  | Y_{\left\{  p_{k,K-i+2 },\cdots , p_{k,K} \right\}  } \right)  + (\delta-1)BL \label{eq: step 1, proof of Rzsigma>=K/B-1, converse}\\
&= \sum_{i=1}^{K-\delta B}
\left( H\left( Y_{\delta  B+i }| Y_{ 1: \delta B+i-1    }   \right)+ \sum_{k=2}^B   H\left(Y_{p_{k, \delta B+i   }}  | Y_{\left\{ p_{k, \delta B+i+1 },\cdots , p_{k,K} \right\}} \right)  
\right)  + (\delta -1)BL\label{eq: step 1.5, proof of Rzsigma>=K/B-1, converse}\\
&= \sum_{i=1}^{K-\delta B }
\left( H\left( Y_{p_{1,\delta  B+i}  }| Y_{\{p_{1,1}, \cdots, p_{1,\delta B+i-1  }  \} }   \right)+ 
\sum_{k=2}^B   H\left(Y_{p_{k, \delta B+i   }}  | Y_{\left\{ p_{k, \delta B+i+1 },\cdots , p_{k,K} \right\} } \right)
\right)   + (\delta-1)BL,\label{eq: step 2, proof of Rzsigma>=K/B-1, converse}
\end{align}
\end{subequations}
where in (\ref{eq: step -2, proof of Rzsigma>=K/B-1, converse}), the chain rule of joint entropy is applied. 
Specifically, 
$H(Y_1,\cdots, Y_K)$ is first partitioned into $\delta$ blocks---namely, $Y_{1:B}, Y_{B+1:2B}, \cdots, Y_{(\delta-1)B+1: \delta B}$---along with the $K-\delta B$ remaining terms $Y_{\delta B+1}, \cdots, Y_K$. Chain rule is then applied first across the blocks and subsequently across the remaining terms. Similarly,  for each $k\in [2:K]$, 
the permuted sequence
$Y_{p_{k,1}}, \cdots, Y_{p_{k,K}} $ is decomposed into $ \delta$ blocks $Y_{\{p_{k,(\delta-1)B+1}, \cdots,p_{k,\delta B}\}}$, $\cdots$, $Y_{\{p_{k, B+1},\cdots, p_{k,2B}\}}$,   $Y_{\{p_{k,1},\cdots, p_{k,B}\}}    $, \awa $K-\delta B$ remaining terms  $Y_{p_{k, \delta  B +1}  }, \cdots, Y_{p_{k, K   }}$. In this case, chain rule is  first applied to the remaining terms and then to the blocks, in reverse order.
(\ref{eq: step 0, proof of Rzsigma>=K/B-1, converse}) is  because in (\ref{eq: step -1, proof of Rzsigma>=K/B-1, converse}), user $(i-1)B+1$ does not appear  in $ \Uc_{[1: (i-1)B]}$, and therefore the \msg block $Y_{1:(i-1)B}$ does not depend on user $(i-1)B+1$. Consequently, $Y_{1:(i-1)B}$ can be recovered from $\{W_{k'},Z_{k'}\}_{k'\in [K]\backslash \{(i-1)B+1\} }$.
 Similarly, observe that $ \Bc_{p_{k, (\delta-i )B+1}  }=\left\{  p_{k, (\delta-i )B+1},\cdots ,
p_{k, (\delta-i+1 )B}
\right\} $, which has no overlap with $\left\{ p_{k, (\delta-i+1 )B+1},\cdots ,p_{k,K}  \right\}$. Therefore, 
$Y_{\left\{ p_{k, (\delta-i+1 )B+1},\cdots ,p_{k,K} \right \}  }$ does not depend on user $p_{k, (\delta-i )B+1}$, and can be recovered from  
$ \{W_{k'}, Z_{k'}\}_{k'  \in [K] \backslash \{p_{k, (\delta-i )B+1} \}  }$. 
In (\ref{eq: step 1, proof of Rzsigma>=K/B-1, converse}), Lemma \ref{lemma: input cutset bound on X} is applied. Finally,
in (\ref{eq: step 2, proof of Rzsigma>=K/B-1, converse}), we used the fact that $\pv_1=(1,\cdots,K)$.

Now we show  that each term  in the summation of (\ref{eq: step 2, proof of Rzsigma>=K/B-1, converse}) is no smaller than $L$. Note that the permutation vectors satisfy
\begin{subequations}
\begin{align}
p_{k+1,i} & = \modular(p_{k, i},K) +1, \; \forall i\in[K], k\in [B-1]\label{eq: permutation property 1} \\
p_{k,i+1}  & =p_{k+1,i},\; \forall i\in [K-1], k\in [B-1]\label{eq: permutation property 2} 
\end{align}
\end{subequations}
Consequently,
\be 
\label{eq: Bc_k def, converse}
\Bc_{k}= 
\left\{ p_{1,k}, 
p_{2,k},\cdots,
p_{B,k}
\right\},\;  k\in [K]
\ee 
according to (\ref{eq: def Bc_i, problem formulation}).
For each $i\in[1: K- \delta B    ]$, define
\begin{align}
\Gc_i & \eqdef  \left(   \bigcup_{j\in [\delta B+i+2:K  ] } \Bc_{j}\right) \cup \{p_{B, \delta B+i+1  }    \}\backslash  \Bc_{\delta B+i}
,\label{eq: def Gi}\\
 \Hc_i & \eqdef \{p_{1,1},p_{1,2},\cdots, p_{1,\delta B+i-1}\} = \bigcup_{j\in [(\delta-1)B+i]} \Bc_j.\label{def: Hi}
\end{align}
For each $i\in[K- \delta B    ]   $, we have
\begin{subequations}
\label{eq: proof of each summand>=L, converse. Rzsigma>=K/B-1}
\begin{align}
& H\left( Y_{p_{1,\delta  B+i}  }| Y_{\Hc_i }   \right)+ 
\sum_{k=2}^B   H\left(Y_{p_{k, \delta B+i   }}  | Y_{\left\{p_{k, \delta B+i+1 },\cdots, p_{k,K}\right \}  } \right) \notag\\
&\quad \overset{(\ref{eq: Bc_k def, converse})}{\ge}    H\left( Y_{p_{1,\delta  B+i}  }| Y_{\Hc_i }   \right) +
\sum_{k=2}^{B-1}
H\left(Y_{p_{k, \delta B+i}}  | Y_{p_{k, \delta B+i+1}}, Y_{\Gc_i} \right) + 
H\left(Y_{p_{B, \delta B+i}}  | Y_{\Gc_i} \right)\\
& \quad \overset{\trm{(\ref{eq: permutation property 2})}}{=}    H\left( Y_{p_{1,\delta  B+i}  }| Y_{\Hc_i }   \right) +
\sum_{k=2}^{B-1}
H\left(Y_{p_{k, \delta B+i}}  | Y_{p_{k+1, \delta B+i}}, Y_{\Gc_i} \right) + 
H\left(Y_{p_{B, \delta B+i}}  | Y_{\Gc_i} \right)\label{eq: step 0, proof of each summand>=L, converse. Rzsigma>=K/B-1}\\
& \quad \ge 
H\left( Y_{p_{1,\delta  B+i}  }| Y_{\Hc_i }   \right) +
\sum_{k=2}^{B-1}
H\left(Y_{p_{k, \delta B+i}}  | Y_{p_{k+1, \delta B+i}}, \cdots, Y_{p_{B, \delta B+i} }  , Y_{\Gc_i} \right) + 
H\left(Y_{p_{B, \delta B+i}}  | Y_{\Gc_i} \right)\\
& \quad =     H\left( Y_{p_{1,\delta  B+i}  }| Y_{\Hc_i }   \right) +
H\left(\left\{ Y_{p_{k, \delta B+i}}\right\}_{k \in [2:B-1]}  | Y_{\Gc_i} \right) + 
H\left(Y_{p_{B, \delta B+i}}  | Y_{\Gc_i} \right)\label{eq: step 1, proof of each summand>=L, converse. Rzsigma>=K/B-1}\\
& \quad\ge     H\left( Y_{p_{1,\delta  B+i}  }| Y_{\Hc_i }   \right) +
H\left(\left\{ Y_{p_{k, \delta B+i}}\right\}_{k \in [2:B]}  | Y_{\Gc_i} \right)\label{eq: step 2, proof of each summand>=L, converse. Rzsigma>=K/B-1}\\
& \quad\ge     H\left( Y_{p_{1,\delta  B+i}  }| Y_{\Hc_i \cap \{p_{k,\delta B+i}  \}_{k\in[2:B] }}, Y_{\left(\Hc_i  \backslash \{p_{k,\delta B+i}  \}_{k\in[2:B] }\right)\cup  \Gc_i      }   \right) \notag\\
& \qquad +
H\left(\left\{ Y_{p_{k, \delta B+i}}\right\}_{k \in [2:B]}  | 
Y_{\left(\Hc_i  \backslash \{p_{k,\delta B+i}  \}_{k\in[2:B] }\right)\cup  \Gc_i      } 
\right)\\
& \quad\ge  
H\left(\left\{ Y_{p_{k, \delta B+i}}\right\}_{k \in [1:B]}  | 
Y_{\left(\Hc_i  \backslash \{p_{k,\delta B+i}  \}_{k\in[2:B] }\right)\cup  \Gc_i      } 
\right)\\
& \quad   \overset{(\ref{eq: Bc_k def, converse})}{=}  
H\left( Y_{\Bc_{\delta B+i}}  | 
Y_{\left(\Hc_i  \backslash \{p_{k,\delta B+i}  \}_{k\in[2:B] }\right)\cup  \Gc_i      } 
\right)\\
& \quad \ge  
H\left( Y_{\Bc_{\delta B+i}}  | 
Y_{\left(\Hc_i  \backslash \{p_{k,\delta B+i}  \}_{k\in[2:B] }\right)\cup  \Gc_i}, \{W_k, Z_k\}_{k\in [K]\backslash \{ \delta B+i \}}   
\right)\label{eq: step 3, proof of each summand>=L, converse. Rzsigma>=K/B-1}\\
& \quad =  
H\left( Y_{\Bc_{\delta B+i}}  | \{W_k, Z_k\}_{k\in [K]\backslash \{ \delta B+i \}}   
\right)\label{eq: step 4, proof of each summand>=L, converse. Rzsigma>=K/B-1}\\
& \quad  \overset{(\ref{eq: Y_i cutset bound, converse})}{\ge }
L, 
\end{align}
\end{subequations}
where (\ref{eq: step 0, proof of each summand>=L, converse. Rzsigma>=K/B-1}) is due to the property of the permutation vectors as shown in (\ref{eq: permutation property 2}). (\ref{eq: step 4, proof of each summand>=L, converse. Rzsigma>=K/B-1}) is due to $ \Bc_{p_{\delta B+i}  } \cap \left(\Hc_i  \backslash \{p_{k,\delta B+i}  \}_{k\in[2:B] }\right) = \emptyset$, which implies that $Y_{\left(\Hc_i  \backslash \{p_{k,\delta B+i}  \}_{k\in[2:B] }\right)}$ can be recovered from $\{W_k, Z_k\}_{k\in [K]\backslash \{ \delta B+i \}}$. 

Continuing with (\ref{eq: step 2, proof of Rzsigma>=K/B-1, converse}), we have
\begin{align}
BH(\zsigma) &  \ge (K-\delta B)L + (\delta-1)BL\notag \\
&  =(K-B)L, \\
& \Rightarrow \rzsigma \ge \rzsigmah \eqdef \frac{H(\zsigma)}{L} \ge  \frac{K}{B}-1.
 \end{align}
This completes the proof of  $\rzsigma \ge  \max\{1,K/B-1\}$.
It is worth noting that the core idea of the above proof lies in the careful selection of conditional entropy terms whose combinations give rise to new inputs. A complete illustration of this converse argument is provided in the following example.

\begin{example}[Illustration of Converse Technique]
\label{example: illustration of converse technique for proving Rzsigma>=K/B-1}
\begin{figure}
    \centering
\includegraphics[width=0.5\linewidth]{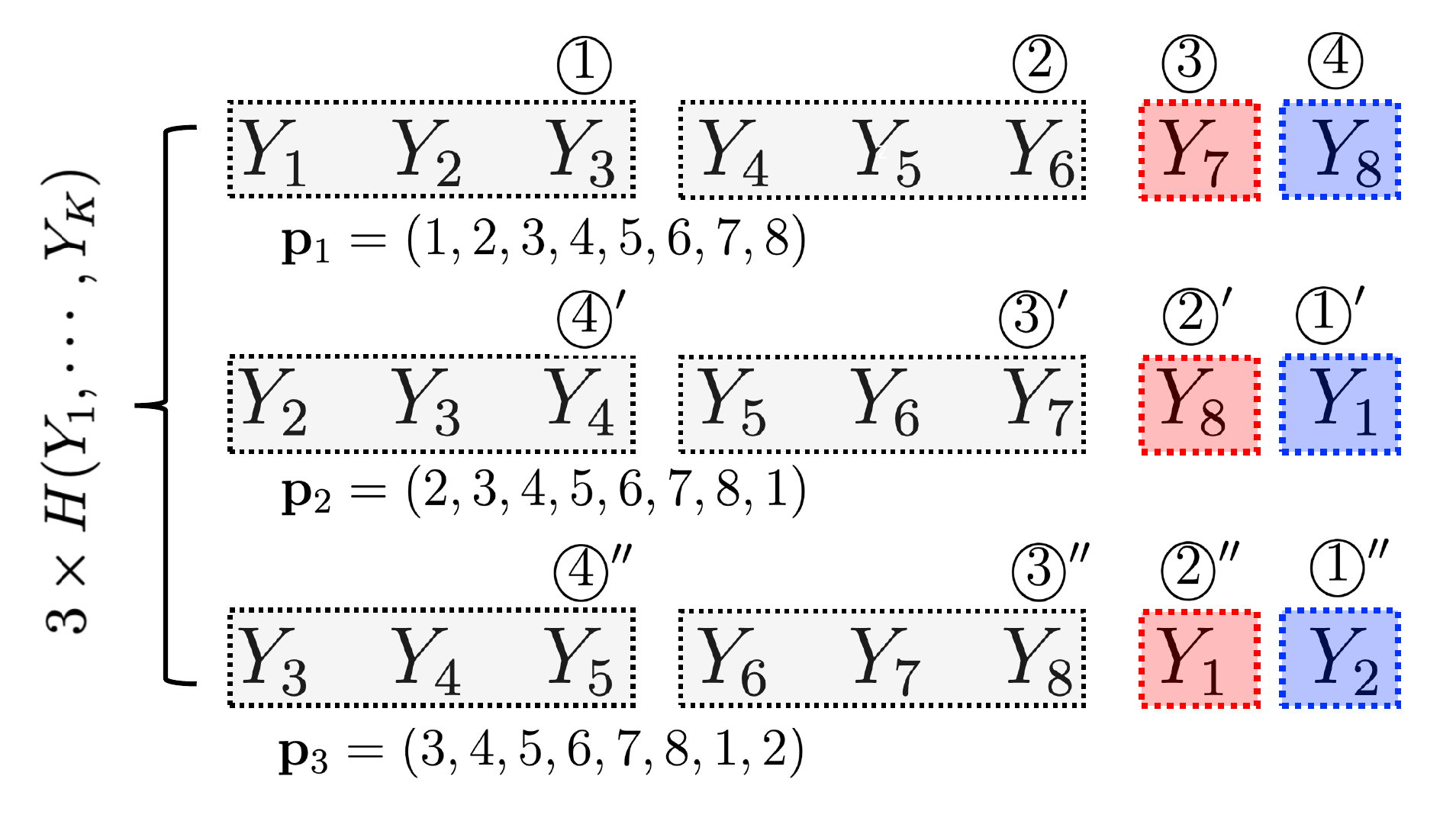}
    \caption{\small Message grouping under chain rule of conditional entropy, $(K,B)=(8,3)$.   }
    \label{fig: K=8,B=3 converse example}
\end{figure}
Consider $K=8$ and $B=3$. In Fig. \ref{fig: K=8,B=3 converse example}, each row represents a specific arrangement of the \msgs $Y_1, \cdots,Y_8$ according to the permutation vectors $\pv_1=(1,2,\cdots,8), \pv_2 =(2,3,\cdots,8,1)$ and $\pv_3=(3,4,\cdots, 8,1,2)$, respectively. These permutations represent 3 different ways to apply the chain rule of entropy to decompose $H(Y_1, \cdots,Y_8)$ into 4 smaller conditional entropy terms. The circled number above each \msg indicates its position in the ordering used for applying the chain rule.
\Ip, we write
\begin{align}
H(Y_{1:8}) & = H\left(Y_{\{1,2,3\}}\right) + H\left(Y_{\{4,5,6\}}|Y_{\{1,2,3\}}\right) + {H\left(Y_7|Y_{1:6}\right)}+ {H\left(Y_8|Y_{1:7}\right)},\label{eq: eq 1, converse example}\\
H(Y_{1:8}) & = {H(Y_1)} +  {H(Y_8|Y_1)} + H\left(Y_{\{5,6,7\}}|Y_{\{1,8\}}\right)+ H\left(Y_{\{2,3,4\}}|Y_{\{1,5,6,7,8\}}\right),\label{eq: eq 2, converse example}\\
H(Y_{1:8}) & =  {H(Y_2)} +  {H(Y_1|Y_2)} + H\left(Y_{\{6,7,8\}}|Y_{\{1,2\}}\right)+ H\left(Y_{\{3,4,5\}}|Y_{\{1,2,6,7,8\}}\right).\label{eq: eq 3, converse example}
\end{align}
{The key idea is that, \tit{by using the $3$ carefully designed permutations $\pv_1, \pv_2$ and $\pv_3$, each of the conditional entropy terms represented by the gray blocks  contributes one input (\ie, of size $L$), while the three red and blue blocks together contribute one additional input, \resp.}}  
Following (\ref{eq: proof of Rzsigma>=K/B-1, converse}), we have
\begin{subequations}
\label{eq: proof Rzsigma>=5/3, converse example}
\begin{align}
3H(\zsigma) & \ge 3H(Y_{1:8})-3L \\
& = (\ref{eq: eq 1, converse example}) + (\ref{eq: eq 2, converse example})+(\ref{eq: eq 3, converse example})-3L\\
& \ge H\lef(Y_{1,2,3}|\{W_k,Z_k\}_{k\in [8]\backslash \{1\}  }\rig) + H\left(Y_{\{4,5,6\}}|Y_{\{1,2,3\}},\{W_k,Z_k\}_{k\in [8]\backslash \{4\}   }\right)\notag\\
& \quad + H\left(Y_{\{5,6,7\}}|Y_{\{1,8\}}, 
\{W_k,Z_k\}_{k\in [8]\backslash \{5\}    }\right)+ H\left(Y_{\{2,3,4\}}|Y_{\{1,5,6,7,8\}}, \{W_k,Z_k\}_{k\in [8]\backslash \{2\}  }  \right)\notag\\
& \quad  + H\left(Y_{\{6,7,8\}}|Y_{\{1,2\}}, 
 \{W_k,Z_k\}_{k\in [8]\backslash \{6\}  }    \right)+ H\left(Y_{\{3,4,5\}}|Y_{\{1,2,6,7,8\}},\{W_k,Z_k\}_{k\in [8]\backslash \{3\}  }  \right)\notag \\
 & \quad +  {\left( H\left(Y_8|Y_{1:7}\right) + H(Y_1) + H(Y_2)\right) } + 
  { \left( H\left(Y_7|Y_{1:6}\right) + H(Y_8|Y_1) +H(Y_1|Y_2) \right)}-3L\\
 &  = \sum_{k\in [1:6]   } H\left( Y_{\Bc_k}| \{W_{k'},Z_{k'}\}_{k'\in [8]\backslash\{k\}  }  \right) +  {\left( H\left(Y_8|Y_{1:7}\right) + H(Y_1) + H(Y_2)\right)} \notag \\
 & \quad + 
 { \left( H\left(Y_7|Y_{1:6}\right) + H(Y_8|Y_1) +H(Y_1|Y_2) \right)}-3L\label{eq: step 0, proof Rzsigma>=5/3, converse example}\\
 & \ge \sum_{k\in  [1:6 ]  } H\left( Y_{\Bc_k}| \{W_{k'},Z_{k'}\}_{k'\in [8]\backslash\{k\}  }  \right) +
 H\left(Y_{\{8,1,2\}}| Y_{3:7} \right )+ 
  H\left(Y_{\{7,8,1\}}| Y_{2:6} \right )-3L\\
  & \ge \sum_{k\in  [1:6  ]  } H\left( Y_{\Bc_k}| \{W_{k'},Z_{k'}\}_{k'\in [8]\backslash\{k\}  }  \right) +
 H\left(Y_{\{8,1,2\}}| Y_{3:7}, \{W_k,Z_k\}_{k\in 
 [8]\backslash \{8\}  }      \right ) \notag\\
 & \quad + 
  H\left(Y_{\{7,8,1\}}| Y_{2:6} ,\{W_k,Z_k\}_{k\in 
 [8]\backslash \{7\}  }    \right )-3L\\
 & = \sum_{k\in  [1:8 ]   } H\left( Y_{\Bc_k}| \{W_{k'},Z_{k'}\}_{k'\in [8]\backslash\{k\}  }  \right)-3L\label{eq: step 1, proof Rzsigma>=5/3, converse example}\\
 & \overset{(\ref{eq: Y_i cutset bound, converse})}{\ge} 5L,\\
 & \Rightarrow \rzsigma \ge 5/3,
\end{align}
\end{subequations}
where (\ref{eq: step 0, proof Rzsigma>=5/3, converse example}) and (\ref{eq: step 1, proof Rzsigma>=5/3, converse example}) are because the $Y$ \msgs being conditioned in the relevant terms can be recovered by the corresponding set of inputs and keys. For example, $H\left(Y_{\{4,5,6\}}|Y_{\{1,2,3\}},\{W_k,Z_k\}_{k\in [8]\backslash \{4\}   }\right)$ is equal to $ H\left(Y_{\{4,5,6\}}|\{W_k,Z_k\}_{k\in [8]\backslash \{4\}   }\right) $ because $\{W_k,Z_k\}_{k\in [8]\backslash \{4\}   }\Rightarrow \{Y_1, Y_2,Y_3\}  $ since  user 4 is associated with relays $4,5$ and  $6$ and does not contributed to $Y_1, Y_2$ and $Y_3$. 
\hfill $\lozenge$
\end{example}

\subsection{Lower Bound on \Indiv Key Rate} 
\label{subsec: indiv key rate}
We prove $\rz\ge 1/B$ \af:
\begin{subequations}
\label{eq: proof of Rz>=1/B}
\begin{align}
KBL_Z & \ge \sum_{k=1}^K \sum_{i\in \Bc_k } H(Z_k)\\
& \overset{(\ref{eq: H(Zki,all i|Zk)=0})}{\ge}   \sum_{k=1}^K \sum_{i\in \Bc_k } H(Z_{k,i})\\
& = \sum_{k=1}^K \sum_{i\in \Uc_k } H(Z_{i,k})\label{eq: step 0, proof of Rz>=1/B}\\
&  \ge  \sum_{k=1}^KH\left(\{Z_{i,k}\}_{i\in \Uc_k }  \right)   \label{eq: step 1, proof of Rz>=1/B} \\
& \overset{\trm{(\ref{eq: step 0, proof of source key rate})-(\ref{eq: step 3, proof of source key rate})}  }{\ge} KL,\label{eq: step 2, proof of Rz>=1/B}\\
& \Rightarrow \rz \eqdef \frac{L_Z}{L} \ge \frac{1}{B},
\end{align}
\end{subequations}
where in (\ref{eq: step 0, proof of Rz>=1/B}) we changed the order  of the summation.
(\ref{eq: step 2, proof of Rz>=1/B}) followed
the same steps (\ref{eq: step 0, proof of source key rate})-(\ref{eq: step 3, proof of source key rate}) as in the proof of $ \rzsigma \ge 1$ (see Section~\ref{subsec: source key rate}).

\section{Discussion: Impossibility Proof of $\rz=1/2$ for $B=K=2$}
\label{sec:discussion}
\Thm~\ref{thm: main result} fully characterizes  the \optml rate region for all $B\le K-1$. When $B=K$, \ie, when each user is connected to all relays, and \achvb scheme with rates   $(\rx=1, \ry=1/(K-1), \rz=1/(K-1), \rzsigma=1)$ is provided. 
In Section~\ref{subsec: indiv key rate}, we derived a lower bound on the \indiv key rate, namely $ \rz \ge  1/B$, which is tight when $B\le K-1$.  However, for \(B=K\), it remains open whether the bound  \( \rz=1/K \) is \achvb. 
In this section, we demonstrate via the example $B=K=2$ that the rate tuple \( (\rx=1,\, \ry=1,\, \rz=1/2,\, \rzsigma=1) \) is, in fact, \emph{not} achievable.
This observation motivates the conjecture that  for any $B=K\ge 3$, the rate tuple  $(\rx=1, \ry=1/(K-1), \rz=1/K, \rzsigma=1)$ is likewise not \achvb.

\begin{lemma}
\label{lemma:impossibility for B=K=2}\emph{
For $B=K=2$, the rate tuple \( (\rx=1,\, \ry=1,\, \rz=1/2,\, \rzsigma=1) \) is not \achvb. }
\end{lemma}

\begin{IEEEproof}
We prove by contradiction. Suppose each input contains $L$ \iid uniform symbols. 
Let $W_{k,1}$ and $W_{k,2}$ (\resp, $Z_{k,1}$ and $Z_{k,2}$) denote the symbols of $W_k$ (\resp, $Z_k$) that appear---regardless of how they are encoded---in the \msgs $X_{k,1}$ and $X_{k,2}$, \resp. \Msp, we have
\begin{align}
\label{eq0:B=K=2 impossibility proof}
& H(W_{k,1}, W_{k,2}|W_k)=0, \notag \\
& H(Z_{k,1}, Z_{k,2}|Z_k)=0, \notag\\
& H(X_{k,i}|W_{k,i},Z_{k,i})=0, \;k,i=1,2\notag\\
& H(Y_i|X_{1,i}, X_{2,i})=0, i=1,2
\end{align}

Suppose $\rz=1/2$ is achievable, \ie, $H(Z_1)=H(Z_2) \le  \lz = \rz L=  L/2$. By \rsec (at both relays) and Shannon's one-time pad theorem, we know that
\begin{align}
\label{eq1:B=K=2 impossibility proof}
& H(W_{k,1}) \le  H(Z_k) \le L/2, \notag\\
& H(W_{k,2}) \le  H(Z_k) \le L/2,\notag\\
& \Rightarrow H(W_{k,1}) + H(W_{k,2}) \le  L,\; k=1,2.
\end{align}
By a simple cut-set bound from User $k$ to both relays, and ultimately to the server, we obtain $H(W_{k,1},W_{k,2}) \ge L$ as shown in the  following lemma:
\begin{lemma}
\label{lemma: H(Wk1,Wk2)>=L, B=K=2 impossibility proof}\emph{
It holds that
\be
\label{eq: H(Wk1,Wk2)>=L, B=K=2 impossibility proof}
H(W_{k,1},W_{k,2}) \ge H(W_k)=L, \; k=1,2
\ee}\end{lemma}

\begin{IEEEproof}
Denote $\bar{k}\eqdef [2]\bkslsh \{k\}, \forall k=1,2$. Then
\begin{subequations}
\begin{align}
H(W_{k,1},W_{k,2}) & \overset{(\ref{eq: key input independence})}{=} H(W_{k,1},W_{k,2}|W_{ \bar{k}  },Z_k, Z_{\bar{k}})\\
& \ge  I( W_{k,1},W_{k,2}; W_1 +W_2|  W_{ \bar{k}  },Z_k, Z_{\bar{k}}     ) \\
& \overset{(\ref{eq: key input independence})}{=} H(W_k) 
- H(W_1+W_2 | W_{ \bar{k}  },Z_k, Z_{\bar{k}}, W_{k,1},W_{k,2}     )\\
& \overset{  (\ref{eq: H((X_ki)_i|W_k,Z_k)=0}),(\ref{eq0:B=K=2 impossibility proof})  } {=}  H(W_k) -
H(W_1+W_2 | W_{ \bar{k}  },Z_k, Z_{\bar{k}}, W_{k,1},W_{k,2}, X_{k,1},X_{k,2}, X_{\bar{k},1},X_{\bar{k},2}    )\\
& \ge  H(W_k) -
H(W_1+W_2 | X_{k,1},X_{k,2}, X_{\bar{k},1},X_{\bar{k},2}    )\\
& \overset{(\ref{eq: H(Yi|(X_{ki})_k)=0})}{=}  H(W_k) -
H(W_1+W_2 | X_{k,1},X_{k,2}, X_{\bar{k},1},X_{\bar{k},2},Y_1,Y_2    )\\
& \ge H(W_k) -
H(W_1+W_2 |Y_1,Y_2    )\\
& \overset{(\ref{eq: recovery constraint})}{=} H(W_k)=L,
\end{align}
which implies $H(W_{k,1},W_{k,2}) \ge H(W_k), k=1,2$.
\end{subequations}
\end{IEEEproof}

Combining (\ref{eq1:B=K=2 impossibility proof}) and (\ref{eq: H(Wk1,Wk2)>=L, B=K=2 impossibility proof}), we have 
\be
\label{eq:temp0, discussion}
H(W_{k,1})=H(W_{k,2})=L/2,\; k=1,2
\ee
The equality in (\ref{eq:temp0, discussion}) is intuitive. On one hand, the key rate $\rz=1/2$ limits the maximum number of input symbols that can be protected from each relay to at most $L/2$; On the other hand, due to the input sum recovery requirement (\ref{eq: recovery constraint}) by the server, a total of at least $L = H(W_k)$ input symbols must be transmitted through the two outgoing links from User $k$. These two aspects together result in the equality in (\ref{eq:temp0, discussion}).

By \rsec, it holds that
\begin{align}
\label{eq2:B=K=2 impossibility proof}
& H(X_{k,i})  \le H(Z_{k,i})\le H(Z_k)=L/2,\;k,i=1,2
\end{align}
which is because, for example,
\begin{align}
H(X_{1,1}) &= H(X_{1,1}|W_{1}) + \underbrace{I(X_{1,1}; W_{1}) }_{=0\trm{ (\Rsec  at Relay 1)}}\notag \\
& = H(X_{1,1}|W_1,W_{1,1})\notag\\
& \le H(X_{1,1}|W_{1,1})\notag\\
& \le H(X_{1,1}, Z_{1,1}|W_{1,1})\notag\\
& =  H(Z_{1,1}|W_{1,1}) + \underbrace{H(X_{1,1}|W_{1,1}, Z_{1,1}  ) }_{ \overset{(\ref{eq0:B=K=2 impossibility proof})}{ =}0  
 }\notag\\
& = H(Z_{1,1}|W_{1,1})\notag\\
& \le H(Z_{1,1}) \le H(Z_1)=L/2.
\end{align}
Because
$H(X_{1,1}, X_{2,1}) \le H(X_{1,1}) + H(X_{2,1}) \overset{(\ref{eq2:B=K=2 impossibility proof})}{\le } H(Z_{1}) + H(Z_{2})=L $ and similarly $H(X_{1,2}, X_{2,2})\le L$, we have
\begin{align}
 H(X_{1,1}, X_{2,1}|Y_1) &= 
\underbrace{H(Y_1|X_{1,1}, X_{2,1})}_{\overset{(\ref{eq0:B=K=2 impossibility proof})}{=}0  }
+ H( X_{1,1}, X_{2,1}  ) - H(Y_1) \notag  \\
& = H( X_{1,1}, X_{2,1}  )  - H(Y_1) \overset{(\ref{eq2:B=K=2 impossibility proof})}{\le } L- H(Y_1),\label{eq3:B=K=2 impossibility proof}\\
 H(X_{1,2}, X_{2,2}|Y_2) & \le L-H(Y_2).\label{eq4:B=K=2 impossibility proof}
\end{align}
Also because $ \ryh \ge 1$, \ie, $H(Y_1) + H(Y_2)\ge 2L$ as shown in (\ref{eq: proof of Ry>=1/B, converse}), we have
\begin{subequations}
\begin{align}
 H(X_{1,1}, X_{2,1}|Y_1) +  H(X_{1,2}, X_{2,2}|Y_2) \overset{(\ref{eq3:B=K=2 impossibility proof}), (\ref{eq4:B=K=2 impossibility proof})}{\le } 2L - (H(Y_1) + H(Y_2))  \le   0, 
\end{align}
\end{subequations}
which implies 
\be 
\label{eq5:B=K=2 impossibility proof}
H(X_{1,k}, X_{2,k}|Y_k)=0,\;k=1,2.
\ee 
This, together with (\ref{eq0:B=K=2 impossibility proof}), implies that there exists a  one-to-one correspondence between $Y_i$ and $(X_{1,i},X_{2,i}),i=1,2$ \resp.
As a result, 
\begin{subequations}
\label{eq: contradiction proof, ss I>0}
\begin{align}
0 & \overset{(\ref{eq: server security constraint})}{=} I(Y_1, Y_2; W_1,W_2|W_1+W_2)\notag\\
& \overset{(\ref{eq5:B=K=2 impossibility proof})}{=}   { 
I(Y_1, X_{1,1}, Y_2, X_{1,2}; W_1,W_2|W_1+W_2)}\\
& \ge I( X_{1,1},  X_{1,2}; W_1|W_1+W_2)\\
& = H(W_1|W_1+W_2) - H(W_1|W_1+W_2, X_{1,1},  X_{1,2} )\\
& \ge  H(W_1) - H(W_1|X_{1,1},  X_{1,2} )\\
& = I(W_1; X_{1,1},  X_{1,2}  )\\
& = H( X_{1,1},  X_{1,2} ) -  H( X_{1,1},  X_{1,2}|W_1 )\\
&  \overset{(\ref{eq: {X_ij}_j cutset bound, converse})  }{\ge }    H(W_1 ) -  H( X_{1,1},  X_{1,2,}|W_1 )\label{eq: step 0, proof of impossibility, B=K=2} \\
& \ge  H(W_1 ) -  H( X_{1,1}, Z_{1,1},  X_{1,2}, Z_{1,2}|W_1 )\\
& = H(W_1 ) -  H(Z_{1,1}, Z_{1,2}|W_1 )- 
\underbrace{ H( X_{1,1},  X_{1,2}|W_1, Z_{1,1},Z_{1,2} ) }_{ \overset{(\ref{eq0:B=K=2 impossibility proof})}{=}0   }
\\
& \ge   H(W_1 ) - {H(Z_{1,1}, Z_{1,2})}
\\
& \overset{(\ref{eq0:B=K=2 impossibility proof})}{\ge } L -H(Z_1)\ge L/2,
\end{align}
\end{subequations}
where (\ref{eq: step 0, proof of impossibility, B=K=2}) is due to the cut-set bound on $W_1$ given in Lemma \ref{lemma: input cutset bound on X}. We reached a contradiction that $0\ge L/2$, which violates \ssec, indicating the rate tuple $(\rx=1, \ry=1, \rz=1/2, \rzsigma=1)$ is not achievable. Note that in the above impossibility proof, apart from the security constraints,  only the condition $\ry \ge 1$ was used; 
neither  $\rx=1$ nor $\rzsigma=1$ was invoked. This implies that relaxing $\rx$ or $\rzsigma$ (\ie, allowing higher rates) does not resolve the contradiction in (\ref{eq: contradiction proof, ss I>0}). Consequently, $\rz=1/2$ is not \achvb. 
\end{IEEEproof}

\section{Conclusion}
\label{sec:conclusion}
In this work, we studied \secagg  in a \hie network with cyclic wrap-around  association, where users connect to the \agg server via an intermediate layer of relays. 
A fundamental trade-off between the association number $B$ and the \comm and key rates was identified. Specifically, when $B\le  K-1$, we completely characterized the  optimal rate region, which shows that the optimal \rts \comm rate and the \indiv key rate are inversely proportional to $B$, the optimal \utr \comm  remains constant, and the source key rate also decreases with $B$ but is lower-bounded by one. Our contribution consists of a novel security key and \comm scheme design, as well as a matching converse proof.
Compared to the existing single-association HSA, allowing multiple associations enables each user to distribute its input across multiple relays, creating coding opportunities among the relays that enhance both communication and key generation rates.
Without the security  constraints, such coding opportunities have been explored in the context of communication-efficient gradient coding (GC)~\cite{ye2018communication}. Our results show that when $B\le K-1$, the same relay-to-server communication rate as in GC can be achieved under a non-trivial key design, implying that security does not incur additional communication overhead. 
However, when $B=K$, the optimal \rts \comm rate cannot be reduced below $\frac{1}{K-1}$---which is strictly higher than the $1/K$ rate in GC---indicating an extra cost incurred by enforcing security. 
To summarize, we characterized the optimal rate region for all $B\le K-1$ and proposed an achievable scheme for the case $B=K$. 

Several future directions can be considered: 1) More general association patterns, such as allowing each user to associate with an arbitrary number of relays—possibly subject to a minimum cardinality constraint, as studied in~\cite{11154963,lu2024capacity}. 
{Arbitrary \assocn patterns  introduce asymmetric input access across  relays, rendering  the coding  at the relay layer more challenging.}
2) Incorporating collusion resilience to address scenarios where a subset of users may collude with the relays and/or the server to compromise security. More generally, collusion among relays and the server can also be considered. {Such scenarios might be challenging due to the intricate coding at the relay layer, which makes it less obvious how collusion affects the optimal  key rates.}

\appendices

{

\section{Proof of $ \widehat{\bm{\theta}}^{(k)}_1(j)\ne 0,\forall j\in[B], k\in [K]$  }
\label{sec:proof of lambdaUkj not zero, appendix}
\begin{lemma}
\label{lemma:proof of lambdaUkj not zero, appendix}
\tit{
Let $\mbfTheta$ and $\mbfTheta_{\Uc_k}$ be defined in (\ref{eq: def Theta matrix, theta^i}) and (\ref{eq: ThetaUk, scheme 2}), \resp, where $\theta_1 , \cdots,\theta_K$ are chosen as $K$ nonzero and distinct \eles from $\Fmbbq$. Denote the inverse of $\mbfTheta_{\Uc_k}$ as  $ \mbf{\Theta}_{\Uc_k}^{-1}\eqdef \big[ \widehat{\bm{\theta}}^{(k)}_1; \cdots; \widehat{\bm{\theta}}^{(k)}_B \big ]    $.
Then $\widehat{\bm{\theta}}^{(k)}_1(j)\ne 0, \forall j\in  [B] ,\forall k\in[K]$.
}
\end{lemma}

\begin{IEEEproof}
$\mbfTheta_{\Uc_k}$ is a $B\times B$
 \Vander matrix built from the $B$ evaluation points $ \{ \theta_{i_1}, \cdots, \theta_{i_B} \} $: 
\be
\renewcommand{\arraystretch}{0.8}
\mbf{\Theta}_{\Uc_k} \eqdef 
\begin{bmatrix}
1  & \theta_{i_1} & \cdots &  \theta_{i_1}^{B-1}\\
\vdots  & \vdots & \ddots &  \vdots\\
1  & \theta_{i_B} & \cdots &  \theta_{i_B}^{B-1} 
\end{bmatrix}\in  \mbb{F}_q^{B\times B}.\notag 
\ee
Consider the  Lagrange basis \polys $L_j(x)$, 
\be 
L_j(x) =\prod_{m \in  [B]\bkslsh \{j\}  }
\frac{x- \theta_{i_m}}{ \theta_{i_j} - \theta_{i_m}},
\ee
the \coeffts (in the monomials $1,x, \cdots, x^{B-1}$) of which form the \jth column of $ \mbf{\Theta}_{\Uc_k}^{-1}$. The constant term  of $L_j(x)$, \ie, $L_j(0)$, equals the $(1,j)^{\rm th}$ entry of $ \mbf{\Theta}_{\Uc_k}^{-1}$, which is exactly $\widehat{\bm{\theta}}^{(k)}_1(j)$:
\be 
\widehat{\bm{\theta}}^{(k)}_1(j) = L_j(0)=\prod_{m \in  [B]\bkslsh \{j\}  }
\frac{-\theta_{i_m}}{ \theta_{i_j} - \theta_{i_m}},
\ee 
which is nonzero because $\theta_1, \cdots, \theta_K$ are all nonzero and distinct.
\end{IEEEproof}

\section{Proof of Lemma \ref{lemma:existence of bmLambda and H,scheme 1}}
\label{sec:proof of lemma:existence of bmLambda and H,scheme 1,appendix}

We show  that there exists a proper $g\in  \Fmbbqnozero$---as long as  $q$ is sufficiently large---\suth  $\bmLambda_g$ is invertible, and $\Hm_g =(\bmLambda_g)^{-1}\Qm$ has the MDS property.

\subsubsection{Proof of $\exists g$ such that $\bmLambda_g$ is invertible}
Given $\bmLambda_g$ in (\ref{eq:bmLambda=circ(q1,...,qB), scheme 1}), let the circulant \poly  be defined as
\be
p_g(x) \eqdef \sum_{j=0}^{B-1} g^j x^j.
\ee 
Then the eigenvalues of $\bmLambda_g$ are $p_g(\omega^\ell), \ell  \in [0:K-1]$.
Hence, $\bmLambda_g$ being invertible is equivalent to $p_g(\omega^\ell)\ne 0,\forall  \ell$.
Using the geometric-sum identity, $p_g(\omega^\ell)$ is equal to  $ \frac{ 1 -(g\omega^\ell)^B  }{1- g\omega^\ell}     $ if $g\omega^\ell\ne 1$, and $B$ if $g\omega^\ell=1$.
It suffices to choose $g$ \suth 
\be
\label{eq:g two conditions, Lambdag invertible, appendix}
g\omega^\ell\ne 1,  \quad (g\omega^\ell)^B \ne 1, \; \forall  \ell \in[0:K-1]
\ee 

The first condition $g\omega^\ell\ne 1, \forall \ell $ holds whenever $g^K\ne 1$. We prove by contradiction: assume $g^K\ne 1$ and $ \exists \ell_0  \in  [0:K-1]$ \suth $g\omega^{\ell_0}=1$. Then $g= \omega^{-\ell_0}$ and thus $g^K=(\omega^{-\ell_0})^K=(\omega^K)^{-\ell_0}=1^{-\ell_0}=1$  since $\omega$ is a primitive $K$-th root of unity. This contradicts our assumption that $g^K\ne 1$. \Thf, $g^K \ne  1 \Rightarrow  g\omega^\ell\ne 1, \forall \ell $. 
In addition, such a $g$ exists in $ \Fmbbq $ whenever $q>K+1$. The proof is \af. First note that  $g^K  =1  \Leftrightarrow g\in \langle\omega \rangle \eqdef \{ \omega^\ell : \ell \in[0:K-1]  \} $ where $\langle\omega \rangle,|\langle\omega \rangle| =K $ denotes the order-$K$ cyclic subgroup generated by $\omega$. Such an  equivalence is proved \af:
\begin{itemize}
    \item ($\Leftarrow$) If $ g\in \langle\omega \rangle $, then $g=\omega^\ell $  for some $ \ell  \in [0:K-1]$, hence $g^K=(\omega^\ell)^K=(\omega^K)^\ell=1$;

    \item ($\Rightarrow$) If $g^K=1$, then  the order $\ord(g)$ divides $K$. In a cyclic  group (here $  \Fmbbqnozero$), every element whose order divides $K$ lies in the unique subgroup of order  $K$, namely $\langle\omega \rangle$. Thus $g\in  \langle\omega \rangle$.  
\end{itemize}
\Thf, $ g^K  \ne  1$ implies $ g\notin  \langle\omega \rangle$.  \Aar, there exists $g\in  \Fmbbqnozero$ with $g^K\ne 1$ iff $ \Fmbbqnozero$ has at least one \ele outside $\langle\omega \rangle$, \ie, 
$
|\Fmbbqnozero|>|\langle\omega \rangle| \Leftrightarrow q> K+1
$.

The second condition in (\ref{eq:g two conditions, Lambdag invertible, appendix}) is  \equi to  $g^B \ne \omega^{-\ell B},\forall  \ell $, \ie, $g$ avoids the size-$\frac{K}{\gcd(K,B)}  $ set $\{  \omega^{-\ell B}:\forall  \ell  \}$.
\Thf, the  set of $g\in \Fmbbqnozero$ violating 
(\ref{eq:g two conditions, Lambdag invertible, appendix}) is finite. \Ip, note that $g^B=\omega^{-\ell B}$ has at most 
$\gcd(B, q-1)$ roots. Denoting
$ \Mc \eqdef \lef\{g  \in \Fmbbqnozero:\exists \ell \in [0:K-1], (g\omega^\ell)^B=1  \rig\}     $ as the bad set, we have
\be
\label{eq: Mc size upper bound,scheme 1,appendix}
|\Mc|
\le  \gcd(B,q-1)\cdot  \frac{K}{\gcd(K,B)} \le KB.
\ee
Combining both conditions, the number of bad $g$'s that render $\bmLambda_g$ noninvertible is limited  in size, \ie,
$|\Mc \cup \langle \omega \rangle|\le K(B+1)$.
\Thf, for sufficiently large $q$, there exists $g\in \Fmbbqnozero$ satisfying (\ref{eq:g two conditions, Lambdag invertible, appendix}).

\subsubsection{Proof of $\exists g$ such that $\Hm_g$ has the MDS property}
Suppose $\bmLambda_g$  is \invtb and $\Hm_g =(\bmLambda_g)^{-1}\Qm$. We show that there exists infinitely many $g$ (as $q\to \infty$) \suth every $(K-B) \times (K-B)$ submatrix of  $ \Hm_g$ has full rank.

For each $\Sc \in \binom{[K]}{K-B} $, define the minor determinant  
$\Delta_\Sc(g) \eqdef \det\lef(\Hm_{g, \Sc,:}  \rig)$.
Using the adjugate identity, $ (\bmLambda_g^T)^{-1} ={\adj(\bmLambda_g^T)}/{\det(\bmLambda_g)}  $. Hence, $ \Hm_g = \frac{1}{\det(\bmLambda_g)} \adj(\bmLambda_g^T)\Qm $ and 
$ \Hm_{g,\Sc,:} = \frac{1}{\det(\bmLambda_g)} \lef(\adj(\bmLambda_g^T)\Qm\rig)_{\Sc,:} $.
The minor determinant is thus equal  to 
$\Delta_\Sc(g) =\det(\bmLambda_g)^{-(K-B)}\cdot 
\det\lef( \big(\adj(\bmLambda_g^T)\Qm\rig)_{\Sc,:} \big)$.
Define the denominator-cleared \poly as
\be 
P_\Sc(g) \eqdef \det(\bmLambda_g)^{(K-B)}\Delta_\Sc(g) =
\det\lef( \lef(\adj(\bmLambda_g^T)\Qm\rig)_{\Sc,:} \rig).
\ee 
Then, whenever $\det(\bmLambda_g)\ne 0$, $\Delta_\Sc(g) \ne0 \Leftrightarrow P_\Sc(g)\ne 0    $.
The degree of $P_\Sc(g)$ is bounded by $\deg(P_\Sc(g)) \le (K-B)(K-1)(B-1)$, which is proved \af. First note that 1) each entry of $\bmLambda_g$ has degree at most $B-1$ (in $g$) since the entry is either $0$ or $g^t$ with $0\le t \le B-1$, and 2) any $(K-1)\times (K-1)$ minor of $\bmLambda_g$ has determinant degree at most $(K-1)(B-1)$. \Thf, each entry of $\adj(\bmLambda_g^T)$ has degree at most $(K-1)(B-1)$. Since $\Qm$ is constant in $g$, each entry of $\adj(\bmLambda_g^T)\Qm$  still has degree at most $(K-1)(B-1)$. Finally, \bcuz  $P_\Sc(g)$ is an $(K-B)\times (K-B)$ determinant, we have
\be
\label{eq:deg Ps(g) upper bound, scheme 1, appendix}
\deg(P_\Sc(g)) \le  (K-B)(K-1)(B-1), \;\forall  \Sc  \in  \binom{[K]}{K-B}
\ee 
Further define the product \poly that involves  determinants of all minors $
P(g) \eqdef \prod_{ \Sc \in  \binom{[K]}{K-B}  }
P_\Sc(g)$,
which has degree at most 
\be 
\label{eq:deg P(g) upper bound, scheme 1,appendix}
\deg(P(g)) \le \binom{K}{B} (K-B)(K-1)(B-1).
\ee 
Note that $P(g)\ne 0$ implies that every $(K-B)\times (K-B)$ submatrix of $\Hm_g$ has full rank. 
Since $P(g)\equiv 0$, $P(g)=0$ has at most $\deg(P(g))$ roots in $\Fmbbq$. Avoiding these roots ensures the MDS property  of $\Hm_g$.

Taking the union bound on the  bad sets of $g$ for both $\bmLambda_g$ and $\Hm_g$ (see (\ref{eq: Mc size upper bound,scheme 1,appendix}) and (\ref{eq:deg P(g) upper bound, scheme 1,appendix})), we conclude that there are at most  
$|\Mc| + \deg(P(g))\le \binom{K}{B} (K-B)(K-1)(B-1)  $ bad points of $g\in  \Fmbbqnozero$.
Define $ \varphi(K,B) \eqdef \binom{K}{B}(K-B)(K-1)(B-1)+BK+2$.
Then  $q-1 \ge  \varphi(K,B)-1   $, \ie, $q \ge \varphi(K,B)$ suffices to guarantee the existence of at least one valid $g$, and the probability that a randomly drawn $g\in  \Fmbbq$ is valid is at least $ 1- \frac{\varphi(K,B)}{q}$. This completes the proof of Lemma \ref{lemma:existence of bmLambda and H,scheme 1}.

} 

\bibliographystyle{IEEEtran}
\bibliography{references_secagg.bib}
\end{document}

%% file: macros.tex
\long\def\comment#1{}

\newtheorem{example}{Example}
\newtheorem{theorem}{Theorem}
\newtheorem{lemma}{Lemma}
\newtheorem{remark}{Remark}

\DeclareMathOperator{\modular}{mod}
\DeclareMathOperator{\rank}{rank}  
\DeclareMathOperator{\nulll}{Null}

\DeclareMathOperator{\spann}{Span}

\DeclareMathOperator{\ord}{ord}
\DeclareMathOperator{\adj}{adj}

\def \mbfTheta{\mathbf{\Theta}}

\def \bmgamma{\bm{\gamma}}
\def \bmlambda{\bm{\lambda}}
\def \bmLambda{\bm{\Lambda}}
\def \rzsigma{{R_{Z_{\Sigma}}}}
\def \rzsigmastar{{R_{Z_{\Sigma}}^*}}
\def \rx{{R_X}} 
\def \ry{{R_Y}}
\def \rz{{R_Z}} 
\def \rxh{{R_X^{H}}} 
\def \ryh{{R_Y^{H}}}
\def \rzh{{R_Z^{H}}} 
\def \rzsigmah{{R_{Z_{\Sigma}}^H}}
\def \lz{{L_Z}} 

\def \rxstar{{R_X^*}} 
\def \rystar{{R_Y^*}}
\def \rzstar{{R_Z^*}}
\def  \zsigma{{Z_{\Sigma}}}
\def  \lzsigma{{L_{Z_{\Sigma}}}}
\def  \lzsigmastar{{L_{Z_{\Sigma}}^*}}

\let\bkslsh\backslash
\let\tbf\textbf
\let\tit\textit

\let\mbf\mathbf
\let\mbb\mathbb

\let\trm\textrm
\let \mrm \mathrm

\def \Fmbb{{\mbb{F}}}
\def \Fmbbq{{\mbb{F}_q}}
\def \Fmbbqnozero{{\mbb{F}_q\backslash \{0\}  }}

\newcommand{\arbily}{arbitrarily\xspace}

\newcommand{\invtb}{invertible\xspace}
\newcommand{\equi}{equivalent\xspace}

\newcommand{\bcuz}{because\xspace}
\newcommand{\nece}{necessary\xspace}
\newcommand{\eval}{evaluation\xspace}
\newcommand{\Thf}{Therefore\xspace}
\newcommand{\coord}{coordinate\xspace}
\newcommand{\eles}{elements\xspace}
\newcommand{\ele}{element\xspace}

\newcommand{\nz}{nonzero\xspace}
\newcommand{\compos}{components\xspace}

\newcommand{\Bef}{Before\xspace}

\newcommand{\secty}{security\xspace}
\newcommand{\soth}{so that\xspace}
\newcommand{\suth}{such that\xspace}
\newcommand{\Vander}{Vandermonde\xspace}

\newcommand{\awa}{as well as\xspace}

\newcommand{\crspdg}{corresponding\xspace}

\newcommand{\lcb}{linear combination\xspace}

\newcommand{\reqs}{requirements\xspace}
\newcommand{\req}{requirement\xspace}

\newcommand{\het}{heterogeneous\xspace}

\newcommand{\skr}{source key rate\xspace}
\newcommand{\polys}{polynomials\xspace}
\newcommand{\poly}{polynomial\xspace}

\newcommand{\genn}{generation\xspace}

\newcommand{\coeffts}{coefficients\xspace}
\newcommand{\coefft}{coefficient\xspace}

\newcommand{\assocn}{association\xspace}
\newcommand{\Assocn}{Association\xspace}
\newcommand{\assoced}{associated\xspace}

\newcommand{\optml}{optimal\xspace}

\newcommand{\eg}{e.g.\xspace}
\newcommand{\utr}{user-to-relay\xspace}
\newcommand{\rts}{relay-to-server\xspace}

\newcommand{\ie}{i.e.\xspace}
\newcommand{\msg}{message\xspace}
\newcommand{\msgs}{messages\xspace}

\newcommand{\Hie}{Hierarchical\xspace}
\newcommand{\hie}{hierarchical\xspace}
\newcommand{\rsec}{relay security\xspace}
\newcommand{\Rsec}{Relay security\xspace}
\newcommand{\ssec}{server security\xspace}
\newcommand{\Ssec}{Server security\xspace}
\newcommand{\Msp}{More specifically\xspace}
\newcommand{\Ip}{In particular\xspace}

\newcommand{\af}{as follows\xspace}

\newcommand{\Aar}{As a result\xspace}
\newcommand{\resp}{respectively\xspace}
\newcommand{\iid}{i.i.d.\xspace}

\newcommand{\Thm}{Theorem\xspace}

\newcommand{\info}{information\xspace}

\newcommand{\itic}{information-theoretic\xspace}
\newcommand{\Itic}{Information-theoretic\xspace}
\newcommand{\ITic}{Information-Theoretic\xspace}

\newcommand{\simuly}{simultaneously\xspace}
\newcommand{\etal}{\textit{et al.}\xspace}

\newcommand{\agg}{aggregation\xspace}
\newcommand{\secagg}{secure aggregation\xspace}
\newcommand{\Secagg}{Secure aggregation\xspace}

\newcommand{\Agg}{Aggregation\xspace}

\newcommand{\diff}{different\xspace}

\newcommand{\indep}{independent\xspace}

\newcommand{\indeply}{independently\xspace}

\newcommand{\indiv}{individual\xspace}
\newcommand{\Indiv}{Individual\xspace}
\newcommand{\comm}{communication\xspace}
\newcommand{\Comm}{Communication\xspace}

\newcommand{\achvb}{achievable\xspace}

\newcommand{\distn}{distribution\xspace}
\newcommand{\Distn}{Distribution\xspace}

\newfont{\bbb}{msbm10 scaled 700}

\newfont{\bb}{msbm10 scaled 1100}

\newcommand{\av}{{\mathbf{a}}}

\newcommand{\dv}{{\mathbf{d}}} 
\newcommand{\ev}{{\mathbf{e}}}

\newcommand{\hv}{{\mathbf{h}}}

\newcommand{\pv}{{\mathbf{p}}} 

\newcommand{\rv}{{\mathbf{r}}}

\newcommand{\wv}{{\mathbf{w}}}

\newcommand{\ith}{{$i^{\rm th}$ }}
\newcommand{\jth}{{$j^{\rm th}$ }}
\newcommand{\kth}{{$k^{\rm th}$ }}

\newcommand{\Am}{{\bf A}}
\newcommand{\Bm}{{\bf B}}

\newcommand{\Hm}{{\bf H}}

\newcommand{\Qm}{{\bf Q}}
\newcommand{\Rm}{{\bf R}}

\newcommand{\Ac}{{\cal A}}
\newcommand{\Bc}{{\cal B}}

\newcommand{\Gc}{{\cal G}}
\newcommand{\Hc}{{\cal H}}

\newcommand{\Mc}{{\cal M}}

\newcommand{\Rc}{{\cal R}}
\newcommand{\Sc}{{\cal S}}

\newcommand{\Uc}{{\cal U}}

\newcommand{\eqdef}{\stackrel{\Delta}{=}}

\newcommand{\be}{\begin{equation}}
\newcommand{\ee}{\end{equation}}
\newcommand{\bea}{\begin{eqnarray}}
\newcommand{\eea}{\end{eqnarray}}



%% file: author_journal.tex
\author{
Xiang~Zhang,~\IEEEmembership{Member,~IEEE},
Zhou Li,~\IEEEmembership{Member,~IEEE},
Kai Wan,~\IEEEmembership{Member,~IEEE}, Hua Sun,~\IEEEmembership{Member,~IEEE},
Mingyue Ji,~\IEEEmembership{Member,~IEEE},
and Giuseppe Caire,~\IEEEmembership{Fellow,~IEEE}

\thanks{Part of this work was presented in the 2025 IEEE International Symposium on Information Theory, Ann Arbor, Michigan, USA.  }

\thanks{X. Zhang and G. Caire are with the Department of Electrical Engineering and Computer Science, Technical University of Berlin, 10623 Berlin, Germany (e-mail: \{xiang.zhang, caire\}@tu-berlin.de).
}

\thanks{Z. Li is with the School of Computer, Electronics and Information, 
Guangxi University, Nanning 530004, China (e-mail: lizhou@gxu.edu.cn).}

\thanks{
K. Wan is with the School of Electronic Information and Communications,
Huazhong University of Science and Technology, Wuhan 430074, China
(e-mail: kai\_wan@hust.edu.cn).}

\thanks{
H. Sun is with the Department of Electrical Engineering, University of North Texas, Denton, TX 76207 USA (e-mail: hua.sun@unt.edu).  }

\thanks{
M. Ji is with the Department of Electrical and Computer Engineering, University of Florida, Gainesville, FL 32611, USA
(e-mail: mingyueji@ufl.edu).}
}